%% file: main.tex
\documentclass[acmtog,nonacm]{acmart}
\usepackage[utf8]{inputenc}

\acmSubmissionID{papers\_552}
\usepackage[utf8]{inputenc}
\usepackage[ruled,vlined,linesnumbered]{algorithm2e}   % For pseudo-code
\usepackage{listings}                           % For code
\usepackage{newtxmath}                          % For blackboard letters
\usepackage {enumitem}
\usepackage{tikz}
\newlength{\unit}
\usepackage{overpic}
\newcommand*{\Scale}[2][4]{\scalebox{#1}{$#2$}}%
\DeclareMathOperator*{\argmin}{arg\,min}

\usepackage{tikz}
\usepackage{pgfplots}

\author{Abdalla G. M. Ahmed}
\orcid{0000-0002-2348-6897}
\affiliation{%
  \institution{KAUST}
  \country{KSA}}
\email{abdalla_gafar@hotmail.com}

\author{Jing Ren}
\orcid{0000-0003-3114-3517}
\affiliation{%
  \institution{KAUST}
  \country{KSA}}
\email{jing.ren@kaust.edu.sa}

\author{Peter Wonka}
\orcid{0000-0003-0627-9746}
\affiliation{%
  \institution{KAUST}
  \country{KSA}}
\email{pwonka@gmail.com}

\setcopyright{none}

\title{Gaussian Blue Noise}

\date{June 2022}

\begin{abstract}
Among the various approaches for producing point distributions with blue noise spectrum, we argue for an optimization framework using Gaussian kernels.
We show that with a wise selection of optimization parameters, this approach attains unprecedented quality, provably surpassing the current state of the art attained by the optimal transport (BNOT) approach.
Further, we show that our algorithm scales smoothly and feasibly to high dimensions while maintaining the same quality, realizing unprecedented high-quality high-dimensional blue noise sets.
Finally, we show an extension to adaptive sampling.
\end{abstract}

\begin{teaserfigure}
  {\scriptsize
  \begin{tabular*}{1\textwidth}{@{}c@{\extracolsep{\fill}}c@{\extracolsep{\fill}}c@{\extracolsep{\fill}}c@{\extracolsep{\fill}}c@{\extracolsep{\fill}}c@{\extracolsep{\fill}}c@{}}
  \centering
  \begin{tikzpicture}
    \node[inner sep=0, anchor=south west] at (0, 1.05\unit) {
      \includegraphics[height=1\unit]{images/teaser/bnot-points.pdf}
    };
    \node[inner sep=0, anchor=south west] at (0, 0) {
      \includegraphics[height=1\unit]{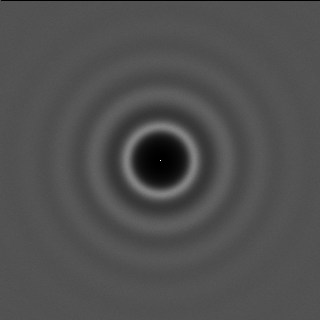}
    };
  \end{tikzpicture}&
  \begin{tikzpicture}
    \node[inner sep=0, anchor=south west] at (0, 1.05\unit) {
      \includegraphics[height=1\unit]{images/teaser/kdm-points.pdf}
    };
    \node[inner sep=0, anchor=south west] at (0, 0) {
      \includegraphics[height=1\unit]{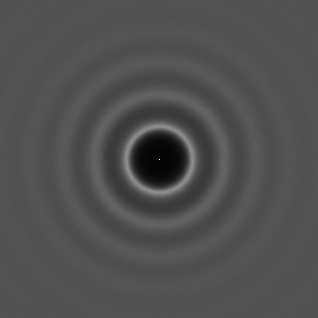}
    };
  \end{tikzpicture}&
  \begin{tikzpicture}
    \node[inner sep=0, anchor=south west] at (0, 1.05\unit) {
      \includegraphics[height=1\unit]{images/teaser/vnc-points.pdf}
    };
    \node[inner sep=0, anchor=south west] at (0, 0) {
      \includegraphics[height=1\unit]{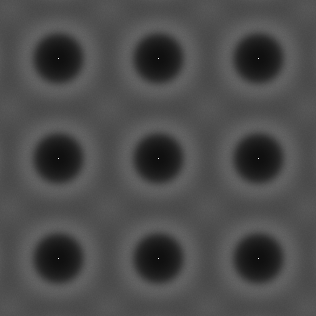}
    };
  \end{tikzpicture}&
  \begin{tikzpicture}
    \node[inner sep=0, anchor=south west] at (0, 1.05\unit) {
      \includegraphics[height=1\unit]{images/teaser/fpo-points.pdf}
    };
    \node[inner sep=0, anchor=south west] at (0, 0) {
      \includegraphics[height=1\unit]{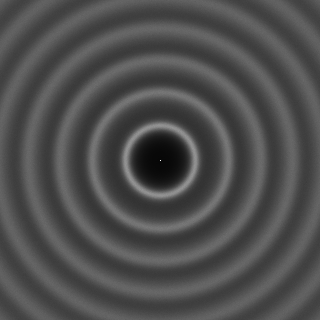}
    };
  \end{tikzpicture}&
  \begin{tikzpicture}
    \node[inner sep=0, anchor=south west] at (0, 1.05\unit) {
      \includegraphics[height=1\unit]{images/teaser/bluenets-points.pdf}
    };
    \node[inner sep=0, anchor=south west] at (0, 0) {
      \includegraphics[height=1\unit]{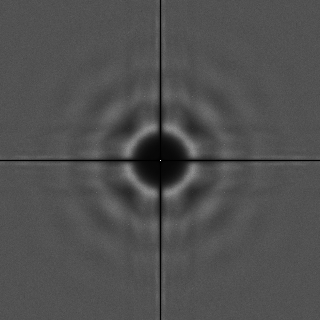}
    };
  \end{tikzpicture}&
  \begin{tikzpicture}
    \node[inner sep=0, anchor=south west] at (0, 1.05\unit) {
      \includegraphics[height=1\unit]{images/teaser/theta-points.pdf}
    };
    \node[inner sep=0, anchor=south west] at (0, 0) {
      \includegraphics[height=1\unit]{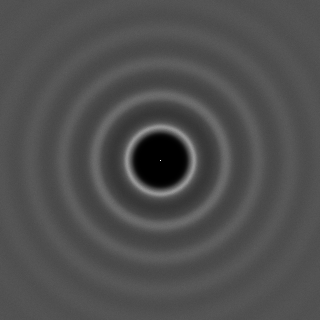}
    };
  \end{tikzpicture}&
  \includegraphics[height=2.05\unit]{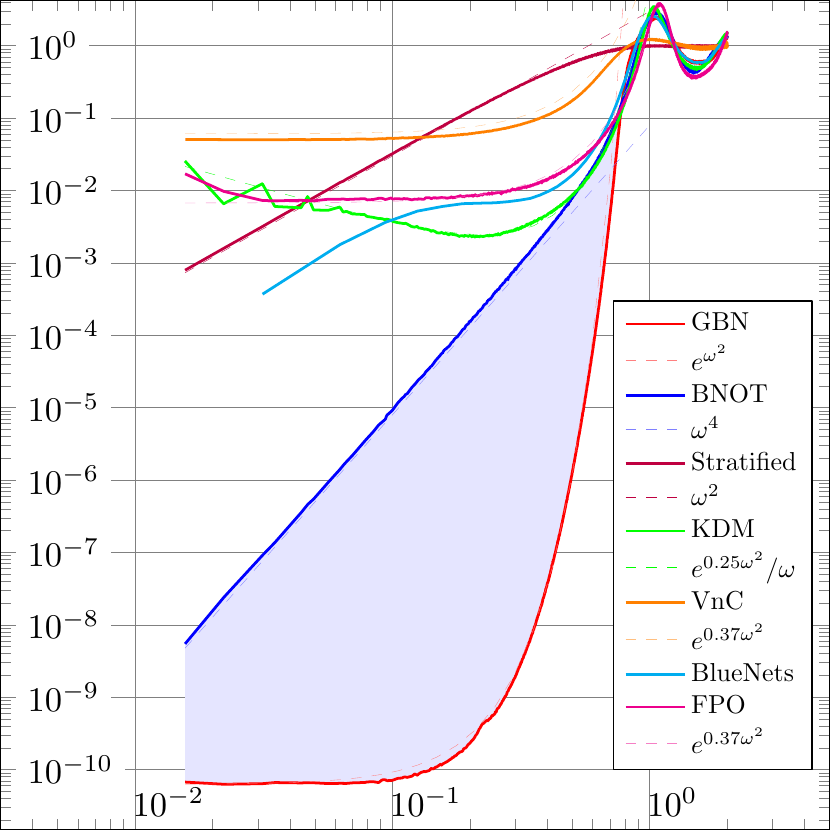}\\[1mm]
    (a) BNOT & (b) KDM & (c) VnC & (d) FPO & (e) BlueNets & (f) GBN (Ours) & (g) Spectral Profiles
  \end{tabular*}}\vspace{-9pt}
  \caption{\label{fig:teaser}%
    Point distributions (1k sets) and power spectra (1k realizations) of different blue-noise optimization techniques including:
      (a) BNOT \protect{\cite{Goes12Blue}},
      (b) KDM \protect{\cite{Fattal2011}}, 
      (c) VnC \protect{\cite{Ulichney93Void}},
      (d) FPO \protect{\cite{Schloemer2011}},
      (e) BlueNets \protect{\cite{Ahmed2021Optimizing}}, and
      (f) GBN (Ours).
      While the frequency power spectral plots may look similar, a closeup view reveals that the low-frequency energy in the middle is perfectly black in GBN, indicating very low low-frequency noise.
      (g) A log-log plot of the radial power spectra, using 1000 realizations of 4k point sets.
      Stratified sets are also included for comparison.
      We may distinguish two families of spectral profiles: polynomial (BNOT and Stratified), appearing as straight lines in the log-log scale, and exponential for GBN, VNC, KDM, and notably FPO.
  }
\end{teaserfigure}

\begin{document}

\maketitle

%%%%%%%%%%%%%%%%%%%%%%%%%%%%%%%%%%%%%%%%%%%%%%%%%%%%%%%%%%%%%%%%%%%%%%%%%%%%%
%%%%%%%%%%%%%%%%%%%%%%%%%%%%%%%%%%%%%%%%%%%%%%%%%%%%%%%%%%%%%%%%%%%%%%%%%%%%%
%%%%%%%%%%%%%%%%%%%%%%%%%%%%%%%%%%%%%%%%%%%%%%%%%%%%%%%%%%%%%%%%%%%%%%%%%%%%%

\section{Introduction}

Point sampling is an essential process in computer graphics (CG) that is employed in areas such as rendering, machine-learning, simulation, geometry processing, halftoning, and stippling.

Regular grid-based sampling is known to suffer quickly in high dimensions \cite{Kuipers74Uniform}, but even in the 2D plane that characterizes graphical images, it was established early in the works of Dipp{\'e} and Wold \shortcite{Dippe85Antialiasing} and Cook \shortcite{Cook86Stochastic} that regular sampling leads to excessive aliasing artifacts in rendering, while a Poisson (random) distribution of samples leads to excessive amounts of noise.
Hence, Poisson-disk sampling emerged as a reasonable compromise, where a minimal spacing, known as the \emph{conflict} or \emph{Poisson-disk} radius $r_\mathrm{f}$, is required between otherwise random samples.
Jittering of the regular grid, also known as stratification, was suggested as a cheaper alternative.
In the context of halftoning, Ulichney \shortcite{Ulichney88Dithering} concluded the superiority of the more-or-less same distribution, for which he coined the name \emph{blue noise} (BN) to describe a distribution of samples whose frequency power spectrum is characterised by a low-energy low-frequency band, a sharp transition towards a small peak (corresponding to $r_\mathrm{f}$), followed by a flat spectrum in higher frequencies.

The loose definition of blue noise, and the lack of a deterministic mean for generating it, lead to a substantial amount of literature devoted to describing algorithms for approximating the blue-noise distribution, as well as means for evaluating its quality, as we will detail in the following Section~\ref{sec:related work}.
The very nature of blue noise, however, remained unclear, lacking a mathematical theory to characterize this important distribution.

Among the wide variety of blue noise generation algorithms, we may broadly identify two distinct approaches: cellular and kernel-based.
Cellular methods are based on a divide-and-conquer principle that assigns one-and-only sample point to represent a partition of the domain, while kernel-based techniques try to maintain a uniform density of kernels placed at the sample point.
For a long time, cellular techniques, flag-shipped by BNOT \cite{Goes12Blue}, were considered the reference methods, while kernel-based techniques were thought of as just an alternative mean of achieving comparable results.
Indeed, in addition to its noticeable higher quality, BNOT also offers a plausible theoretical justification rooted in least-square fitting, in contrast to the heuristic motivation in known kernel-based methods.

Quite recently, Ahmed and Wonka \shortcite{Ahmed2021Optimizing} presented a formulation of a loss function for Gaussian-kernel-based blue noise, and used it for optimization in a specific discretized context, namely dyadic nets.
In this paper we extend this approach to continuous domains and high-dimensional spaces, and show that, with the appropriate design choices, kernel-based methods outperform cellular methods in terms of scalability in dimensions, coding complexity, and even quality.
The key insight is that cellular methods work by modeling interactions between points and their immediate neighbors, whereas kernel-based methods enable modeling a longer range of interactions.

%%%%%%%%%%%%%%%%%%%%%%%%%%%%%%%%%%%%%%%%%%%%%%%%%%%%%%%%%%%%%%%%%%%%%%%%%%%%%

\noindent\paragraph{\textbf{Contributions}}
In summary, our main contributions are:
\begin{enumerate}[leftmargin=*]
    \item We derive an analytical formulation of the power spectrum of Gaussian-kernel-based blue noise, demonstrate that it is provably superior to BNOT, and show empirical results to support the claim. 
    \item Starting from the loss function in \cite{Ahmed2021Optimizing}, and following a series of objective design choices, we present a robust algorithm for our Gaussian blue noise (GBN) optimization over a uniform toroidal domain of any number of dimensions, and an adaptation for bounded domains.
    \item We present a new algorithm for adaptive sampling that improves the current state of the art, along with a reconstruction algorithm.
\end{enumerate}

\paragraph{\textbf{Paper Structure}}
The paper is organized as follows.
We start by reviewing related literature in Section~\ref{sec:related work}.
In Section~\ref{sec:ideal bn} we discuss the theoretical foundation of GBN, and in Section~\ref{sec:realization} we discuss practical details of how to realize it, and present two efficient algorithms for uniform and adaptive blue-noise optimization. 
We then showcase actual results in Section~\ref{sec:results}, and compare to state-of-the-art methods, before making concluding remarks in Section~\ref{sec:conclusion}.

%%%%%%%%%%%%%%%%%%%%%%%%%%%%%%%%%%%%%%%%%%%%%%%%%%%%%%%%%%%%%%%%%%%%%%%%%%%%%
%%%%%%%%%%%%%%%%%%%%%%%%%%%%%%%%%%%%%%%%%%%%%%%%%%%%%%%%%%%%%%%%%%%%%%%%%%%%%
%%%%%%%%%%%%%%%%%%%%%%%%%%%%%%%%%%%%%%%%%%%%%%%%%%%%%%%%%%%%%%%%%%%%%%%%%%%%%

\section{Related Work\label{sec:related work}}

Thanks to the special nature of sampling in CG discussed in the introduction, blue noise is recognized as an important local product of the graphics community, and received a lot of attention evident in the large bulk of related literature. In the following subsections we briefly outline the most related work to this paper.

%%%%%%%%%%%%%%%%%%%%%%%%%%%%%%%%%%%%%%%%%%%%%%%%%%%%%%%%%%%%%%%%%%%%%%%%%%%%%

\subsection{Generating Blue Noise}

Most of the literature on blue noise is devoted to presenting generation and optimization algorithms of blue-noise point sets.
There are algorithms that use dart throwing \cite{Dippe85Antialiasing,Cook86Stochastic,McCool1992,Gamito2009,Ebeida2011,Wei2008Parallel,Yan2013GPA,Yuksel2015Sample} or advancing front \cite{Jones2006,Dunbar2006,Bridson07Fast,Mitchell18Spoke} techniques for the direct non-iterative generation of sample points that maintain a Poisson-disk property, and hence bear a blue noise spectrum. While some of these algorithms are very fast, the quality of the point distributions is relatively poor.
There are also general spectral tailoring algorithms \cite{Zhou2012,Oztireli2012,Heck2013,Kailkhura2016Stair,Oeztireli20Comprehensive} that may be used for the production of blue noise, but these algorithms are typically costly, and are mainly of theoretical importance.

Our concern in this paper is on optimization algorithms aimed at producing high-quality blue noise point sets. There is a wide range of such algorithms, but they can be grouped into two distinct categories.
In one category we have cellular methods that associate each point with a partition of the domain: a Voronoi or power cell, and optimize the local neighborhood of the points \cite{McCool1992,Deussen2000Floating,Secord2002Weighted,Schloemer2011,Ostromoukhov1993Pseudo,Ahmed2016-PPO,Balzer2009,Chen2012,Xu2011,Goes12Blue,Xin2016Centroidal,Paulin2020Sliced}. Then we have kernel-based methods that use a decaying kernel to model the influence of sample points \cite{Ulichney93Void,Hanson03Quasi,Hanson05Halftoning,Schmaltz2010,Fattal2011,Oztireli10Spectral,Jiang2015,Ahmed2021Optimizing}.
While most of these methods were developed heuristically, they are closely related to kernel density estimation \cite{Terrell1992variable}.

Since its introduction, blue noise through optimal transport (BNOT) \cite{Goes12Blue} --- a cellular method --- has widely been accepted as the reference algorithm for best-quality blue noise, while kernel-based methods are thought of as secondary alternatives. Subsequent research tried to find faster implementations, e.g., \cite{Xin2016Centroidal}, port it to higher dimensions, e.g., \cite{Paulin2020Sliced}, or emulate it with faster methods, e.g., \cite{Jiang2015}, but we are not aware of published works that claimed improved quality.
The fact that the algorithm could not be improved upon, until now, is a testament to the amazing quality of BNOT.
In this paper we demonstrate that, with informed choices of optimization parameters and settings, kernel-based methods can actually outperform BNOT.

%%%%%%%%%%%%%%%%%%%%%%%%%%%%%%%%%%%%%%%%%%%%%%%%%%%%%%%%%%%%%%%%%%%%%%%%%%%%%

\subsection{Distributing Blue Noise Samples}

The high cost of generating blue noise samples makes it not suitable for direct generation on demand. Instead, blue noise samples are typically generated offline and stored in lookup tables \cite{Glassner1995}.
Beyond the direct storage of a fixed list of samples, different techniques were proposed to distribute arbitrary numbers of blue-noise samples, including Wang tiling techniques \cite{Cohen2003,Lagae2006Alternative,Kopf06Recursive}, self-similar tiling \cite{Ostromoukhov2004,Ostromoukhov07Polyominoes,Wachtel14Fast,Lagae2006SSTile,Ahmed17ART,Ahmed2019Pinwheel}, and AA~Patterns \cite{Ahmed15AA}.

While our focus in this paper is on generating blue noise, a good understanding of how the samples are eventually distributed is important in favoring a generation technique.
Notably, the kernel-based methodology we are advocating is more versatile for optimizing the samples in tiling techniques thanks to the localized definition of the energy target.

%%%%%%%%%%%%%%%%%%%%%%%%%%%%%%%%%%%%%%%%%%%%%%%%%%%%%%%%%%%%%%%%%%%%%%%%%%%%%

\subsection{Evaluation}

The third line of research on blue noise is devoted to developing measures for evaluating and comparing blue noise distributions.
The primary tool for assessing blue noise is the frequency power spectrum of the point process that generates it, typically estimated empirically by averaging the periodograms of a reasonable number of realizations.
This tool was introduced by Ulichney \shortcite{Ulichney88Dithering,Ulichney87Digital}, who also defined two radially averaged plots for summaries: radial power, obtained by averaging the frequency power over rings of different radii, and the anisotropy, which measures the variance over each frequency ring, and detects directional bias and regularity.
Schl\"{o}emer et al. \shortcite{Schloemer11PSA} developed a tool, PSA, that standardizes the computation and presentation of frequency spectra. We extended that tool to work with high dimensions, so all the plots in this paper use exact evaluation of the periodograms.

Many alternative and additional tools were developed subsequently to assess the quality of blue noise sets, including the  autocorrelation plots \cite{Wei11Differential} for anisotropic and adaptive point processes, and their 1D profiles for stationary processes \cite{Oztireli2012,Heck2013}.
In addition to these plot, there are also scalar means to assess blue noise point sets.
%Poisson disk radius measures the minimum spacing between the points \cite{Lagae08Comparison}.
Conflict radius \cite{Lagae08Comparison} prevailed before the advent of capacity-constrained cellularization by Balzer et al. \shortcite{Balzer2009}.
Heck et al. \shortcite{Heck2013} then introduced three additional measures: effective Nyquist rate $\nu_\mathrm{eff}$, oscillation $\mathrm{\Omega}$, and bond orientation order $Q_6$, to assess the quality of blue noise for reconstruction.

More recently, Ahmed and Wonka \shortcite{Ahmed2021Optimizing} derived an energy term for evaluating blue noise based on a Gaussian kernel that filters it.
This is the most relevant work to our current work, and we take it further to characterize the frequency spectrum of kernel-based blue noise processes, and to study its realizability in 2D and its extensibility to higher dimensions.
We actually show in the supplementary materials that blue noise is realizeable even in one dimension, and we discuss the effect of the number of dimensions on the generated blue noise.

%%%%%%%%%%%%%%%%%%%%%%%%%%%%%%%%%%%%%%%%%%%%%%%%%%%%%%%%%%%%%%%%%%%%%%%%%%%%%

\subsection{Integration with Blue Noise}

While blue noise was originally intended for reconstructing a visual signal from samples to minimize coherent aliasing, there has been a long-standing curious question about its utility for Monte Carlo integration.
Hanson \shortcite{Hanson03Quasi} empirically demonstrated the advantage of blue noise sets over Halton low-discrepancy distribution. The seminal report by Durand \shortcite{Durand11Frequency} on the frequency analysis of numerical integration established a theoretical base, followed up by Ramamoorthi et al. \shortcite{Ramamoorthi2012Theory}, Subr and Kautz \shortcite{Subr13Fourier}, and Pilleboue et al. \shortcite{Pilleboue15Variance}.
\"{O}ztireli \shortcite{Oztireli2016} took a different path by studying the same problem in the spatial domain, using auto-correlation.
In this paper we show analytical cues and present empirical data that confirms the suitability of blue noise for numerical integration, especially in higher dimensions, where it is less hit by the curse of dimensionality than the competing alternatives.

%%%%%%%%%%%%%%%%%%%%%%%%%%%%%%%%%%%%%%%%%%%%%%%%%%%%%%%%%%%%%%%%%%%%%%%%%%%%%
%%%%%%%%%%%%%%%%%%%%%%%%%%%%%%%%%%%%%%%%%%%%%%%%%%%%%%%%%%%%%%%%%%%%%%%%%%%%%
%%%%%%%%%%%%%%%%%%%%%%%%%%%%%%%%%%%%%%%%%%%%%%%%%%%%%%%%%%%%%%%%%%%%%%%%%%%%%
\section{A Theory of Gaussian Blue Noise\label{sec:ideal bn}}

The idea of using a Gaussian kernel for blue noise optimization dates back to Ulichney's void-and-cluster algorithm \shortcite{Ulichney93Void}, and was re-introduced at least three times thereafter by Hanson \shortcite{Hanson03Quasi,Hanson05Halftoning}, \"{O}ztireli et al. \shortcite{Oztireli10Spectral}, and Fattal \shortcite{Fattal2011}, with a different motivation each time.
The idea is to place Gaussian kernels at the points, and optimize the placement of the points so that a uniform density is maintained everywhere.
The concept was mostly developed heuristically, but is closely related to kernel density estimation, as can be seen by comparing \cite[Eq.~(1)]{Fattal2011} to \cite[Eq.~(1.7)]{Terrell1992variable}.
Ahmed and Wonka \shortcite{Ahmed2021Optimizing} have recently derived an analytical formulation,
%%%%%%%%%%%%%%%%%%%%%%%%%%%%%%%%%%%%%
%%%%%%%%%%%%%%%%%%%%%%%%%%%%%%%%%%%%%
\begin{equation}
    \mathrm{Var}\big(A(\mathbf{X})\big)
        = \frac{\pi\sigma^{2}}{N}\sum_{k=1}^{N}\sum_{l=1}^{N}\exp\left(-\frac{\left\Vert \mathbf{x}_{k}-\mathbf{x}_{l}\right\Vert ^{2}}{4\sigma^{2}}\right)-\left(2\pi\sigma^{2}\right)^{2}\,, \label{eq:variance uniform}
\end{equation}
%%%%%%%%%%%%%%%%%%%%%%%%%%%%%%%%%%%%%
%%%%%%%%%%%%%%%%%%%%%%%%%%%%%%%%%%%%%
that underlies these methods, defined as the variance
%%%%%%%%%%%%%%%%%%%%%%%%%%%%%%%%%%%%
%%%%%%%%%%%%%%%%%%%%%%%%%%%%%%%%%%%%
\begin{equation}
    \mathrm{Var}\left(A(\mathbf{x})\right)=E\left(A^{2}(\mathbf{x})\right)-\big(E\left(A(\mathbf{x})\right)\big)^{2}\label{eq:variance}
\end{equation}
%%%%%%%%%%%%%%%%%%%%%%%%%%%%%%%%%%%%
%%%%%%%%%%%%%%%%%%%%%%%%%%%%%%%%%%%%
of a sum
%%%%%%%%%%%%%%%%%%%%%%%%%%%%%%%%%%%%%
%%%%%%%%%%%%%%%%%%%%%%%%%%%%%%%%%%%%%
\begin{equation}
    A\left(\mathbf{X}\right)
        = g(\mathbf{x}) \ast \delta(\mathbf{X})
        = \sum_{k=1}^{N}\exp\left(-\frac{\left\Vert \mathbf{x}-\mathbf{x}_{k}\right\Vert ^{2}}{2\sigma^{2}}\right) \label{eq:A(X)}
\end{equation}
%%%%%%%%%%%%%%%%%%%%%%%%%%%%%%%%%%%%%
%%%%%%%%%%%%%%%%%%%%%%%%%%%%%%%%%%%%%
of Gaussian kernels
%%%%%%%%%%%%%%%%%%%%%%%%%%%%%%%%%%%%%
%%%%%%%%%%%%%%%%%%%%%%%%%%%%%%%%%%%%%
\begin{equation}
    g(\mathbf{x}) = \exp\left( -\frac{\left\Vert \mathbf{x}\right\Vert ^{2}}{2\sigma^{2}} \right) \label{eq:filter}
\end{equation}
%%%%%%%%%%%%%%%%%%%%%%%%%%%%%%%%%%%%%
%%%%%%%%%%%%%%%%%%%%%%%%%%%%%%%%%%%%%
placed at a set
%%%%%%%%%%%%%%%%%%%%%%%%%%%%%%%%%%%%%
%%%%%%%%%%%%%%%%%%%%%%%%%%%%%%%%%%%%%
\begin{equation}
    \mathbf{X} = \left\{ \mathbf{x}_{k}\right\} _{k=1}^{N}
\end{equation}
%%%%%%%%%%%%%%%%%%%%%%%%%%%%%%%%%%%%%
%%%%%%%%%%%%%%%%%%%%%%%%%%%%%%%%%%%%%
of sample points.

In this section, we derive an analytical formula for the power spectrum of a point distribution that minimizes this variance to serve as a theoretical reference for all Gaussian-kernel-based algorithms, and we discuss its inherent superiority over cellular-based methods.

%%%%%%%%%%%%%%%%%%%%%%%%%%%%%%%%%%%%%%%%%%%%%%%%%%%%%%%%%%%%%%%%%%%%%%%%%%%%%

\subsection{Frequency Analysis}

We start by analyzing the variance in Eq.~\eqref{eq:variance} directly in the frequency domain.
As already established by Durand \shortcite[Eq.~(15)]{Durand11Frequency}, ``the variance is the integral of the power spectrum except at the DC,'':
%%%%%%%%%%%%%%%%%%%%%%%%%%%%%%%%%%%%
%%%%%%%%%%%%%%%%%%%%%%%%%%%%%%%%%%%%
\begin{equation}
    \mathrm{Var}\left(A(\mathbf{X})\right)=
        \int \vert\hat{A}\vert^2(\pmb{\omega})\, d\pmb{\omega} - \vert\hat{A}\vert^2(0)\,,\label{eq:variance frequency}
\end{equation}
%%%%%%%%%%%%%%%%%%%%%%%%%%%%%%%%%%%%
%%%%%%%%%%%%%%%%%%%%%%%%%%%%%%%%%%%%
which follows directly from Parseval's theorem.
The right-side terms of Eq.~\eqref{eq:variance frequency} correspond directly to their counterparts in Eq.~\eqref{eq:variance uniform}. The DC term is invariant, and minimizing the variance in Eq.~\eqref{eq:variance uniform} is therefore equivalent to attenuating the whole power spectrum except for the DC, which is intuitive, since our goal is to come close to a constant.
An idealized optimization process would not have spectral bias, and would therefore bring the power spectrum of the filtered point set below some level
%%%%%%%%%%%%%%%%%%%%%%%%%%%%%%%%%%%%
%%%%%%%%%%%%%%%%%%%%%%%%%%%%%%%%%%%%
\begin{equation}
    \vert\hat{A}\vert^2(\pmb{\omega}) \leq \epsilon \label{eq:noise floor}
\end{equation}
%%%%%%%%%%%%%%%%%%%%%%%%%%%%%%%%%%%%
%%%%%%%%%%%%%%%%%%%%%%%%%%%%%%%%%%%%
that manifests as random bumping on the surface of the filtered signal $A(\mathbf{x})$.

We further analyze this power spectrum. Applying the kernels in the spatial domain is a convolution process that translates into a multiplication in the frequency domain; hence
%%%%%%%%%%%%%%%%%%%%%%%%%%%%%%%%%%%%
%%%%%%%%%%%%%%%%%%%%%%%%%%%%%%%%%%%%
\begin{equation}
    \vert\hat{A}\vert^2(\pmb{\omega}) =
        \vert\hat{g}\vert^2(\pmb{\omega}) \cdot \vert\mathcal{F}\vert^2(\pmb{\omega})\,,\label{eq:variance factors}
\end{equation}
%%%%%%%%%%%%%%%%%%%%%%%%%%%%%%%%%%%%
%%%%%%%%%%%%%%%%%%%%%%%%%%%%%%%%%%%%
where $\hat{g}$ is the Fourier transform of the kernel $g$, and
%%%%%%%%%%%%%%%%%%%%%%%%%%%%%%%%%%%%%
%%%%%%%%%%%%%%%%%%%%%%%%%%%%%%%%%%%%%
\begin{equation}
    \mathcal{F}(\pmb{\omega}) = \sum_{k=1}^{N} \exp(-i \pmb{\omega}\cdot\mathbf{x}_k) \label{eq:Fourier spectrum}
\end{equation}
%%%%%%%%%%%%%%%%%%%%%%%%%%%%%%%%%%%%%
%%%%%%%%%%%%%%%%%%%%%%%%%%%%%%%%%%%%%
is the frequency spectrum of the point set $\mathbf{X}$.
This second factor in Eq.~\eqref{eq:variance factors} is, by definition, the power spectrum
%%%%%%%%%%%%%%%%%%%%%%%%%%%%%%%%%%%%%
%%%%%%%%%%%%%%%%%%%%%%%%%%%%%%%%%%%%%
\begin{equation}
    P(\pmb{\omega}) = \mathcal{F}(\pmb{\omega})\cdot\mathcal{F}^*(\pmb{\omega}) \label{eq:P(f)}
\end{equation}
%%%%%%%%%%%%%%%%%%%%%%%%%%%%%%%%%%%%%
%%%%%%%%%%%%%%%%%%%%%%%%%%%%%%%%%%%%%
of the point set.
Combining Eqs.~(\ref{eq:noise floor}, \ref{eq:variance factors}, \ref{eq:P(f)}) gives
%%%%%%%%%%%%%%%%%%%%%%%%%%%%%%%%%%%%%
%%%%%%%%%%%%%%%%%%%%%%%%%%%%%%%%%%%%%
\begin{equation}
    P(\pmb{\omega}) \leq \epsilon \vert\hat{g}\vert^{-2}(\pmb{\omega})\,. \label{eq:P and g}
\end{equation}
%%%%%%%%%%%%%%%%%%%%%%%%%%%%%%%%%%%%%
%%%%%%%%%%%%%%%%%%%%%%%%%%%%%%%%%%%%%
This equation characterizes the power spectrum of kernel-based methods, and applies to any square-integrable kernel used to filter the point set. The only assumption is that the optimization process is not frequency biased; otherwise the constant $\epsilon$ would have to be replaced by a frequency profile of the optimization process.

For the case of a Gaussian kernel $g(\mathbf{x})$ in Eq.~\eqref{eq:filter}, the Fourier transform is another Gaussian:
%%%%%%%%%%%%%%%%%%%%%%%%%%%%%%%%%%%%%
%%%%%%%%%%%%%%%%%%%%%%%%%%%%%%%%%%%%%
\begin{equation}
    \hat{g}(\pmb{\omega})
        = \exp\left( -\frac{\sigma^2}{2} \left\Vert \pmb{\omega} \right\Vert^2 \right) \label{eq:g(f)}\,.
\end{equation}
%%%%%%%%%%%%%%%%%%%%%%%%%%%%%%%%%%%%%
%%%%%%%%%%%%%%%%%%%%%%%%%%%%%%%%%%%%%
Substituting Eq.~\eqref{eq:g(f)} in Eq.~\eqref{eq:P and g} gives
%%%%%%%%%%%%%%%%%%%%%%%%%%%%%%%%%%%%%
%%%%%%%%%%%%%%%%%%%%%%%%%%%%%%%%%%%%%
\begin{equation}
\boxed{
    P(\pmb{\omega}) \leq \epsilon e^{\sigma^2 \left\Vert \pmb{\omega} \right\Vert^2}\,. \label{eq:ideal spectrum}
}
\end{equation}
%%%%%%%%%%%%%%%%%%%%%%%%%%%%%%%%%%%%%
%%%%%%%%%%%%%%%%%%%%%%%%%%%%%%%%%%%%%

\subsection{Feasibility}

This Eq.~\eqref{eq:ideal spectrum} characterizes the frequency spectrum of an idealized Gaussian-kernel-based blue noise when the filtered set is close to a constant, which is the target of minimizing the variance in Eq.~\eqref{eq:variance uniform}.
As can be seen in Fig.~\ref{fig:teaser}(e), such a frequency spectral profile is actually realizable, and is faithfully attained by Gaussian-based methods like KDM~\cite{Fattal2011} and VnC~\cite{Ulichney93Void}.
It is interesting that FPO~\cite{Schloemer2011} bears a similar profile and the same exponent as VnC, which possibly comes from the fact that they both search for fartherst points.
Cellular-based methods, in contrast, follow a fundamentally different, polynomial power profile, which is already noted in the literature \cite{Pilleboue15Variance}.
BlueNets~\cite{Ahmed2021Optimizing} exhibit a mixed behavior, reflecting the combination between their stratified nature and the kernel-based optimization.

A very important note about the Gaussian-based spectrum is that it is not only ideal for reconstruction, but also for numerical integration.
Indeed, Pilleboue et al. \shortcite{Pilleboue15Variance} characterized the variance of numerical integration of (semi-)stochastic point sets by the growth-rate of the frequency power spectrum of the point process, and they favored a higher polynomial degree for the curve of the spectrum.
The idealized blue noise model in Eq.~\eqref{eq:ideal spectrum}, however, has an exponential growth, or even better, \emph{quadrexponential},
which means that, starting from the same noise floor $\epsilon$ at the lowest frequency, it will always stay for a wide range below the polynomially-shaped power spectra, as can be visualized by comparing the linearly-sloped spectra in Fig.~\ref{fig:teaser} with the curved ones.

This intrinsic advantage of Gaussian-based blue noise, however, remained undiscovered in the past possible because the known algorithms all stop at a very shallow noise floor $\epsilon$, as seen in Fig.~\ref{fig:teaser}(e), either due to an inherent limit of the algorithm, e.g. FPO, or a numerical limit, e.g. VnC, or an arbitrary stop down to avoid developing regular patterns.
BNOT, in contrast, is able to attain a very low noise floor without suffering quality issues.
We recall, however, that BNOT was proceeded by Lloyd's algorithm \cite{McCool1992,Lloyd1982Least}, which suffered from similar problems to the mentioned ones in kernel-based methods.
It was not until Balzer et al. \shortcite{Balzer2009} introduced the capacity constraint, 17 years later, and de Goes et al. developed a theory of it, that the cellular-based approach was able to unlock its full potential and reach the BNOT quality.
Analogously, the preceding analysis furnishes as theoretical basis to develop an algorithm that is able to unlock the full potential of kernel-based optimization and reach its extent, and in the following section we go through the many practical aspects that need to be taken into consideration.
Through these objective design choices we were able to reach a 10 orders of magnitudes lower noise floor than the state-of-the-art kernel-based algorithms, as seen in Fig.~\ref{fig:teaser}(e), and even surpass BNOT itself by two orders of magnitudes. The actual improvement over BNOT is all the highlighted volume in Fig.~\ref{fig:teaser}(e), which  manifests as a visibly lower noise in the the point plots in Fig.~\ref{fig:teaser}(a, f) and later Figs.~(\ref{fig:spatial}, \ref{fig:adaptive}, \ref{fig:adaptive convergence}), or as a reduction in the numerical integration variance, as will be demonstrated in Section~\ref{sec:numerical integration}.

%%%%%%%%%%%%%%%%%%%%%%%%%%%%%%%%%%%%%%%%%%%%%%%%%%%%%%%%%%%%%%%%%%%%%%%%%%%%%
%%%%%%%%%%%%%%%%%%%%%%%%%%%%%%%%%%%%%%%%%%%%%%%%%%%%%%%%%%%%%%%%%%%%%%%%%%%%%
%%%%%%%%%%%%%%%%%%%%%%%%%%%%%%%%%%%%%%%%%%%%%%%%%%%%%%%%%%%%%%%%%%%%%%%%%%%%%

\section{Realization\label{sec:realization}}

Kernel-based optimization for blue noise consists of placing a set of identical kernels $h$ on the sample points, and optimizing the point locations towards minimizing the variance of the sum, as defined in Eq.~\eqref{eq:variance uniform} for a Gaussian kernel. As we have seen in the preceding section, an idealized, fair, frequency-neutral optimization process would shape the power spectrum of the point set by the inverse square $\vert \hat{h} \vert^{-2}$ of the frequency transform of the kernel, as in Eq.~\eqref{eq:P and g}.
There are already many known kernel-based algorithms \cite{Ulichney93Void,Hanson03Quasi,Oztireli10Spectral,Schmaltz2010,Fattal2011,Jiang2015,Ahmed2021Optimizing} that differ in design parameters. In this section we discuss important design choices, and make informed decisions in the light of the preceding discussion about the target frequency spectrum.

%%%%%%%%%%%%%%%%%%%%%%%%%%%%%%%%%%%%%%%%%%%%%%%%%%%%%%%%%%%%%%%%%%%%%%%%%%%%%

\subsection{Choice of Kernel}

Our focus in this paper is on Gaussian kernels, but it is worthwhile having a brief discussion, in the light of our theoretical model, about the merits of this specific kernel among alternatives such as the inverse distance \cite{Schmaltz2010} or SPH \cite{Jiang2015}.
A great advantage of the Gaussian kernel is that it has a well-defined parametric frequency transform, and this transform happens to be quite favorable in having a very fast decaying rate towards high-frequencies, making it possible to band-limit the optimization to the more important low-frequency range, as will be discussed in the following subsection.
In contrast, an inverse distance kernel, for example, is not even square-integrable to fit the theory, and any shaping to escape the singularity adds complexity to the frequency profile.
We are not claiming optimality, though; only that the Gaussian kernel leads to an excellent blue noise profile; specifically, superior to the polynomial profile of BNOT.

Besides this analytical merit, there is another very important unique advantage of the Gaussian kernel, that it is separable in dimensions, making the optimization scales with dimension in a feasible manner.

%%%%%%%%%%%%%%%%%%%%%%%%%%%%%%%%%%%%%%%%%%%%%%%%%%%%%%%%%%%%%%%%%%%%%%%%%%%%%

%%%%%%%%%%%%%%%%%%%%%%%%%%%%%%%%%%%%%
%%%%%%%%%%%%%%%%%%%%%%%%%%%%%%%%%%%%%
\begin{figure}
  \setlength{\unit}{0.48\columnwidth}        % This will make all the gaps 0.05 unit
  \centering
  {\scriptsize
    \begin{tabular*}{1\columnwidth}{@{}c@{\extracolsep{\fill}}c@{\extracolsep{\fill}}c@{}}
        \begin{tikzpicture}
            \node[inner sep=0, anchor=south west] at (0, 0.51\unit) {%
                \includegraphics[height=0.49\unit]{images/sigma/g05-i30-points.pdf}
            };
            \node[inner sep=0, anchor=south west] at (0, 0) {%
                \includegraphics[height=0.49\unit]{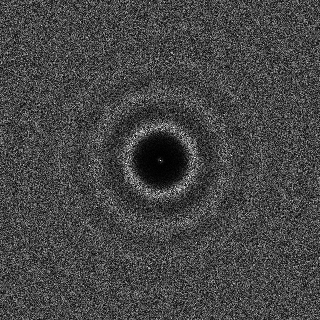}
            };
        \end{tikzpicture}&%
        \begin{tikzpicture}
            \node[inner sep=0, anchor=south west] at (0, 0.51\unit) {%
                \includegraphics[height=0.49\unit]{images/sigma/g11-i1M-points.pdf}
            };
            \node[inner sep=0, anchor=south west] at (0, 0) {%
                \includegraphics[height=0.49\unit]{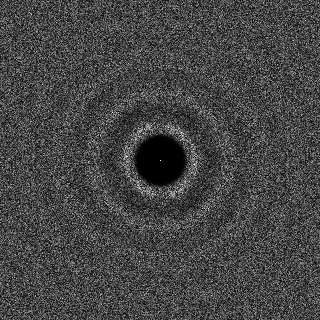}
            };
        \end{tikzpicture}&%
        \includegraphics[height=1\unit]{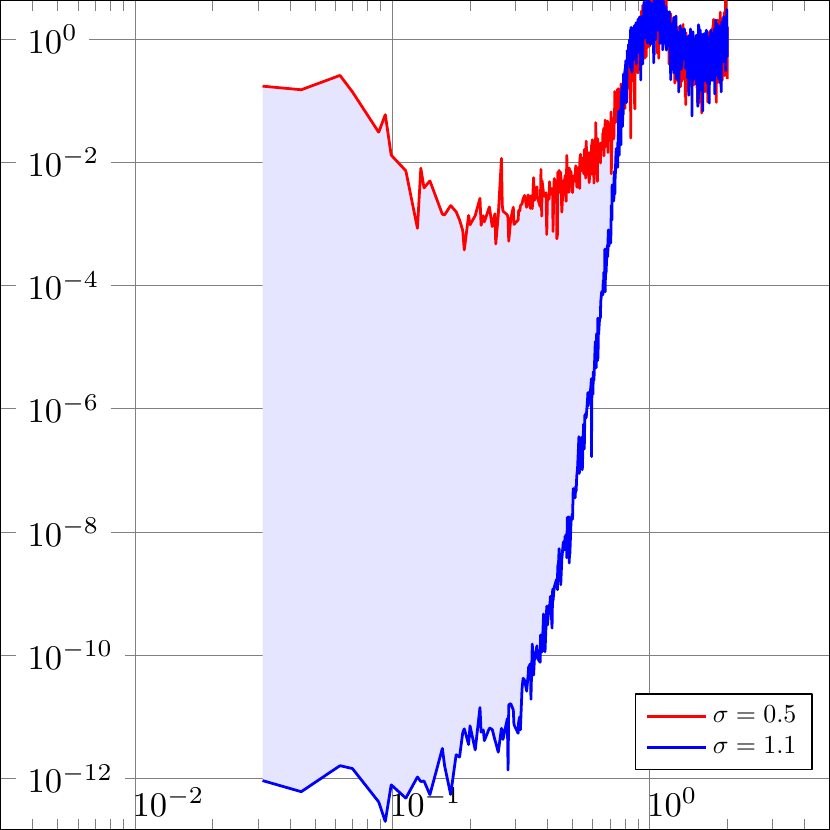}\\[2mm]
    (a) $\sigma = 0.5$, 30 iterations & (b) $\sigma = 1.1$, 1M iterations. & (c) Radial Power
    \end{tabular*}}\vspace{-9pt}
    \caption{
        Two different settings leading to visually similar distributions, but very different spectral profiles.
    }
    \label{fig:sigma}
\end{figure}
%%%%%%%%%%%%%%%%%%%%%%%%%%%%%%%%%%%%%
%%%%%%%%%%%%%%%%%%%%%%%%%%%%%%%%%%%%%

\subsection{Optimization Method\label{sec:optimization method}}

Deriving a gradient-descent minimization algorithm from Eq.~\eqref{eq:variance uniform} is straightforward.
We first write the Gaussian in a standardized form \cite[Eq.~(36)]{Ahmed2021Optimizing}:
%%%%%%%%%%%%%%%%%%%%%%%%%%%%%%%%%%%%
%%%%%%%%%%%%%%%%%%%%%%%%%%%%%%%%%%%%
\begin{equation}
    \sigma^2 \leftarrow 2\sigma^2_\mathrm{Filtering}\,.
\end{equation}
%%%%%%%%%%%%%%%%%%%%%%%%%%%%%%%%%%%%
%%%%%%%%%%%%%%%%%%%%%%%%%%%%%%%%%%%%
The second (DC) term is invariant, so our target energy becomes
%%%%%%%%%%%%%%%%%%%%%%%%%%%%%%%%%%%%%
%%%%%%%%%%%%%%%%%%%%%%%%%%%%%%%%%%%%%
\begin{equation}
    \mathcal{E}\big(\mathbf{X}\big) =
        \frac{\pi\sigma^{2}}{2N}
        \sum_{k=1}^{N}
        \sum_{l=1}^{N}
        \exp\left(-\frac{\left\Vert \mathbf{x}_{k}-\mathbf{x}_{l}\right\Vert ^{2}}{2\sigma^{2}}\right)\,. \label{eq:energy whole}
\end{equation}
%%%%%%%%%%%%%%%%%%%%%%%%%%%%%%%%%%%%%
%%%%%%%%%%%%%%%%%%%%%%%%%%%%%%%%%%%%%
We can then extract a loss function 
%%%%%%%%%%%%%%%%%%%%%%%%%%%%%%%%%%%%%
%%%%%%%%%%%%%%%%%%%%%%%%%%%%%%%%%%%%%
\begin{equation}
    \mathcal{E}(\mathbf{x}_{k}) = 
        \frac{\pi\sigma^{2}}{N}
        \sum_{l\neq k}
        \exp\left(
            - \frac{ \left\Vert \mathbf{x}_k - \mathbf{x}_l \right\Vert^{2} }{ 2\sigma^2 }
        \right) \label{eq:energy granular}
\end{equation}
%%%%%%%%%%%%%%%%%%%%%%%%%%%%%%%%%%%%%
%%%%%%%%%%%%%%%%%%%%%%%%%%%%%%%%%%%%%
for each point $\mathbf{x}_k$ such that
%%%%%%%%%%%%%%%%%%%%%%%%%%%%%%%%%%%%%
%%%%%%%%%%%%%%%%%%%%%%%%%%%%%%%%%%%%%
\begin{equation}
    \argmin_{\mathbf{X}} \text{Var}\left(A\left(\mathbf{X}\right)\right) =
        \argmin_{\mathbf{x}} \sum_{k=1}^{N} \mathcal{E}(\mathbf{x}_k)\,.
\end{equation}
%%%%%%%%%%%%%%%%%%%%%%%%%%%%%%%%%%%%%
%%%%%%%%%%%%%%%%%%%%%%%%%%%%%%%%%%%%%
Note that each point counts twice in Eq.~\eqref{eq:energy whole}: once as $\mathbf{x}_{k}$ and once as $\mathbf{x}_{l}$, hence the factor of two in Eq.~\eqref{eq:energy granular}.
Finally, we compute the gradient
%%%%%%%%%%%%%%%%%%%%%%%%%%%%%%%%%%%%%
%%%%%%%%%%%%%%%%%%%%%%%%%%%%%%%%%%%%%
\begin{align}
    \nabla \mathcal{E}(\mathbf{x}_{k})
         & =-\frac{\pi}{N}\sum_{l\neq k}^{N}\exp\left(-\frac{\left\Vert \mathbf{x}_{k}-\mathbf{x}_{l}\right\Vert ^{2}}{2\sigma^{2}}\right)\left(\mathbf{x}_{k}-\mathbf{x}_{l}\right) \label{eq:grad1}
\end{align}
%%%%%%%%%%%%%%%%%%%%%%%%%%%%%%%%%%%%%
%%%%%%%%%%%%%%%%%%%%%%%%%%%%%%%%%%%%%
of the loss function.
This was already derived by {\"O}ztireli \shortcite{Oztireli10Spectral} from a different approach.
Note that the energy is halved between each point and the remaining set.

The gradient formulation in Eq.~\eqref{eq:grad1} suggests a quadratic complexity, and it is tempting to look for alternative implementations that would accelerate convergence, or otherwise reduce computational complexity.
Unfortunately, a Hessian matrix of the energy function is inherently singular, ruling out second order methods such as Newton's method.
Another choice is discretization, as in \cite{Ulichney93Void,Fattal2011,Ahmed2021Optimizing,Hanson03Quasi}, but these methods typically converge to a shallow minimum, as can be seen in Fig.~\ref{fig:teaser}(g).
For best quality we therefore recommend a simple gradient-descent algorithm, which comes also with other advantages including GPU parallelization and low coding complexity.

%%%%%%%%%%%%%%%%%%%%%%%%%%%%%%%%%%%%%%%%%%%%%%%%%%%%%%%%%%%%%%%%%%%%%%%%%%%%%

\subsection{Kernel Parameter\label{sec:kernel parameter}}

The choice of $\sigma$ is an essential design decision.
Values used in previous methods include 1.5 \cite{Ulichney93Void,Fattal2011}, $\approx$1 \cite{Oztireli10Spectral}, and 0.5 \cite{Ahmed2021Optimizing}.
These values were chosen experimentally for the different algorithms, coordinated with other parameters.

Only considering the \emph{spatial} domain can lead to wrong choices of $\sigma$.
Specifically, it is tempting to set a small $\sigma$ which will concentrate the energy in the nearby neighborhood, leading to a visually quick convergence towards a large Poisson-disk radius in the spatial domain, as demonstrated in Fig.~\ref{fig:sigma}(a).
However, a more thorough investigation of the radial power, i.e., in the \emph{spectral} domain, would reveal that the spectra of these point sets are quite shallow.

\begin{figure}
  \setlength{\unit}{0.49\columnwidth}        % This will make all the gaps 0.05 unit
  \centering
  {\scriptsize
    \begin{tabular*}{1\columnwidth}{@{}c@{\extracolsep{\fill}}c@{}}
    \includegraphics[width=1\unit]{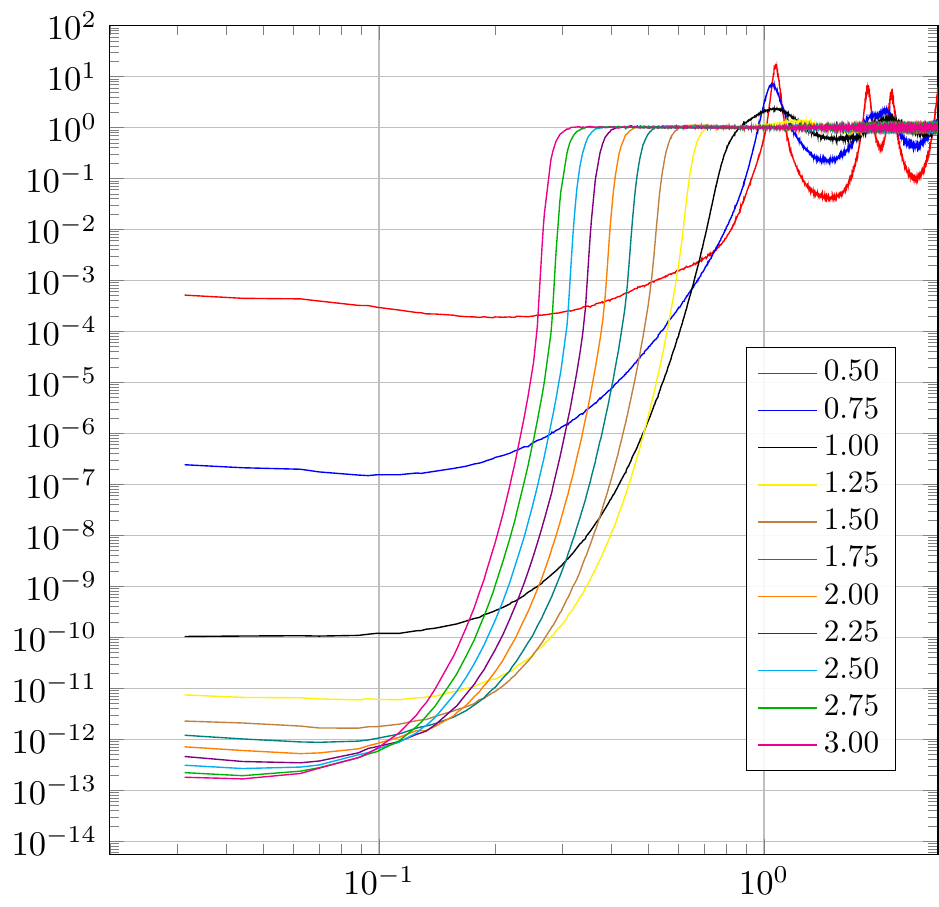}&%
    \includegraphics[width=1\unit]{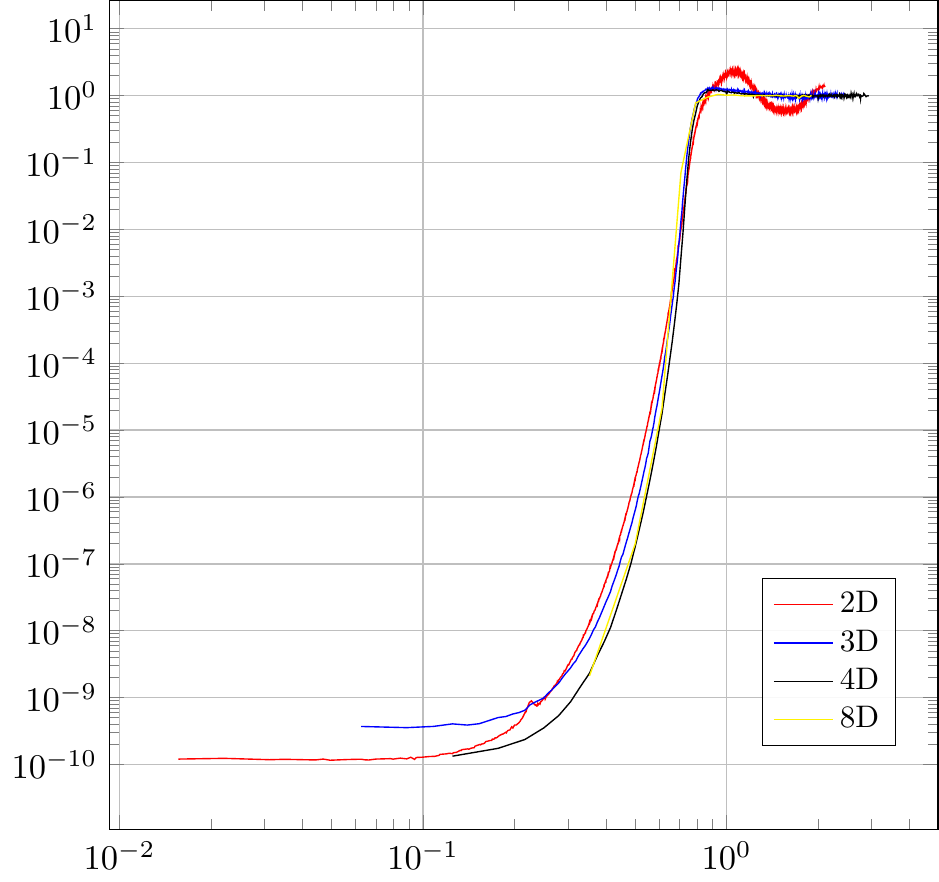}\\[0mm]
    (a) & (b)
    \end{tabular*}}\vspace{-9pt}
    \caption{
        Actual frequency spectra we obtained for (a) various values of $\sigma$ in 2D, using 1k point sets and 10k iterations, and (b) different dimensions, with \mbox{$\sigma = 1$}, using 4k point sets to provide a better frequency resolution.
    }
    \label{fig:spectrum}
\end{figure}

The key insight to guide the choice of a kernel width $\sigma$ is that ``you get what you pay for''.
The point locations are the only degree of freedom we have for optimization; that is, $2N$ real parameters for a set of $N$ points in 2D, for example.
Thus, we have a limited budget of information bits to use for shaping the spectrum.
A rough measure for the attenuated energy is to count the number of zeros to the right of the fractional point in the frequency range (0, 1), which is proportional to the area below 1 in a log-linear scale of the spectrum. 
In Fig.~\ref{fig:spectrum}(a) we show the spectra of our obtained GBN with different choices of $\sigma$.
%%%%%%%%%%%%%%%%%%%%%%%%%%%%%%%%%%%%%
%%%%%%%%%%%%%%%%%%%%%%%%%%%%%%%%%%%%%
%%%%%%%%%%%%%%%%%%%%%%%%%%%%%%%%%%%%%
%%%%%%%%%%%%%%%%%%%%%%%%%%%%%%%%%%%%%
We note that for a small $\sigma$ the frequency spectrum is distorted significantly from the spectrum of an idealized blue noise model, having residual low-frequency content and too-much oscillation in the high-frequency range.
A possible explanation is that a narrow kernel transforms into a wide one in the frequency domain, assigning substantial weights to higher frequencies where our algorithm tries to optimize worthlessly.
On the other hand, optimizing for a large $\sigma$ gives too much improvement in the low-frequency range, at the cost of narrowing the low-energy band.
We found that \emph{$\sigma = 1$}, relative to the nominal grid $N^{-1/d}$ spacing between the points, gives an excellent trade-off, so we recommend it.
With $\sigma=1$, the energy weight of the nominal grid spacing drops to $e^{-(2\pi)^2}\approx7.2\times10^{-18}$ relative to the DC weight.
With a proper implementation, this guards the optimization from seeking regular or triangular structures, and leads to a very neat formula
%%%%%%%%%%%%%%%%%%%%%%%%%%%%%%%%%%%%%
%%%%%%%%%%%%%%%%%%%%%%%%%%%%%%%%%%%%%
\begin{equation}
\boxed{
    P(\pmb{\omega}) = \epsilon e^{\left\Vert \pmb{\omega} \right\Vert^2} \label{eq:ref spectrum}
}
\end{equation}
%%%%%%%%%%%%%%%%%%%%%%%%%%%%%%%%%%%%%
%%%%%%%%%%%%%%%%%%%%%%%%%%%%%%%%%%%%%
for our reference blue-noise power-spectrum profile shown in Fig~\ref{fig:teaser}.

We can also realize high-dimensional (e.g., 8D) Gaussian Blue Noise with comparable noise floor to 2D GBN as illustrated in Fig.~\ref{fig:spectrum}(b).
%In Fig.~\ref{fig:spectrum}(b) we see that the same favorable profile, with at least the same low noise floor, is realizable in higher dimensions.
An interesting, unexpected result we obtained is that higher dimensions actually converge faster towards the target noise floor, or, alternatively, attain even lower noise levels.
Our first guess is that this is due to the availability of more degrees of freedom with dimensions (i.e., more budget for information bits), but this needs further investigation.

%%%%%%%%%%%%%%%%%%%%%%%%%%%%%%%%%%%%%%%%%%%%%%%%%%%%%%%%%%%%%%%%%%%%%%%%%%%%%
%%%%%%%%%%%%%%%%%%%%%%%%%%%%%%%%%%%%%
%%%%%%%%%%%%%%%%%%%%%%%%%%%%%%%%%%%%%
\begin{figure}
    \centering
   \begin{overpic}[trim=2.5cm 11cm 1.9cm 11cm,clip,width=1\linewidth,grid=false]{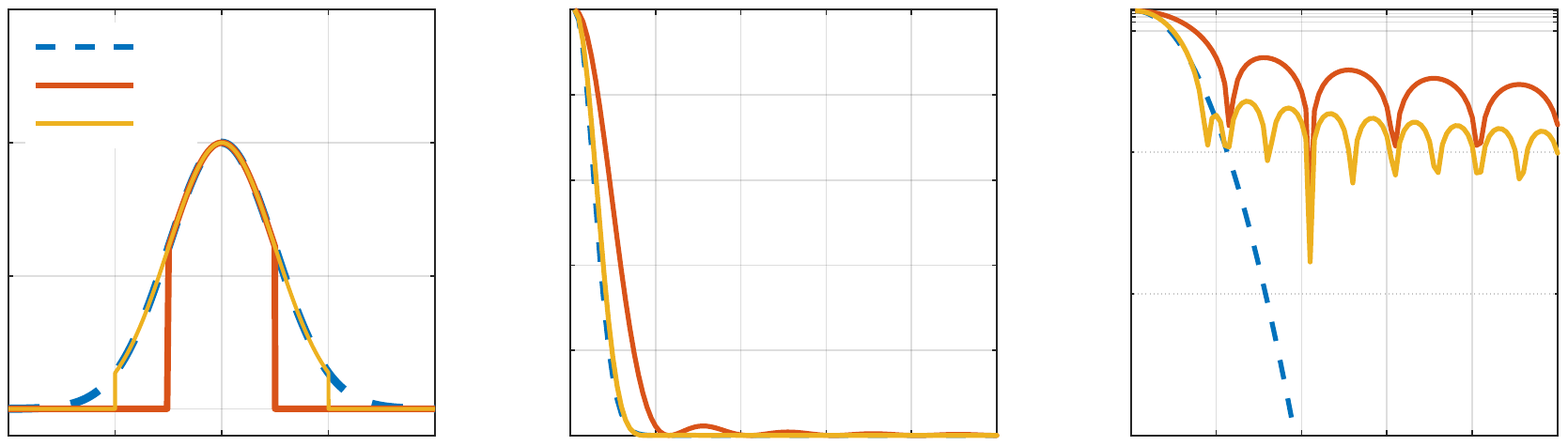}
   \put(1,1){\footnotesize (a) (truncated) Gaussian}
   \put(39,1){\footnotesize (b) FT (linear-scale)}
   \put(76,1){\footnotesize (c) FT (log-scale)}
   \put(10,28){\tiny standard Gaussian}
   \put(10,26.1){\tiny truncated at $[-1,1]$}
   \put(10,23.9){\tiny truncated at $[-2,2]$}
   \end{overpic}\vspace{-9pt}
    \caption{%
    The Fourier transform (shown in (b,c)) of the three (truncated) Gaussians (shown in (a)) including: the standard Gaussian (colored blue), a truncated Gaussian in the range of $[-1, 1]$ (colored red) and in the range of $[-2,2]$ (colored yellow).
    }\vspace{-9pt}
    \label{fig:truncated kernel}
\end{figure}
%%%%%%%%%%%%%%%%%%%%%%%%%%%%%%%%%%%%%
%%%%%%%%%%%%%%%%%%%%%%%%%%%%%%%%%%%%%

%%%%%%%%%%%%%%%%%%%%%%%%%%%%%%%%%%%%%
%%%%%%%%%%%%%%%%%%%%%%%%%%%%%%%%%%%%%
\begin{figure*}
  \setlength{\unit}{0.12\textwidth}        % This will make all the gaps 0.05 unit
  \centering
  {\scriptsize
    \begin{tabular*}{1\textwidth}{@{}c@{\extracolsep{\fill}}c@{\extracolsep{\fill}}c@{\extracolsep{\fill}}c@{\extracolsep{\fill}}c@{\extracolsep{\fill}}c@{\extracolsep{\fill}}c@{}}
        \begin{tikzpicture}
            \node[inner sep=0, anchor=south west] at (0, 1.01\unit) {%
                \includegraphics[height=1\unit]{images/truncated/1-points.pdf}%
            };
            \node[inner sep=0, anchor=south west] at (0, 0) {%
                \includegraphics[height=1\unit]{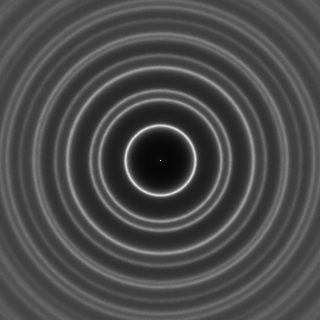}%
            };
        \end{tikzpicture}&
        \begin{tikzpicture}
            \node[inner sep=0, anchor=south west] at (0, 1.01\unit) {%
                \includegraphics[height=1\unit]{images/truncated/2-points.pdf}%
            };
            \node[inner sep=0, anchor=south west] at (0, 0) {%
                \includegraphics[height=1\unit]{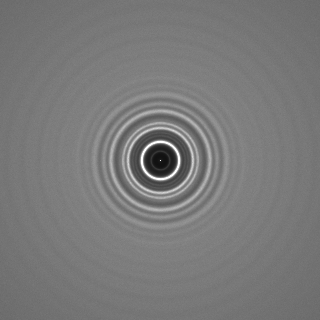}%
            };
        \end{tikzpicture}&
        \begin{tikzpicture}
            \node[inner sep=0, anchor=south west] at (0, 1.01\unit) {%
                \includegraphics[height=1\unit]{images/truncated/3-points.pdf}%
            };
            \node[inner sep=0, anchor=south west] at (0, 0) {%
                \includegraphics[height=1\unit]{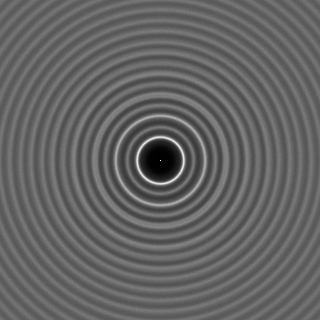}%
            };
        \end{tikzpicture}&
        \begin{tikzpicture}
            \node[inner sep=0, anchor=south west] at (0, 1.01\unit) {%
                \includegraphics[height=1\unit]{images/truncated/4-points.pdf}%
            };
            \node[inner sep=0, anchor=south west] at (0, 0) {%
                \includegraphics[height=1\unit]{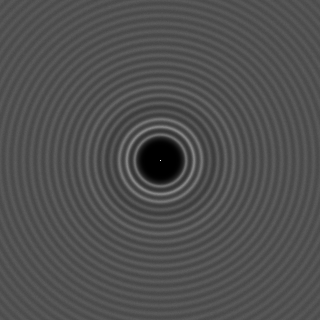}%
            };
        \end{tikzpicture}&
        \begin{tikzpicture}
            \node[inner sep=0, anchor=south west] at (0, 1.01\unit) {%
                \includegraphics[height=1\unit]{images/truncated/5-points.pdf}%
            };
            \node[inner sep=0, anchor=south west] at (0, 0) {%
                \includegraphics[height=1\unit]{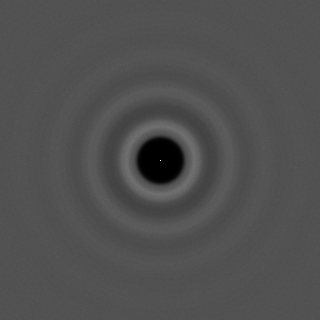}%
            };
        \end{tikzpicture}&
        \begin{tikzpicture}
            \node[inner sep=0, anchor=south west] at (0, 1.01\unit) {%
                \includegraphics[height=1\unit]{images/truncated/6-points.pdf}%
            };
            \node[inner sep=0, anchor=south west] at (0, 0) {%
                \includegraphics[height=1\unit]{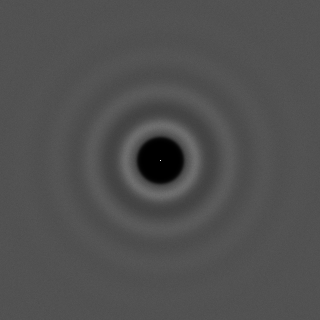}%
            };
        \end{tikzpicture}&
        \includegraphics[height=2.01\unit]{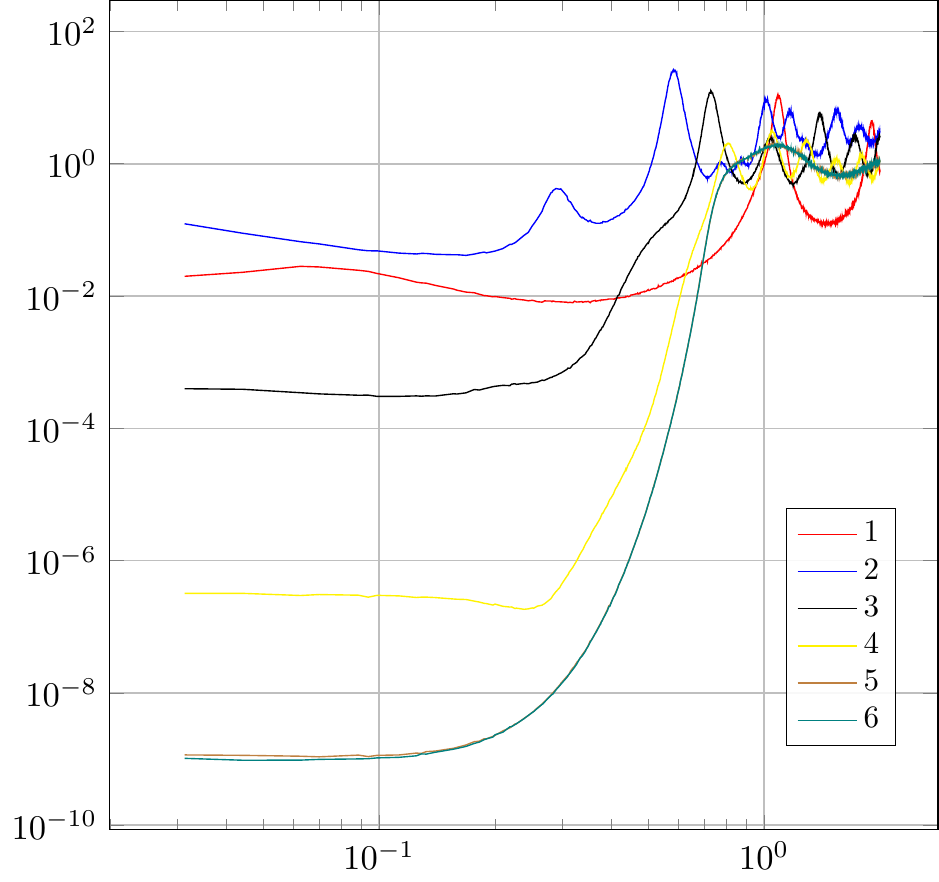}\\[1mm]
        1$\sigma$ & 2$\sigma$ & 3$\sigma$ & 4$\sigma$ & 5$\sigma$ & 6$\sigma$ & Radial Power
    \end{tabular*}}\vspace{-9pt}
    \caption{
        Actual effect of kernel truncation, using the same number of iterations 10K.
    }
    \label{fig:truncated-results}
\end{figure*}
%%%%%%%%%%%%%%%%%%%%%%%%%%%%%%%%%%%%%
%%%%%%%%%%%%%%%%%%%%%%%%%%%%%%%%%%%%%

\subsection{Kernel Support}
Similar to the misleading spatial-domain intuition to use a small kernel width $\sigma$, as discussed in Sec.~\ref{sec:kernel parameter}, it is also tempting to consider only a local neighborhood for optimization: (i) it has linear complexity for optimization compared to the quadratic complexity when considering all the sample points, and (ii) it seems to quickly converge towards a large Poisson-disk radius in the spatial domain~\cite{Oztireli10Spectral,Fattal2011}.
However, in fact, considering only a local neighborhood is detrimental to the optimization, since it brings harmful distortion to the frequency structure of the energy kernel.
Specifically, only considering a local neighborhood is equivalent to truncating the kernel $h$, which can be formulated as multiplying $h$ with a box function.
In the frequency domain, multiplying $h$ with a box function transforms into a convolution between $\hat{h}$ and a sinc function. Note that a sinc function is known to be slowly decaying, which suggests that truncating the kernel would expand its frequency support.
As illustrated in Fig.~\ref{fig:truncated kernel}, the smaller the range of the truncation is, the more are the higher frequencies leaking in, which makes most of the common algorithms tend to settle at patches of regular structures.
Fig.~\ref{fig:truncated-results} shows actual results obtained by truncated kernels, revealing different kinds of distortions.
The 1$\sigma$ support effectively addresses the first ring of Voronoi neighbors, leading to a similar result to Centroidal Voronoi Tessellation (CVT).

To avoid these distortions, we therefore use full kernel support by considering all the point pairs when evaluating the energy term in Eq.~\eqref{eq:energy granular}.
This choice of global optimization is arguably the second most important design choice we had to make, following the choice of $\sigma$.
It leads to quadratic time complexity, but gives more accurate and much higher-quality results.
It is also the gateway to high-dimensional blue noise, as we will see in the following subsection.

There is a limit, though, to the effective kernel support by the numerical precision of the machine, and the mutual energy is effectively zero after 6/9-$\sigma$ steps with float/double data types.
A practical implementation may take advantage of this, especially with a large number of points, but for the point counts used in this paper the saving would not offset the extra coding complexity.
Note that these neighborhoods are still considerably wider than the ranges used in common algorithms, e.g., \cite{Oztireli10Spectral,Fattal2011}.

%%%%%%%%%%%%%%%%%%%%%%%%%%%%%%%%%%%%%%%%%%%%%%%%%%%%%%%%%%%%%%%%%%%%%%%%%%%%%

\subsection{Toroidal Boundary}

Most of the application scenarios favor point sets optimized with toroidal boundaries.
This allows the points to be uniformly distributed over the domain, and enables toroidal shifting to randomize the set \cite{Cranley76Randomization}. Additionally, having toroidal boundaries is also helpful during kernel-based optimizations to keep a balance for the points at the boundaries.

The essence of a toroidal domain is that each point ``sees'' every other point on both sides of each axis, which raises a question about which image of a point to use?
The obvious answer is to take the nearest, which works well as long as the number of points is large enough relative to the effective kernel support of the numeric precision, but is incorrect otherwise: in theory, all replicas, in all dimensions, must be considered to attain the correct frequency profile of the energy kernel.
At first glance this may sound infeasible, but the separable nature of the Gaussian kernel comes in quite handy here.
For example, the correct energy term $\mathcal{E}_{ij}$ at point \mbox{$p_i=(x_i,y_i)$} due to another point \mbox{$p_j = (x_j, y_j)$} in a toroidal domain is
%%%%%%%%%%%%%%%%%%%%%%%%%%%%%%%%%%%%%
%%%%%%%%%%%%%%%%%%%%%%%%%%%%%%%%%%%%%
\begin{align}
     \mathcal{E}_{ij} 
        & = \sum_{k=-\infty}^\infty \sum_{l=-\infty}^\infty \exp \left(
                - \frac{(x_i - x_j -k)^2 + (y_i-y_j-l)^2}{2\sigma^2} \right)\\
        & = \sum_{k=-\infty}^\infty \exp \left(
                -\frac{(x_i - x_j     -k)^2}{2\sigma^2}\right)
            \sum_{l=-\infty}^\infty \exp \left(
                -\frac{(y_i - y_j - l)^2}{2\sigma^2}\right)
     \,, \label{eq:e_ij}
\end{align}
%%%%%%%%%%%%%%%%%%%%%%%%%%%%%%%%%%%%%
%%%%%%%%%%%%%%%%%%%%%%%%%%%%%%%%%%%%%
where we dropped the scaling factors for sake of simplicity.
This formulation of $\mathcal{E}_{ij}$ is helpful in developing scalable algorithms and understanding the frequency spectrum of blue noise in a toroidal domain. 
The first sum of Gaussians in Eq.~\eqref{eq:e_ij} can be shown to satisfy
%Not only does this help in developing scalable algorithms, but it also helps in better understanding the frequency spectrum of blue noise in a toroidal domain.
%If we denote \mbox{$x_{ij} = x_i - x_j$}, then the first sum of Gaussians in Eq.~\eqref{eq:e_ij} may be shown to be
%%%%%%%%%%%%%%%%%%%%%%%%%%%%%%%%%%%%%
%%%%%%%%%%%%%%%%%%%%%%%%%%%%%%%%%%%%%
\begin{align}
    \sum_{k=-\infty}^\infty \exp \left(-\frac{(x_{ij} -k)^2}{2\sigma^2}\right)
         & = \sqrt{2\pi}\sigma \vartheta_3\left(-\pi x_{ij}, \exp(-2 \pi^2 \sigma^2)\right)
    \,, \label{eq:theta}
\end{align}
%%%%%%%%%%%%%%%%%%%%%%%%%%%%%%%%%%%%%
%%%%%%%%%%%%%%%%%%%%%%%%%%%%%%%%%%%%%
where $x_{ij}=x_i-x_j$, and $\vartheta_3$ is a Jacobi theta function, which also satisfies
%%%%%%%%%%%%%%%%%%%%%%%%%%%%%%%%%%%%%
%%%%%%%%%%%%%%%%%%%%%%%%%%%%%%%%%%%%%
\begin{align}
    \vartheta_3\left(-\pi x_{ij}, \exp(-2 \pi^2 \sigma^2)\right)
        = 1 + \sum_{f = -\infty}^\infty e^{-2 \pi^2 \sigma^2 f^2} \cos{(2\pi x_{ij} f)}\,. \label{eq:theta_form2}
\end{align}
%%%%%%%%%%%%%%%%%%%%%%%%%%%%%%%%%%%%%
%%%%%%%%%%%%%%%%%%%%%%%%%%%%%%%%%%%%%
This gives the energy function in terms of harmonics, offering an alternative implementation.
We actually verified empirically that almost identical results are obtained by summing the Gaussians or the Gaussian-weighted cosines.

\begin{figure}
    \centering
    \includegraphics[width=1\columnwidth]{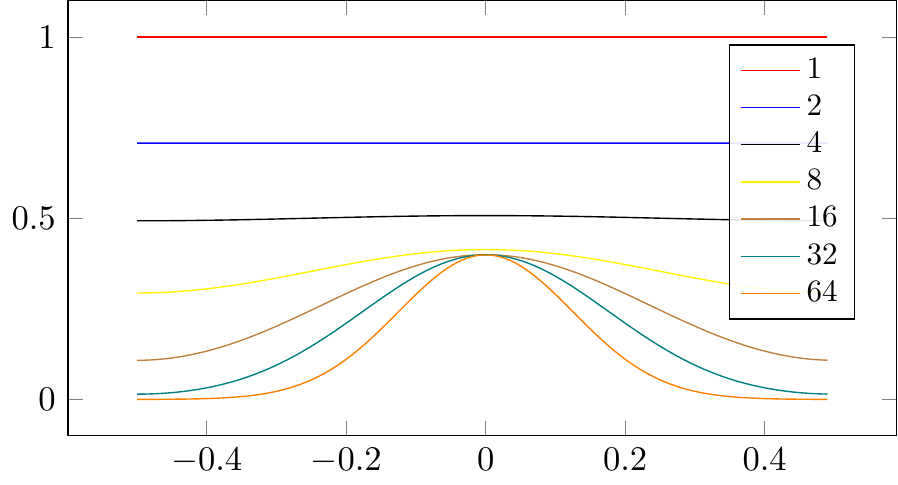}\vspace{-6pt}
    \caption{%
       Energy kernel profiles for different nominal number of points per axis.
       The actual kernel is obtained by multiplying these profiles.
       For example, in 2D, the energy kernel for 64 points is obtained as a Cartesian product of two 8-point profiles.
    }
    \label{fig:theta kernels}
\end{figure}

Thus, even though we are placing Gaussian kernels over the points, to obtain the ideal blue noise spectrum in a toroidal domain the kernels should be folded over the domain boundaries, over and over, to make a theta function profile.
Note that the kernel changes with the number of points, as illustrated in Fig.~\ref{fig:theta kernels}.
% %%%%%%%%%%%%%%%%%%%%%%%%%%%%%%%%%%%%%
% %%%%%%%%%%%%%%%%%%%%%%%%%%%%%%%%%%%%%
% %%%%%%%%%%%%%%%%%%%%%%%%%%%%%%%%%%%%%
% %%%%%%%%%%%%%%%%%%%%%%%%%%%%%%%%%%%%%
Notably, the composite kernel, obtained by multiplying 1D profiles, is not isotropic, but is deformed to cope with the rectangular shape of the domain.
For example, for 64 points in 2D, the energy kernel is a product of two 8-point profiles, which look more like raised cosines than Gaussian because the low-frequency harmonics have much higher weights.
Note also that the kernels are almost flat with low point counts, implying that the actual shape of the distribution does not significantly matter in these cases, as the periodicity of the domain dominates.

%%%%%%%%%%%%%%%%%%%%%%%%%%%%%%%%%%%%%%%%%%%%%%%%%%%%%%%%%%%%%%%%%%%%%%%%%%%%%

\subsection{Bounded Domain}
To optimize the points in a bounded domain we need a strategy to keep them within the domain boundary, since otherwise the minimum energy is obtained just by letting the points go infinitely far apart.
Just restricting the point locations does not work: the points will condensate at the boundaries, since that gives the minimal energy.
The more appropriate way is to give the domain itself an appropriately scaled energy to attract the points.
This is equivalent to the semantically more meaningful model of simulating a fictitious presence of points outside the domain.
Modeling the domain as a continuum of infinitesimal points, the analytic nature of the gradient force in Eq.~\eqref{eq:grad1} offers a neat analytical solution.
For a point $\mathbf{x}_i$ in 2D, the energy gradient due to the domain can be modeled as
%%%%%%%%%%%%%%%%%%%%%%%%%%%%%%%%%%%%
%%%%%%%%%%%%%%%%%%%%%%%%%%%%%%%%%%%%
\begin{equation}
    \Scale[0.95]{\nabla \mathcal{E}(\mathbf{x}_{i})
        =-\int_{0}^{1}\int_{0}^{1}\exp\left(-\frac{(x-x_{i})^{2}+(y-y_{i})^{2}}{4\sigma^{2}}\right)
        \left(\begin{array}{c}x-x_{i}\\y-y_{i}\end{array}\right)\,dx\,dy\,}.
\end{equation}
%%%%%%%%%%%%%%%%%%%%%%%%%%%%%%%%%%%%
%%%%%%%%%%%%%%%%%%%%%%%%%%%%%%%%%%%%
The $x$-axis component can be computed by splitting the integration:
%%%%%%%%%%%%%%%%%%%%%%%%%%%%%%%%%%%%
%%%%%%%%%%%%%%%%%%%%%%%%%%%%%%%%%%%%
\begin{align}
    \Scale[0.95]{\nabla \mathcal{E}_{x}(\mathbf{x}_{i})
         =-\int_{0}^{1}\exp\left(-\frac{\left(y-y_{i}\right)^{2}}{4\sigma^2}\right)\,dy\,\int_{0}^{1}\exp\left(-\frac{\left(x-x_{i}\right)^{2}}{4\sigma^{2}}\right)\left(x-x_{i}\right)\,dx\,},
\end{align}
%%%%%%%%%%%%%%%%%%%%%%%%%%%%%%%%%%%%
%%%%%%%%%%%%%%%%%%%%%%%%%%%%%%%%%%%%
which eventually evaluates to
%%%%%%%%%%%%%%%%%%%%%%%%%%%%%%%%%%%%
%%%%%%%%%%%%%%%%%%%%%%%%%%%%%%%%%%%%
\begin{equation}
    \nabla \mathcal{E}_{x}(\mathbf{x}_{i})
        \propto
            \left(\mathrm{erf}\left(\frac{1-y_{i}}{2\sigma}\right) + 
                \mathrm{erf}\left(\frac{y_{i}}{2\sigma}\right)\right)
            \left(e^{-\left(\frac{1-x_{i}}{2\sigma}\right)^{2}} -
                e^{-\left(\frac{x_{i}}{2\sigma}\right)^{2}}\right)\,.
\end{equation}
%%%%%%%%%%%%%%%%%%%%%%%%%%%%%%%%%%%%
%%%%%%%%%%%%%%%%%%%%%%%%%%%%%%%%%%%%
The second factor is the 1D gradient, which is just the difference between the Gaussian-weighted distance to the edges of the domain, while the first term gives a Gaussian-weighted sum of gradients at different vertical stripes.
The $y$-axis component is computed similarly, and the model scales smoothly to any dimensions just by incorporating the respective erf weights.

Noting that we no longer need to consider replicas of the points, we can see that optimization over a bounded domain is simpler and more efficient than over a toroidal domain. One noteworthy aspect of the resulting distribution is that it will be offset from the domain boundaries by an $\mathcal{O}(\sigma)$ distance.

%%%%%%%%%%%%%%%%%%%%%%%%%%%%%%%%%%%%%%%%%%%%%%%%%%%%%%%%%%%%%%%%%%%%%%%%%%%%%

\subsection{High Dimensions}

All the previous derivations automatically scale to higher dimensions.
A tricky note, though, is that the nominal number of points per axis drops quickly below 2, making the energy profile follow an almost flat kernel as shown on the top of Fig.~\ref{fig:theta kernels}.
Then it is natural to ask whether the blue noise energy is well-defined or meaningful in 20 dimensions for example?
The answer is \emph{yes}, since even a small slope of the energy kernel profile will get scaled up quickly after we multiply it over the dimensions, which will lead to a considerable variance in the energy for different point distributions. 
We then note that the energy function is smooth and continuous, which means that, applying a decent algorithms, the points will continue to move down the valleys of the energy field. As shown in Fig.~\ref{fig:spectrum}(b),
we actually managed to obtain blue noise point sets in higher dimensions of the same quality as the 2D point sets, judging by their frequency spectra.

%%%%%%%%%%%%%%%%%%%%%%%%%%%%%%%%%%%%%%%%%%%%%%%%%%%%%%%%%%%%%%%%%%%%%%%%%%%%%

\subsection{Convergence\label{sec:convergence}}

As mentioned in Section~\ref{sec:optimization method}, the Hessian of the energy function in Eq.~\eqref{eq:energy whole} is inherently singular, which makes any gradient decent algorithm eventually linear, hence non-converging to a complete rest.
The only converging algorithms we are aware of are those applied to discrete domains, e.g., \cite{Ulichney93Void,Hanson05Halftoning,Ahmed2021Optimizing}.
An intuitive explanation to this is that each point should end up at a trough of the Gaussian field induced by the other points. This goes the fastest when a point is at a distance of $\sigma$ from the target trough, which is the peak of the gradient
%%%%%%%%%%%%%%%%%%%%%%%%%%%%%%%%%%%%%
%%%%%%%%%%%%%%%%%%%%%%%%%%%%%%%%%%%%%
\begin{equation}
    \frac{d\,e^{-x^2/2\sigma^2}}{dx} = -\frac{x}{\sigma^2} e^{-x^2/2\sigma^2}
\end{equation}
%%%%%%%%%%%%%%%%%%%%%%%%%%%%%%%%%%%%%
%%%%%%%%%%%%%%%%%%%%%%%%%%%%%%%%%%%%%
of a Gaussian, but becomes rather slow near the trough, which is an inverted peak of a Gaussian, as the gradient goes to 0.

\begin{figure}
  \setlength{\unit}{0.49\columnwidth}        % This will make all the gaps 0.05 unit
  \centering
  {\scriptsize
    \begin{tabular*}{1\columnwidth}{@{}c@{\extracolsep{\fill}}c@{}}
        \includegraphics[width=1\unit]{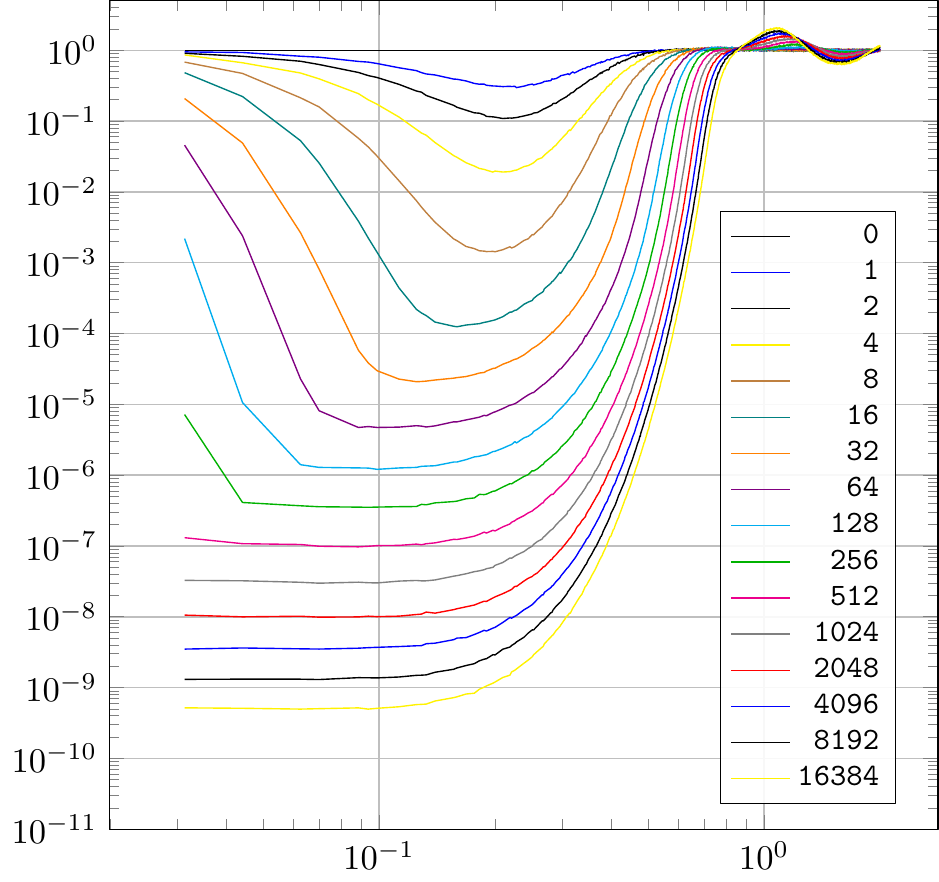}&%
        \includegraphics[width=1\unit]{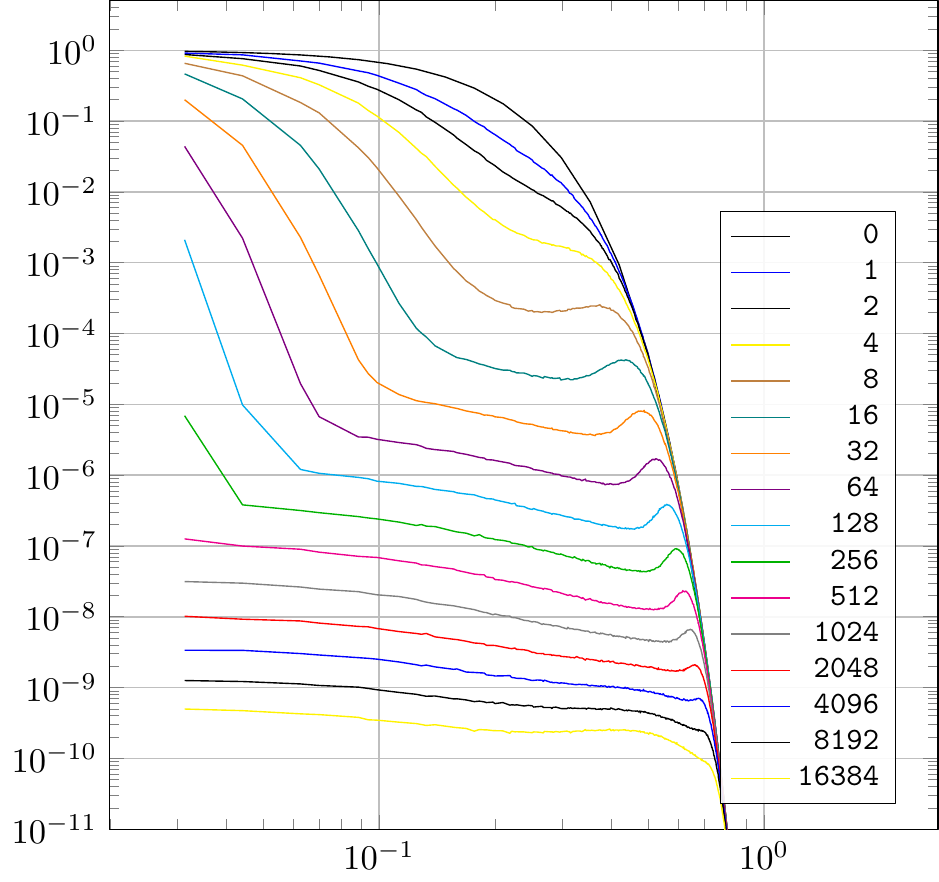}\\[1mm]
        (a) $P(\pmb{\omega})$ & (b) $\vert\hat{A}\vert^2(\pmb{\omega})$
    \end{tabular*}}\vspace{-9pt}
  \caption{\label{fig:convergence}%
    Evolution of the radial power curve of our algorithm at different iterations counts, showing the power spectrum of (a) the discrete points and (b) the Gaussian-filtered set.
  }\vspace{-6pt}
\end{figure}

Unlike common blue noise optimization algorithms, however, observing the preceding considerations makes our model has considerably lower tendency to converge to the global minimum of a triangular grid in the 2D case.
The plot in Fig.~\ref{fig:teaser}(f), for example, uses a million iterations, and still maintains the isotropic blue-noise spectrum.
To demonstrate these insights, in Fig.~\ref{fig:convergence} we show plots of the power spectrum at different iterations during gradient-descent optimization.
%%%%%%%%%%%%%%%%%%%%%%%%%%%%%%%%%%%%%
%%%%%%%%%%%%%%%%%%%%%%%%%%%%%%%%%%%%%
%%%%%%%%%%%%%%%%%%%%%%%%%%%%%%%%%%%%%
%%%%%%%%%%%%%%%%%%%%%%%%%%%%%%%%%%%%%
As expected from a gradient decent algorithm, we see that it starts excavating the energy from the most feasible part somewhere in the middle.
If the parameters are set correctly then this point will be at the peak of the gradient of the energy, which is the $\sigma$ of the kernel image in the frequency domain.
Scaling the gradient could accelerate the optimization process, but may pose some distortion into the frequency spectrum.
The algorithm keeps reducing the energy in the middle until it reaches some level, possibly constrained by the numerical resolution of the points, then starts spreading towards the ends.
As discussed in Section~\ref{sec:kernel parameter}, the outward direction, with exponential decay, carries much less energy than the inward linear-decay direction. This is where the points start to spread evenly globally.
Thus, even though the optimization algorithm is frequency biased, it is securely band-restricted, as seen in Fig.~\ref{fig:convergence}(b), and would eventually lead to a semi-flat spectrum of the filtered point set, thus attaining the favorable quadrexponential profile.

In our experiments we used a parallel implementation on the GPU.
We note that, apart from the time step, the optimization process is deterministic, and the resulting blue noise depends only on the initial distribution.
Looking at the convergence behavior, we note that some of the previous methods may not have used a sufficient number of iterations. For example, \"{O}ztireli \shortcite{Oztireli10Spectral} considered only 10 iterations, while Fattal \shortcite{Fattal2011} uses 15 iterations per scale in a logarithmic subdivision.
This clearly shows the importance of correct parameter settings: the favorable blue noise profile actually appears around 500 iterations, and we recommend 10k iterations.

Finally, we note from Fig.~\ref{fig:convergence} that the algorithm converges linearly at the beginning, and starts to slow down from around 500 iterations, when the spectrum takes its designated shape.
We do not claim any optimally of our algorithm, and better convergence rates might be attainable.

%%%%%%%%%%%%%%%%%%%%%%%%%%%%%%%%%%%%%%%%%%%%%%%%%%%%%%%%%%%%%%%%%%%%%%%%%%%%%

\subsection{Algorithm Outlines}

Observing these guidelines that constrain the frequency behavior of optimization we were able to come close to the idealized model, and reach unprecedented blue-noise quality, as we will show in Section~\ref{sec:results}.
Further, our algorithm progressively converges towards higher quality, and more iterations are consistently better.
Algorithm~\ref{alg:optimization} lists our steps for optimization in a uniform unit torus of any number of dimensions, and we provide our actual implementations in the supplementary materials.
%%%%%%%%%%%%%%%%%%%%%%%%%%%%%%%%%%%%
%%%%%%%%%%%%%%%%%%%%%%%%%%%%%%%%%%%%
%%%%%%%%%%%%%%%%%%%%%%%%%%%%%%%%%%%%
\SetKwInOut{KwIn}{Input}
\SetKwInOut{KwOut}{Output}
\begin{algorithm} [tb]
    \caption{
        Uniform blue-noise optimization. The variable names $\{i,j,k\}$ are used for the nested loop counters for a reference point, another point, and a replica index of the other point. The variable $g$ stands for Gaussian, while $g^{\prime}$ stands for its gradient. The value of ``periods'' is decided by the numeric precision.
    }
    \label{alg:optimization}
    \KwIn{A list $p$ of $N$ point locations in the $d$-dimensional unit torus $[0, 1)^d$.}
    \KwOut{An optimized list of locations that minimizes the BN energy for a filtering kernel size of $\sigma^2$.}
    \Repeat{Optimization criteria met} {
        \For{$i\leftarrow 0$ \KwTo $N - 1$} {
            $g^{\prime}[i] \leftarrow 0$\;
            \For{$j\leftarrow 0$ \KwTo $N - 1$, $j \neq i$}{
                $g_{ij}\leftarrow 0$\;
                $g^{\prime}_{ij}\leftarrow 0$\;
                \For{$dim\leftarrow 0$ \KwTo $Dimensions - 1$}{
                    $x \leftarrow p_i[dim] - p_j[dim]$\;
                    \If{$x < 0$} {
                        $x \leftarrow x + 1$\;
                    }
                    \For{$k\leftarrow 1 - Periods$ \KwTo $Periods$}{
                        $x_k \leftarrow x - k$\;
                        $g_{ij}[dim] \leftarrow g_{ij}[dim] + e^{-\frac{x_k^2}{2\sigma^2}}$\;
                        $g^{\prime}_{ij}[dim] \leftarrow g^{\prime}_{ij}[dim] + x_k \cdot e^{-\frac{x_k^2}{2\sigma^2}}$\;
                    }
                }
                \For{$dim_\mathrm{ref}\leftarrow 0$ \KwTo $Dimensions - 1$}{
                    \For{$dim \leftarrow 0$ \KwTo $Dimensions - 1$}{
                        \If{$dim \neq dim_\mathrm{ref}$}{
                            $g_{ij}^{\prime}[dim_\mathrm{ref}] \leftarrow g_{ij}^{\prime}[dim_\mathrm{ref}] \cdot g_{ij}[dim]$\;
                        }
                    }
                    $g^{\prime}[i][dim_\mathrm{current}] \leftarrow g_{ij}^{\prime}[dim_\mathrm{current}]$\;
                }
            }
        }
        \For{$i\leftarrow 0$ \KwTo $N - 1$} {
            $p[i]  \leftarrow p[i] + g^{\prime}[i]$\;
        }
    }
\end{algorithm}
%%%%%%%%%%%%%%%%%%%%%%%%%%%%%%%%%%%%
%%%%%%%%%%%%%%%%%%%%%%%%%%%%%%%%%%%%
%%%%%%%%%%%%%%%%%%%%%%%%%%%%%%%%%%%%

%%%%%%%%%%%%%%%%%%%%%%%%%%%%%%%%%%%%%%%%%%%%%%%%%%%%%%%%%%%%%%%%%%%%%%%%%%%%%
%%%%%%%%%%%%%%%%%%%%%%%%%%%%%%%%%%%%%%%%%%%%%%%%%%%%%%%%%%%%%%%%%%%%%%%%%%%%%
%%%%%%%%%%%%%%%%%%%%%%%%%%%%%%%%%%%%%%%%%%%%%%%%%%%%%%%%%%%%%%%%%%%%%%%%%%%%%

\subsection{Adaptive Sampling}

Typical contemporary blue-noise samplers are expected to support adaptive and importance sampling, where the sample density is spatially varied in accordance with a given density map. We combine four ideas to extend our model to adaptive sampling.
From Schmaltz et al. \shortcite{Schmaltz2010} we borrow the idea of treating pixels as negatively weighted kernels that attract the sample points. In our model, the idea is to minimize the variance of a zero-mean sum of Gaussians comprising the (positive) points and the (negative) pixels. From Fattal \shortcite[Figure~2]{Fattal2011} we borrow the idea of shaping the kernels in accordance with the local density, so we introduce a shaping factor $a_k$ to shape the normalized energy kernels
%%%%%%%%%%%%%%%%%%%%%%%%%%%%%%%%%%%%
%%%%%%%%%%%%%%%%%%%%%%%%%%%%%%%%%%%%
\begin{equation}
    g_k(\mathbf{x}) = a_k \exp \left( -a_k \frac{\Vert \mathbf{x} - \mathbf{x}_k\Vert^2}{2\sigma^2} \right)
\end{equation}
%%%%%%%%%%%%%%%%%%%%%%%%%%%%%%%%%%%%
%%%%%%%%%%%%%%%%%%%%%%%%%%%%%%%%%%%%
so that they shrink in high-density areas to capture higher frequency details.
We then developed our own Algorithm~\ref{alg:kernel optimization} for adapting the kernel shapes.
%%%%%%%%%%%%%%%%%%%%%%%%%%%%%%%%%%%%
%%%%%%%%%%%%%%%%%%%%%%%%%%%%%%%%%%%%
%%%%%%%%%%%%%%%%%%%%%%%%%%%%%%%%%%%%
\SetKwInOut{KwIn}{Input}
\SetKwInOut{KwOut}{Output}
\begin{algorithm} [!t]
    \caption{
        Optimize shaping factors of normalized kernels to reproduce an adaptively sampled density map.
    }
    \label{alg:kernel optimization}
    \KwIn{A nominal kernel width $\sigma$ and list $\{\mathbf{x}_k\}_{k=1}^N$ of kernel centers.}
    \KwOut{An optimized list $\{a_k\}_{k=1}^N$ of kernel amplitudes.}
    Initialize all amplitudes assuming a uniform density: $a_k \leftarrow 1$\;
    \For{$I$-iterations} {
        Compute accumulated density at each point: $d_k=\sum_{l\neq k} a_l \exp\left(-a_l\frac{\left\Vert \mathbf{x}_k - \mathbf{x}_l \right\Vert^2}{2\sigma^2}\right)$\;
        Set all amplitudes to respective densities: $a_k = d_k$\;
        Normalize $a_k$ so that their squares average to 1\;
    }
\end{algorithm}
%%%%%%%%%%%%%%%%%%%%%%%%%%%%%%%%%%%%
%%%%%%%%%%%%%%%%%%%%%%%%%%%%%%%%%%%%
%%%%%%%%%%%%%%%%%%%%%%%%%%%%%%%%%%%%
% While we conceived this algorithm independently, it is closely connected to variable kernel density estimation \cite{Terrell1992variable}.
Finally, we extend the variance term in Eq.~\eqref{eq:variance uniform} to the adaptive kernels case:
%%%%%%%%%%%%%%%%%%%%%%%%%%%%%%%%%%%%%
%%%%%%%%%%%%%%%%%%%%%%%%%%%%%%%%%%%%%
\begin{multline}
    \mathrm{Var}\left(A(\mathbf{X})\right)
        = \frac{1}{N}\sum_{k=1}^{N}\sum_{l=1}^{N} \frac{2\pi\sigma^{2} a_k a_l}{a_k + a_l}
        \exp\left(-a_k a_l\frac{\left\Vert \mathbf{x}_{k}-\mathbf{x}_{l}\right\Vert ^{2}}{2\sigma^{2}(a_k+a_l)}\right) \\
        -\left(2\pi\sigma^{2}\right)^{2}\,,\label{eq:variance variable}
\end{multline}
%%%%%%%%%%%%%%%%%%%%%%%%%%%%%%%%%%%%%
%%%%%%%%%%%%%%%%%%%%%%%%%%%%%%%%%%%%%
from which we extract a loss function 
%%%%%%%%%%%%%%%%%%%%%%%%%%%%%%%%%%%%%
%%%%%%%%%%%%%%%%%%%%%%%%%%%%%%%%%%%%%
\begin{equation}
    \mathcal{E}(\mathbf{x}_{k}) = 
        \frac{\pi\sigma^2}{N}
        \sum_{l\neq k}
        a_{kl}
        \exp\left(
            - a_{kl} \frac{ \left\Vert \mathbf{x}_k - \mathbf{x}_l \right\Vert^{2} }{ 2\sigma^2 }
        \right) \label{eq:energy granular adaptive}
\end{equation}
and its gradient
%%%%%%%%%%%%%%%%%%%%%%%%%%%%%%%%%%%%%
%%%%%%%%%%%%%%%%%%%%%%%%%%%%%%%%%%%%%
\begin{align}
    \nabla \mathcal{E}(\mathbf{x}_{k})
         & =-\frac{\pi}{N}\sum_{l\neq k}^{N} a_{kl}^2\exp\left(-a_{kl}\frac{\left\Vert \mathbf{x}_{k}-\mathbf{x}_{l}\right\Vert ^{2}}{2\sigma^{2}}\right)\left(\mathbf{x}_{k}-\mathbf{x}_{l}\right) \label{eq:grad}
\end{align}
%%%%%%%%%%%%%%%%%%%%%%%%%%%%%%%%%%%%%
%%%%%%%%%%%%%%%%%%%%%%%%%%%%%%%%%%%%%
for individual points, where
%%%%%%%%%%%%%%%%%%%%%%%%%%%%%%%%%%%%%
%%%%%%%%%%%%%%%%%%%%%%%%%%%%%%%%%%%%%
\begin{equation}
    a_{kl} = \frac{2 a_k a_l}{a_k + a_l}
\end{equation}
%%%%%%%%%%%%%%%%%%%%%%%%%%%%%%%%%%%%%
%%%%%%%%%%%%%%%%%%%%%%%%%%%%%%%%%%%%%
is the mutual shaping factor of two kernels.
Details of these derivations are provided in the supplementary materials. Optimization for adaptive domains then proceeds similar to the uniform case, and observes the same guidelines for kernel width and support, iteration counts, etc., but alternates between variance minimization and kernel shaping.
%As a result
Thus, we propose a novel and efficient auxiliary algorithm for adaptive samples (see Algorithm~\ref{alg:kernel optimization}) that may also be used for reconstruction from adaptive samples.
In practice, we use a single iteration of Algorithm~\ref{alg:kernel optimization} per optimization step, and 10 total iterations in reconstruction.

\begin{figure*}
  \setlength{\unit}{0.19\textwidth}
  {\scriptsize
    \begin{tabular*}{1\textwidth}{@{}c@{\extracolsep{\fill}}c@{\extracolsep{\fill}}c@{\extracolsep{\fill}}c@{\extracolsep{\fill}}c@{}}%
    \includegraphics[width=1\unit]{images/spatial/fpo-points.pdf}&%
    \includegraphics[width=1\unit]{images/spatial/kdm-points.pdf}&%
    \includegraphics[width=1\unit]{images/spatial/bnot-points.pdf}&%
    \includegraphics[width=1\unit]{images/spatial/gbn-points.pdf}&%
    \includegraphics[width=1\unit]{images/spatial/bounded-points.pdf}\\
    \includegraphics[width=1\unit]{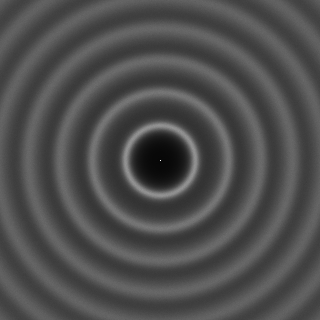}&%
    \includegraphics[width=1\unit]{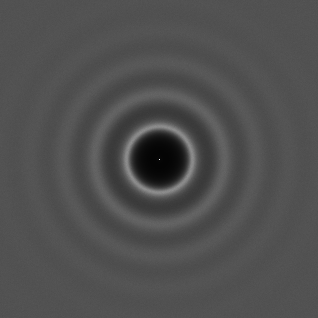}&%
    \includegraphics[width=1\unit]{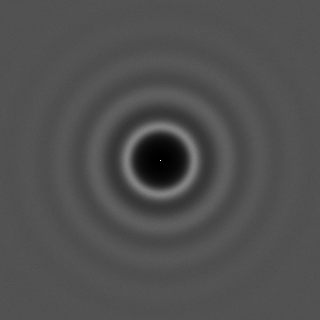}&%
    \includegraphics[width=1\unit]{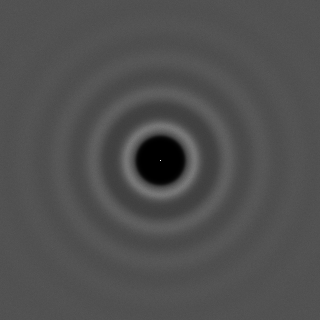}&%
    \includegraphics[width=1\unit]{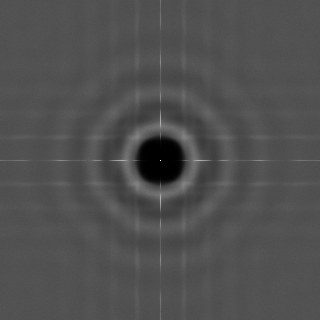}\\
    \includegraphics[width=1\unit]{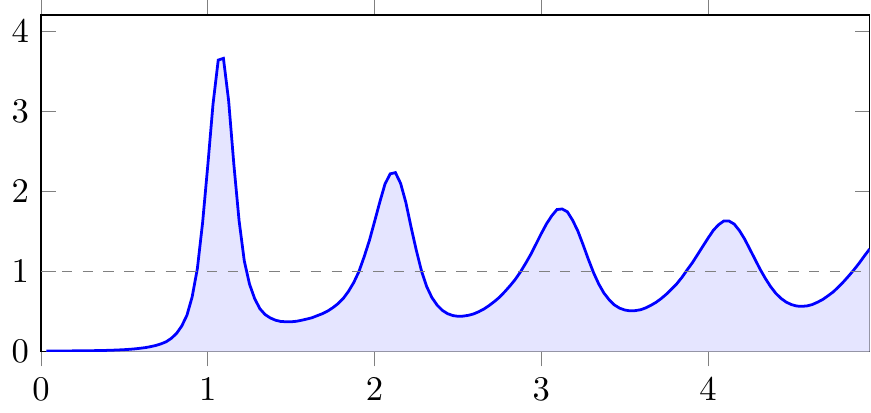}&%
    \includegraphics[width=1\unit]{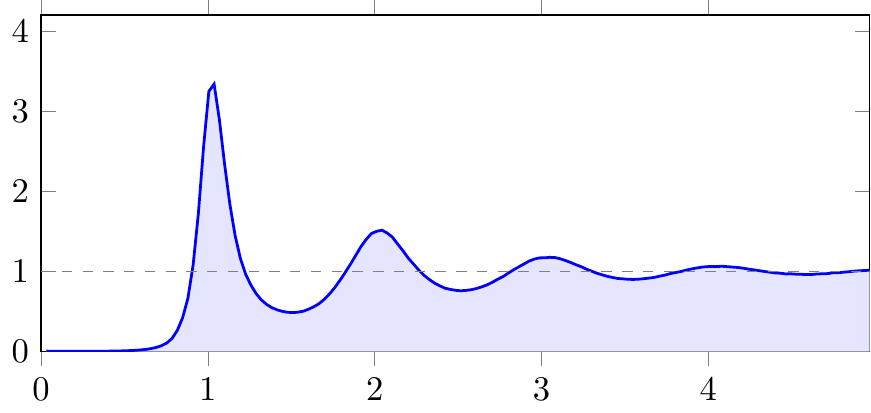}&%
    \includegraphics[width=1\unit]{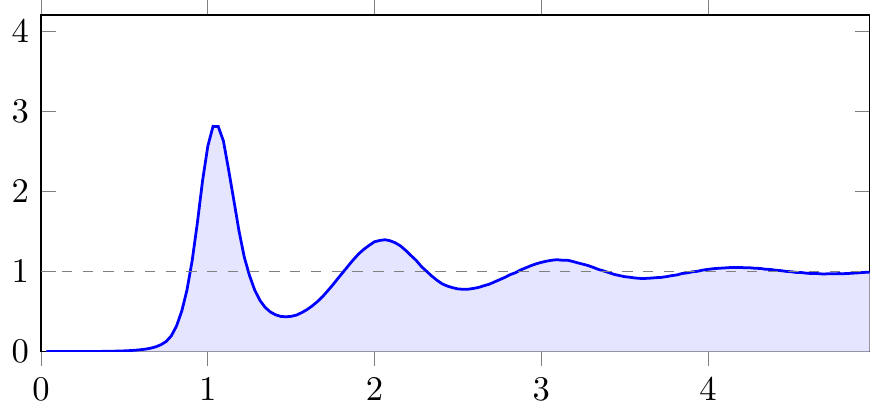}&%
    \includegraphics[width=1\unit]{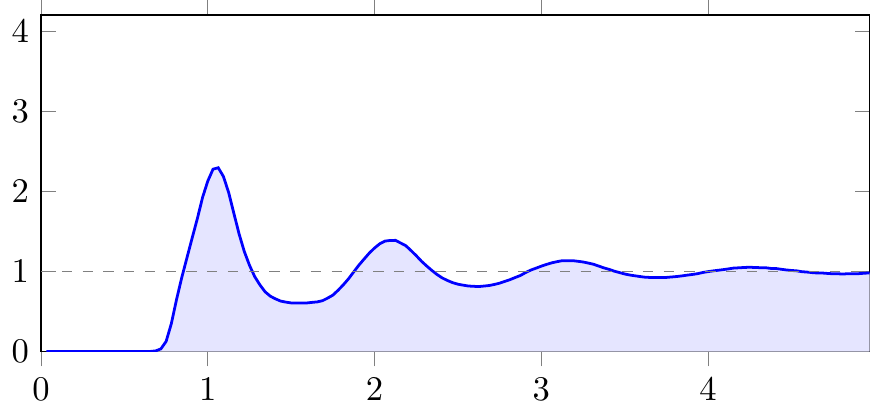}&%
    \includegraphics[width=1\unit]{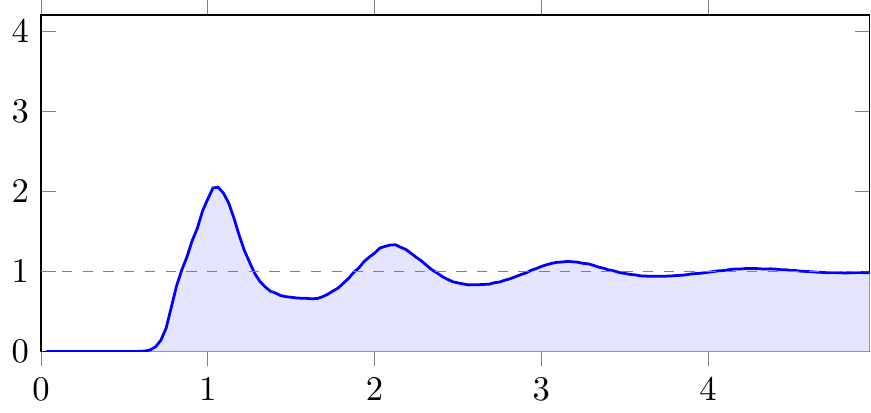}\\
    \includegraphics[width=1\unit]{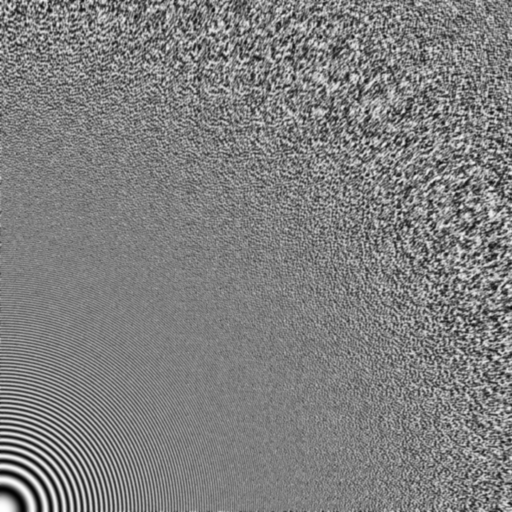}&%
    \includegraphics[width=1\unit]{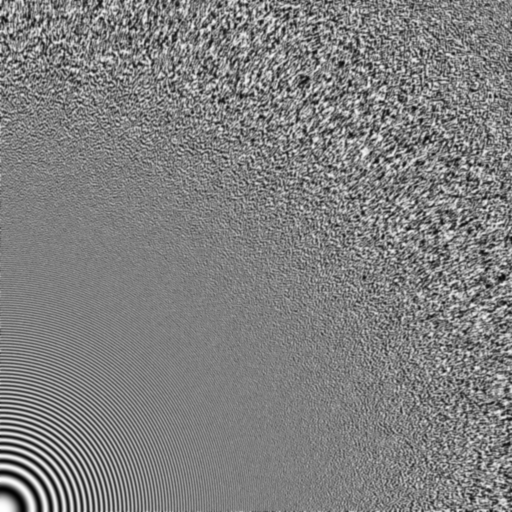}&%
    \includegraphics[width=1\unit]{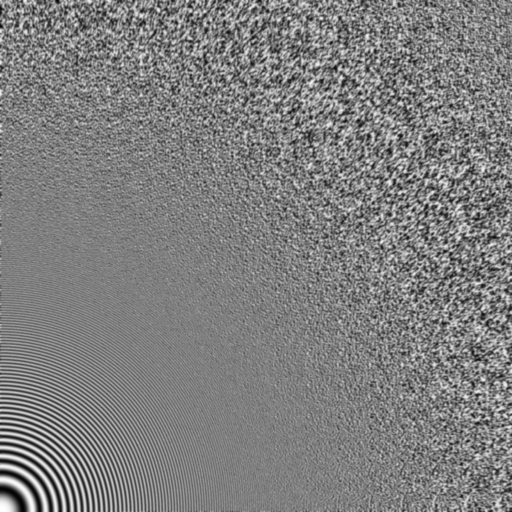}&%
    \includegraphics[width=1\unit]{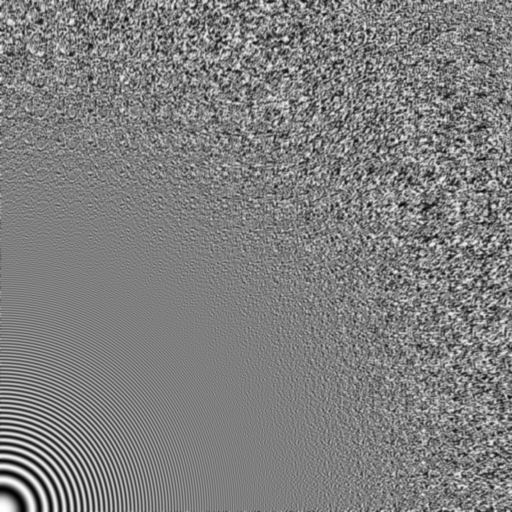}&%
    \includegraphics[width=1\unit]{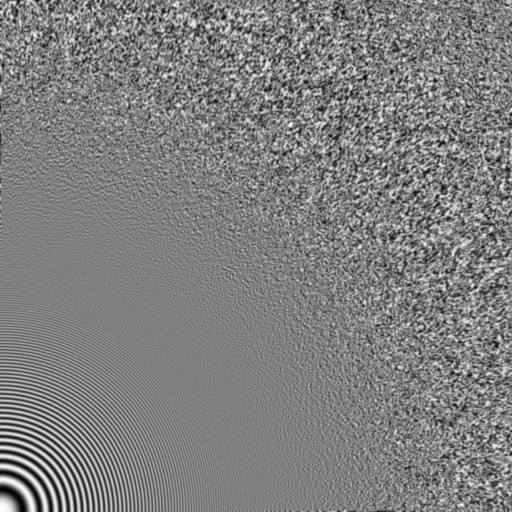}\\
    \includegraphics[width=1\unit]{images/spatial/fpo-valence.pdf}&%
    \includegraphics[width=1\unit]{images/spatial/kdm-valence.pdf}&%
    \includegraphics[width=1\unit]{images/spatial/bnot-valence.pdf}&%
    \includegraphics[width=1\unit]{images/spatial/gbn-valence.pdf}&%
    \includegraphics[width=1\unit]{images/spatial/bounded-valence.pdf}\\[1mm]
    (a) FPO & (b) KDM & (c) BNOT & (d) GBN (Ours) & (e) GBN Bounded Domain
    \end{tabular*}}\vspace{-9pt}
  \centering
    \caption{%
        Common measures of blue noise comparing our algorithm to state of the art techniques.
        The rows respectively show a 1K sample point set, the frequency spectrum and its radial profile averaged over 1K sets, the zone plate, and the valence of the Delaunay triangulation.
    }
    \label{fig:spatial}
\end{figure*}
%%%%%%%%%%%%%%%%%%%%%%%%%%%%%%%%%%%%%%%%%%%%%%%%%%%%%%%%%%%%%%%%%%%%%%%%%%%%%
%%%%%%%%%%%%%%%%%%%%%%%%%%%%%%%%%%%%%%%%%%%%%%%%%%%%%%%%%%%%%%%%%%%%%%%%%%%%%
%%%%%%%%%%%%%%%%%%%%%%%%%%%%%%%%%%%%%%%%%%%%%%%%%%%%%%%%%%%%%%%%%%%%%%%%%%%%%

\section{Results and Comparison\label{sec:results}}

In this section we discuss various practical details of applying the discussed principles and models, and demonstrate the superiority of our method over baselines, achieving higher-quality blue noise, and better results on adaptive sampling for stippling and reconstruction.

%%%%%%%%%%%%%%%%%%%%%%%%%%%%%%%%%%%%%%%%%%%%%%%%%%%%%%%%%%%%%%%%%%%%%%%%%%%%%

\subsection{Spatial Properties} 

We start by benchmarking our GBN against common techniques using the classic measures of blue noise, as shown in Fig.~\ref{fig:spatial}.
Possibly the plots do not reveal a clear difference, except the zoneplate, which manifests the lower noise floor in our point sets in the low-frequency band.
The actual difference is orders of magnitude, however, as revealed in Fig.~\ref{fig:teaser}(g).

Careful inspection of the radial power also reveals that our blue noise has the most flat low-frequency region combined with the smallest peak, which means that it is the least noised and also the least aliased, as reflected in the zoneplate plot.
Our algorithm offers two handles for controlling the noise-aliasing trade-off: the value of $\sigma$ and the number of iterations, as discussed in Section~\ref{sec:kernel parameter} and Section~\ref{sec:convergence}.
For the results in Fig.~\ref{fig:spatial} we set $\sigma=1$ and use 10K iterations.

%%%%%%%%%%%%%%%%%%%%%%%%%%%%%%%%%%%%%%%%%%%%%%%%%%%%%%%%%%%%%%%%%%%%%%%%%%%%%

\subsection{Numerical Integration\label{sec:numerical integration}}

\begin{figure*}
  \setlength{\unit}{0.30\textwidth}
  {\scriptsize
  \begin{tabular*}{1\textwidth}{@{}c@{\extracolsep{\fill}}c@{\extracolsep{\fill}}c@{}}
      \centering
      & (a) Gaussian & (b) Halfspace\\
      \rotatebox{90}{\parbox{1\unit}{\centering{2D}}}&%
      \input{images/integration_tex/gaussians-2d.tex}&%
      \input{images/integration_tex/edge-2d.tex}\\
      \rotatebox{90}{\parbox{1\unit}{\centering{3D}}}&%
      \input{images/integration_tex/gaussians-3d.tex}&%
      \input{images/integration_tex/edge-3d.tex}\\
      \rotatebox{90}{\parbox{1\unit}{\centering{8D}}}&%
      \input{images/integration_tex/gaussians-8d.tex}&%
      \input{images/integration_tex/edge-8d.tex}\\
  \end{tabular*}}\vspace{-9pt}
  \caption{\label{fig:integration}%
    Variance of Monte Carlo integration comparing GBN to common point distributions in two, three, and eight dimensions, using (a) a sum of Gaussians and (b) half-space integrands.
    We average over 100 instances of each family of integrands, 1000 point sets in the 2D case, and 100 sets in 3D and 8D.
  }
\end{figure*}
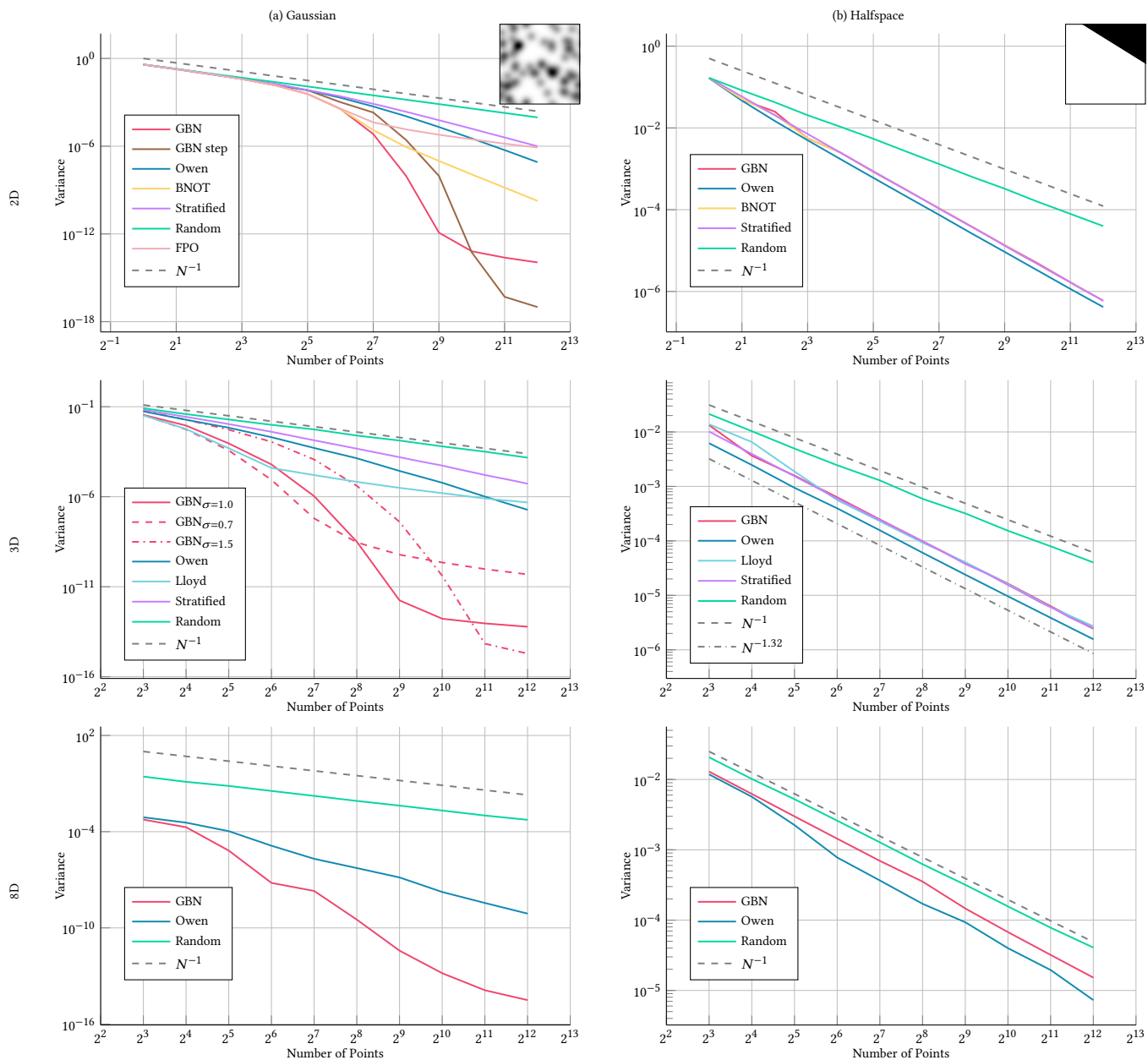

In Fig.~\ref{fig:integration} we show numerical integration comparisons using various sample distributions in two, three, and eight dimensions, using different integrands.
%%%%%%%%%%%%%%%%%%%%%%%%%%%%%%%%%%%%%
%%%%%%%%%%%%%%%%%%%%%%%%%%%%%%%%%%%%%
%%%%%%%%%%%%%%%%%%%%%%%%%%%%%%%%%%%%%
%%%%%%%%%%%%%%%%%%%%%%%%%%%%%%%%%%%%%

As a representative of smooth isotropic functions we use a toroidal sum of 64 Gaussians placed at random points, as illustrated in the top-left corner of the 2D plot.
The spectrum of this integrand is a Gaussian multiplied by white noise; that is, a noisy Gaussian, which is representative of a wide range of signals dominated by a DC level.
We adjust the variance of the Gaussians to the nominal grid frequency of 512 points:
%%%%%%%%%%%%%%%%%%%%%%%%%%%%%%%%%%%%%
%%%%%%%%%%%%%%%%%%%%%%%%%%%%%%%%%%%%%
\begin{equation}
    \sigma^{\prime} = 512^{-1/\mathrm{dimensions}}\,,
\end{equation}
%%%%%%%%%%%%%%%%%%%%%%%%%%%%%%%%%%%%%
%%%%%%%%%%%%%%%%%%%%%%%%%%%%%%%%%%%%%
so that we can see the behavior of the sampling sets in both under-sampling and over-sampling conditions.

We first note that the variance of random sampling faithfully follows the $N^{-1}$ convergence rate that characterizes Monte Carlo integration. This offers a good reference for comparison.
In all dimensions, GBN evidently outperforms the other distributions by orders of magnitudes.
It exhibits an interesting sigmoid curve that faithfully follows analytical prediction.
Indeed, if the integrand is $\mathcal{O}\left(e^{-\alpha x^2}\right)$, and the the spectrum of the point process is $\mathcal{O}\left(e^{\beta x^2}\right)$, as we demonstrated, then Pilleboue et al. \shortcite{Pilleboue15Variance} estimate the variance to be the inner product of the two spectra, which is $\mathcal{O}\left(e^{(\beta-\alpha ) x^2}\right)$. When $\beta$ is smaller than $\alpha$, as is the case for $N < 512$, the variance spectra have an exponential decay, which explains the quick convergence rate at smaller point counts. On the other hand, the variance spectrum grows exponentially for a large number of points, but that is already bounded by the small peak in the points spectrum, no more than 4 times the white noise level, while the decay of the integrand should have already reached a very low level. Thus, GBN offers a very decent behavior in integration, being very competent in under-sampling conditions, and losing its advantage only at already-redundant sampling rates. This conclusion is very interesting, because it implies that blue noise faithfully brings its reconstruction advantage to numerical integration.

As an example of directionally biased signals we use a half-space step by picking a random point and a random direction. All point sets exhibit the same convergence rate that comes closer to white noise with the increase of dimensions, but Owen-scrambled Sobol slightly excels in this test. Our guess is that this improvement arises from the chances where the direction of the step comes aligned with one or another axis.
It is worth noting here that the separability of the Gaussian kernel makes it an excellent block for optimizing on projective sub-domains, but we deem this as an application-specific level of detail.

%%%%%%%%%%%%%%%%%%%%%%%%%%%%%%%%%%%%%%%%%%%%%%%%%%%%%%%%%%%%%%%%%%%%%%%%%%%%%

\subsection{Adaptive Sampling and Reconstruction}

\begin{figure*}
  \setlength{\unit}{0.24\textwidth}
  {\scriptsize
    \begin{tabular*}{1\textwidth}{@{}c@{\extracolsep{\fill}}c@{\extracolsep{\fill}}c@{\extracolsep{\fill}}c@{}}%
    \includegraphics[width=1\unit]{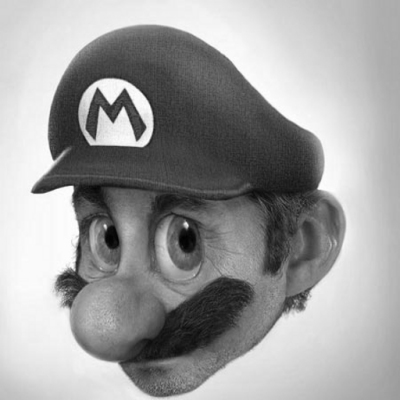}&%
    \includegraphics[width=1\unit]{images/adaptive/mario-bnot.pdf}&%
    \includegraphics[width=1\unit]{images/adaptive/mario-kdm.pdf}&%
    \includegraphics[width=1\unit]{images/adaptive/mario-plus.pdf}\\%
    \includegraphics[width=1\unit]{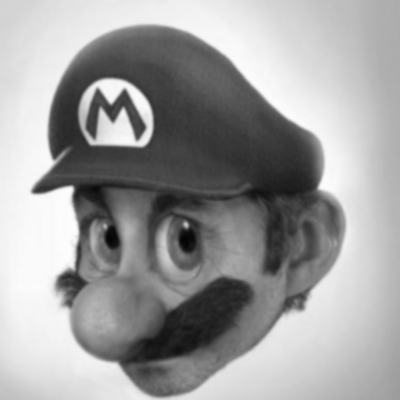}&%
    \includegraphics[width=1\unit]{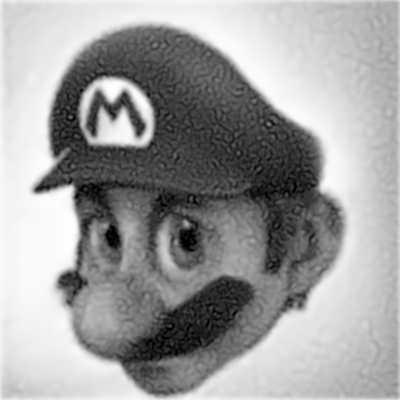}&%
    \includegraphics[width=1\unit]{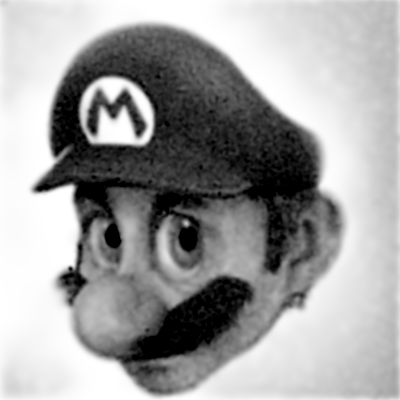}&%
    \includegraphics[width=1\unit]{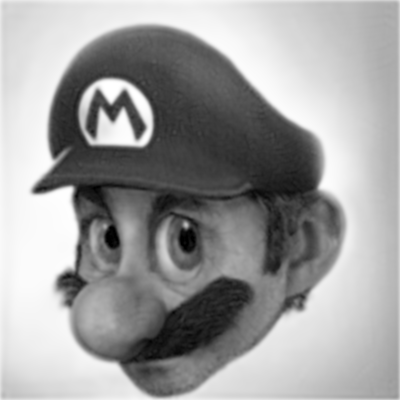}\\%
    (a) Input & (b) BNOT & (c) KDM & (d) GBN (Ours)
    \end{tabular*}}\vspace{-9pt}
  \centering
    \caption{%
        Adaptive sampling a density map, comparing
        (b) BNOT \protect{\cite{Goes12Blue}},
        (c) KDM \protect{\cite{Fattal2011}}, and
        (d) our algorithm.
        The input image is shown in (a), along with a blurred version below.
        The reconstructed images using Algorithm~\ref{alg:kernel optimization} are shown in the bottom row.
    }
    \label{fig:adaptive}
\end{figure*}

Similar to the uniform sampling case, we could obtain superior results for adaptive sampling, as illustrated in Fig.~\ref{fig:adaptive}.
%%%%%%%%%%%%%%%%%%%%%%%%%%%%%%%%%%%%%
%%%%%%%%%%%%%%%%%%%%%%%%%%%%%%%%%%%%%
%%%%%%%%%%%%%%%%%%%%%%%%%%%%%%%%%%%%%
%%%%%%%%%%%%%%%%%%%%%%%%%%%%%%%%%%%%%
As predicted by the spectral profile in Fig.~\ref{fig:teaser}(g), our blue noise is much less noisy than BNOT and KDM.
We also used our auxiliary Algorithm~\ref{alg:kernel optimization} for reconstructing the density map back from the samples, and it proves faithful in capturing the noise visible to the eye.
We also note that our method is structure-aware, making the samples follow the feature lines.
More results are provided in the supplementary materials.
In Fig.~\ref{fig:adaptive convergence} we show the convergence behavior of the optimization process.
It bears close resemblance to the uniform case in Fig~\ref{fig:convergence}, suggesting the correction of our model.
Finally, we note that our method is parametric, enabling uniformity-noise trade-offs by controlling the $\sigma$ parameter and the number of iterations.

%%%%%%%%%%%%%%%%%%%%%%%%%%%%%%%%%%%%%
%%%%%%%%%%%%%%%%%%%%%%%%%%%%%%%%%%%%%
\begin{figure*}[!t]
  \setlength{\unit}{0.245\textwidth}
  {\scriptsize
  \begin{tabular*}{1\textwidth}{@{}c@{\extracolsep{\fill}}c@{\extracolsep{\fill}}c@{\extracolsep{\fill}}c@{}}
      \centering
      \includegraphics[width=1\unit]{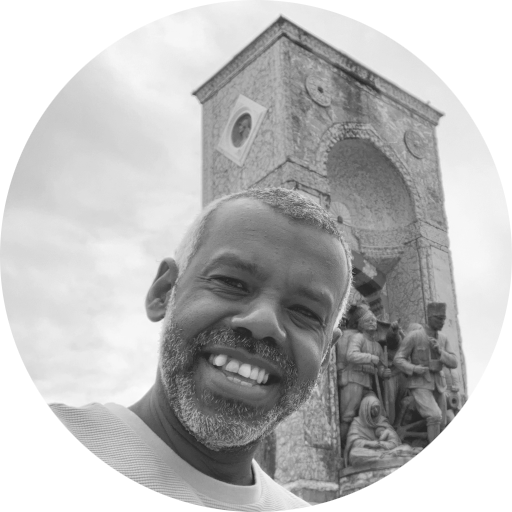}&%
      \includegraphics[width=1\unit]{images/convergence/adaptive/i0000000.pdf}&%
      \includegraphics[width=1\unit]{images/convergence/adaptive/i0000001.pdf}&%
      \includegraphics[width=1\unit]{images/convergence/adaptive/i0000008.pdf}\\%
      Input & 0 & 1 & 8\\[1mm]
      \includegraphics[width=1\unit]{images/convergence/adaptive/i0000064.pdf}&%
      \includegraphics[width=1\unit]{images/convergence/adaptive/i0000512.pdf}&%
      \includegraphics[width=1\unit]{images/convergence/adaptive/i0004096.pdf}&%
      \includegraphics[width=1\unit]{images/convergence/adaptive/i0032768.pdf}\\%
      64 & 512 & 4K & 32K\\[1mm]
      \includegraphics[width=1\unit]{images/convergence/adaptive/i0262144.pdf}&%
      \includegraphics[width=1\unit]{images/convergence/adaptive/i1048576.pdf}&%
      \includegraphics[width=1\unit]{images/convergence/adaptive/trace.pdf}&%
      \includegraphics[width=1\unit]{images/convergence/adaptive/taksim-bnot-n10k.pdf}\\%
      256K & 1M (Fully Converged) & traces & BNOT\\
  \end{tabular*}}\vspace{-9pt}
  \caption{\label{fig:adaptive convergence}%
        Convergence of our method in adaptive sampling, starting from a weighted random initialization \protect{\cite{Goes12Blue}}, and showing the point distributions at doubled iteration steps. BNOT \protect{\cite{Goes12Blue}} is shown for bench-marking.
        We note that the convergence apparently follows the same behavior as the uniform case of Fig.~\ref{fig:convergence}, starting in middle-range frequencies, and starting to attain the favorable spectrum from around 512.
        Notably, this number of iterations takes roughly the same time as BNOT, by which the quality is already significantly better.
        The last plot traces the actual paths of the points along the process.
  }
\end{figure*}
%%%%%%%%%%%%%%%%%%%%%%%%%%%%%%%%%%%%%
%%%%%%%%%%%%%%%%%%%%%%%%%%%%%%%%%%%%%

%%%%%%%%%%%%%%%%%%%%%%%%%%%%%%%%%%%%%%%%%%%%%%%%%%%%%%%%%%%%%%%%%%%%%%%%%%%%%

\subsection{Complexity}

The time complexity of Algorithm~\ref{alg:optimization} is clearly quadratic, which may be taken as a disadvantage.
Thanks to the separability of the Gaussian kernel, however, as well as the associated theta function, the algorithm scales linearly with dimension, rather than exponentially.
Thus, even though the quadratic complexity is inferior to the log-linear performance of BNOT \cite{Goes12Blue} and the linear convergence of KDM \cite{Fattal2011}, our algorithm remains quadratic in all dimensions, while the time complexity of computing the underlying power diagram in BNOT grows exponentially, as well as the space complexity of the computational grid of KDM.

Our algorithm is inherently parallel, and lends itself to a GPU implementation, which gives a significant boost in performance compared to CPU-based algorithms like BNOT and KDM. In fact, the actual speed performance of our implementation is significantly faster than BNOT with reasonably-sized point sets. For example, computing our recommended 10K iterations of a 4K set takes around 5 seconds on a Titan GPU.
Orders of 100K points can still be computed in a reasonable time, but the quadratic time complexity makes it prohibitively slow to work with orders of millions.
In that case the algorithm may be adapted to work within a local neighborhood of 9$\sigma$, which is quite sufficient.

Finally, our algorithm leads the competition when it comes to coding complexity, as we only use elementary functions of the standard C or Cuda library, and simple array data structures.

%%%%%%%%%%%%%%%%%%%%%%%%%%%%%%%%%%%%%%%%%%%%%%%%%%%%%%%%%%%%%%%%%%%%%%%%%%%%%
%%%%%%%%%%%%%%%%%%%%%%%%%%%%%%%%%%%%%%%%%%%%%%%%%%%%%%%%%%%%%%%%%%%%%%%%%%%%%
%%%%%%%%%%%%%%%%%%%%%%%%%%%%%%%%%%%%%%%%%%%%%%%%%%%%%%%%%%%%%%%%%%%%%%%%%%%%%

\section{Conclusion\label{sec:conclusion}}

In this paper, starting from an abstract model of kernel-based blue noise, we derived a simple formula for what we deem as an ideal blue noise spectrum, and we demonstrated its realizability and feasibility with empirical results.
Through the design of an algorithm, following the theoretical model, we obtained unprecedented quality for 2D and high-dimensional sample distributions.

The results in the paper suggest many directions for future research. 
For example, we demonstrated that properly sampled adaptive BN distributions actually retain a lot more information than can be seen by the bare eye looking at a stippling, and we demonstrated a practically working algorithm for retrieving this information. This poses a natural question about the utility of blue noise for data representation and compression.
Our results demonstrated the superiority of high-dimensional blue noise over known alternatives for numerical integration. Is it time, then, to export blue noise to the Monte Carlo communities outside the graphics community?
Furthermore, since the (adaptive) reconstruction algorithm scales with dimension, then an intriguing question we ask is about the possibility of using a blue-noise set of samples as an equivalent of pixels in high dimensions.
But even in 2D, the idea we demonstrated that the mere sampling with a high-quality blue-noise set implies low-pass filtering, hence suppressing moire effects, for example, poses a question about the utility of blue noise for distributing the photosensors in digital cameras.
With the advent of adaptive kernels in blue-noise sampling, pioneered by Fattal \shortcite{Fattal2011}, the Gaussian filtering associated with blue noise does not necessarily result in blurring, as evident in the sharp details we were able to retrieve in Fig.~\ref{fig:adaptive}(b).

Finally, we reply to a long-standing question: are the different families of algorithms to produce blue noise equivalent?
The answer now is a clear No!
Cellular methods are intrinsically different from \emph{diffused} methods, as evident in the polynomial vs. quadrexponential spectral profile.
We have explained the superiority of the later analytically, and demonstrated it empirically, and we conclude by giving an intuitive insight of the difference.
The partitioning in cellular methods immediately introduces quantization noise all over the domain. Intuitively, we are introducing boundaries of our own that do not actually exist in the domain we are sampling.
Kernel-based methods, in contrast, enable the sample points to work collaboratively, transferring the \emph{claim} gradually between the sample points, which is quite more natural.
This discussion applies to sampling in other fields as well.
For example, a cellular mobile network does not actually build walls to restrict the assigned frequency bands to the planned area of a cell, and it is more natural to plan under the assumption of overlapped coverage.
Thus, GBN may be exported to other fields outside the computer graphics.

% Bibliography
\bibliographystyle{ACM-Reference-Format}
\bibliography{main.bib}

%%%%%%%%%%%%%%%%%%%%%%%%%%%%%%%%%%%%%%%%%%%%%%%%%%%%%%%%%%%%%%%%%%%%%%%%%%%%%
%%%%%%%%%%%%%%%%%%%%%%%%%%%%%%%%%%%%%%%%%%%%%%%%%%%%%%%%%%%%%%%%%%%%%%%%%%%%%
%%%%%%%%%%%%%%%%%%%%%%%%%%%%%%%%%%%%%%%%%%%%%%%%%%%%%%%%%%%%%%%%%%%%%%%%%%%%%
%%%%%%%%%%%%%%%%%%%%%%%%%%%%%%%%%%%%%%%%%%%%%%%%%%%%%%%%%%%%%%%%%%%%%%%%%%%%%
%%%%%%%%%%%%%%%%%%%%%%%%%%%%%%%%%%%%%%%%%%%%%%%%%%%%%%%%%%%%%%%%%%%%%%%%%%%%%
%%%%%%%%%%%%%%%%%%%%%%%%%%%%%%%%%%%%%%%%%%%%%%%%%%%%%%%%%%%%%%%%%%%%%%%%%%%%%
%%%%%%%%%%%%%%%%%%%%%%%%%%%%%%%%%%%%%%%%%%%%%%%%%%%%%%%%%%%%%%%%%%%%%%%%%%%%%
%%%%%%%%%%%%%%%%%%%%%%%%%%%%%%%%%%%%%%%%%%%%%%%%%%%%%%%%%%%%%%%%%%%%%%%%%%%%%
%%%%%%%%%%%%%%%%%%%%%%%%%%%%%%%%%%%%%%%%%%%%%%%%%%%%%%%%%%%%%%%%%%%%%%%%%%%%%
% This is for arXiv submission only:

\appendix

\section{Adaptive Gaussian Blue Noise}\label{sec:GBN}
As mentioned in Sec. 4.10 of the main paper, we can compute the power spectrum of a point distribution with different shaped and sized Gaussian kernels, as similarly proposed by Fattal \shortcite{Fattal2011} to support adaptive samplings.
Specifically, we consider a sum of Gaussians in a toroidal domain:
\begin{equation}
    A\left(\mathbf{X}\right) = \sum_{i=1}^{N}a_i\exp\left(-a_i\frac{\left\Vert \mathbf{x}-\mathbf{x}_{i}\right\Vert ^{2}}{2\sigma^{2}}\right), \label{eq:A}
\end{equation}
Our goal is to optimally place the points to reduce the variance of $A\left(\mathbf{X}\right)$, cf. \cite{Fattal2011, Ahmed2021Optimizing}.
\begin{equation}
    \mathrm{Var}\left(A(\mathbf{x})\right)=E\left( A^{2}(\mathbf{x})\right)-\big(E\left(A(\mathbf{x})\right)\big)^{2} \label{eq:variance}
\end{equation}
Since the second term of Eq.~\eqref{eq:variance} is fixed, our task reduces to minimizing the average square of $A(\mathbf{x})$:
\begin{equation}
    E\left(A^{2}(\mathbf{x})\right) = \frac{1}{N} \int_{D}A^{2}(\mathbf{x})\,d\mathbf{x} \label{eq:avg-sq-integral}
\end{equation}
We proceed by expanding the integrand
\begin{align}
    A^{2}(\mathbf{x})
        & =\left(\sum_{i=1}^{N} a_i \exp\left(-a_i\frac{\left\Vert \mathbf{x}-\mathbf{x}_{i}\right\Vert ^{2}}{2\sigma^{2}}\right)\right)^{2}\\
        & =\sum_{i=1}^{N}\sum_{j=1}^{N} a_i a_j \exp\left(-a_i\frac{\left\Vert \mathbf{x}-\mathbf{x}_{i}\right\Vert ^{2}}{2\sigma^{2}}\right)\exp\left(-a_j\frac{\left\Vert \mathbf{x}-\mathbf{x}_{j}\right\Vert ^{2}}{2\sigma^{2}}\right)\nonumber \\
        & =\sum_{i=1}^{N}\sum_{j=1}^{N} a_i a_j \underbrace{\exp\left(-\frac{a_i\left\Vert \mathbf{x}-\mathbf{x}_{i}\right\Vert ^{2}+a_j\left\Vert \mathbf{x}-\mathbf{x}_{j}\right\Vert ^{2}}{2\sigma^{2}}\right)}_{g_{i,j}(\mathbf{x})}\,. \label{eq:A-square}
\end{align}
Each point can be broken down axis-wise, and aggregated back later, so looking only at the x-axis, for example,
\begin{align}
    g_{i,j}(x)
        & =\exp\left(-\frac{a_i(x-x_{i})^{2}+a_j(x-x_{j})^{2}}{2\sigma^{2}}\right)\,\\
        %& =\exp\left(-\frac{\sigma^{2}\left(x^{2}-2xx_{i}+x_{i}^{2}\right)+\sigma^{2}\left(x^{2}-2xx_{j}+x_{j}^{2}\right)}{2\sigma^{2}\sigma^{2}}\right)\\
        & =\exp\left(-\frac{\left(a_i+a_j\right)x^{2}-2x\left(a_jx_{i}+a_ix_{j}\right)+a_jx_{i}^{2}+a_ix_{j}^{2}}{2\sigma^{2}}\right)\,.
\end{align}
After rearranging the terms and completing the square, we get:
\begin{align}
    g_{i,j}(x)
        & =\exp\left(-\frac{a_ia_j\left(x_{i}-x_{j}\right)^{2}}{2\sigma^2\left(a_i+a_j\right)}\right)\exp\left(-\frac{\left(x-\frac{a_jx_{i}+a_ix_{j}}{a_i + a_j}\right)^{2}}{2\sigma^{2}/\left(a_i + a_j\right)}\right)\,.
\end{align}
We can then aggregate the results from the $x,y$ axis and get:
\begin{equation}
    g_{i,j}(\mathbf{x})
        = \exp\left(-\frac{a_i a_j\left\Vert \mathbf{x}_{i}-\mathbf{x}_{j}\right\Vert ^{2}}{2\sigma^2\left(a_i + a_j\right)}\right)\exp\left(-\frac{\left\Vert \mathbf{x}-\frac{a_j\mathbf{x}_{i}+a_i\mathbf{x}_{j}}{a_i + a_j}\right\Vert ^{2}}{2\sigma^{2}/\left(a_i + a_j\right)}\right)\,,\label{eq:g_x(i,j) last}
\end{equation}
which is a multiplication of two Gaussians.
The latter is a Gaussian of variance $\frac{\sigma^{2}}{a_i + a_j}$
and is centered at the average of $\mathbf{x}_i$ and $\mathbf{x}_j$, each weighted by the variance of the other.
Substituting Eq.~\eqref{eq:g_x(i,j) last} in Eq.~\eqref{eq:A-square} gives
\begin{equation}
    \Scale[1]{ A^{2}(\mathbf{x}) 
        = \sum\limits_{i=1}^{N}\sum\limits_{j=1}^{N} a_i a_j \exp\left(-\frac{a_ia_j\left\Vert \mathbf{x}_{i}-\mathbf{x}_{j}\right\Vert ^{2}}{2 \sigma^2\left(a_i + a_j\right)}\right)\exp\left(-\frac{\left\Vert \mathbf{x}-\frac{a_j\mathbf{x}_{i}+a_i\mathbf{x}_{j}}{a_i + a_j}\right\Vert ^{2}}{2\sigma^{2}/\left(a_i + a_j\right)}\right)}
\end{equation}
and substituting this in Eq.~\eqref{eq:avg-sq-integral} gives
\begin{multline}
    E\left(A^{2}(\mathbf{x})\right) 
        =\frac{1}{N}\sum_{i=1}^{N}\sum_{j=1}^{N} a_i a_j
        \exp\left(-\frac{a_i a_j\left\Vert \mathbf{x}_{i}-\mathbf{x}_{j}\right\Vert ^{2}}{2\sigma^2\left(a_i + a_j\right)}\right)
        \cdot \\
        \int_{D}\exp\left(-\frac{\left\Vert \mathbf{x}-\frac{a_j\mathbf{x}_{i}+a_i\mathbf{x}_{j}}{a_i + a_j}\right\Vert ^{2}}{2\sigma^{2}/\left(a_i + a_j\right)}\right)d\mathbf{x}\,.\label{eq:average-sqaure-mid}
\end{multline}
The integrations evaluate to constants that do not depend on the point locations but on the kernel variances; hence,
\begin{equation}
    E\left(A^{2}(\mathbf{x})\right)
        = \frac{1}{N}\sum_{i=1}^{N}\sum_{j=1}^{N} 
        \frac{2\pi a_i a_j\sigma^{2}}{a_i + a_j}
        \exp\left(
            -\frac{a_ia_j\left\Vert \mathbf{x}_{i}-\mathbf{x}_{j}\right\Vert ^{2}}{2\sigma^2\left(a_i+ a_j\right)}
        \right)\,.    \label{eq:average square final}
\end{equation}
Similarly,
\begin{equation}
    E\left(A(\mathbf{x})\right)
        = \frac{1}{N}\sum\limits_{i=1}^{N} a_i \int_{D}\exp\left(-\frac{a_i\left\Vert \mathbf{x}-\mathbf{x}_{i}\right\Vert ^{2}}{2\sigma^{2}}\right)\,d\mathbf{x}
        = 2\pi \sigma^2\,,\label{eq:average}
\end{equation}
And the final term of variance is obtained by substituting Eqs.~(\ref{eq:average square final}, \ref{eq:average}) in Eq.~\eqref{eq:variance}:
\begin{multline}
    \mathrm{Var}\left(A(\mathbf{x})\right)
        = \frac{1}{N}\sum_{i=1}^{N}\sum_{j=1}^{N}\frac{2\pi a_i a_j\sigma^{2}}{a_i + a_j} \exp\left(-\frac{a_ia_j\left\Vert \mathbf{x}_{i}-\mathbf{x}_{j}\right\Vert ^{2}}{2\sigma^2\left(a_i + a_j\right)}\right) \\
        - \left(2\pi \sigma^2\right)^{2}\,.\label{eq:variance final}
\end{multline}
Note that setting all $a_i = \sigma^2$ reduces $\mathrm{Var}\left(A(\mathbf{x})\right) $ to Eq. ~(1) in the main paper.

%----------------------------------------
\begin{figure}[!t]
  \setlength{\unit}{0.48\columnwidth}        % This will make all the gaps 0.05 unit
  \centering
  {\scriptsize
    \begin{tabular*}{1\columnwidth}{@{}c@{\extracolsep{\fill}}c@{\extracolsep{\fill}}c@{}}
        \begin{tikzpicture}
            \node[inner sep=0, anchor=south west] at (0, 0.51\unit) {%
                \includegraphics[height=0.5\unit]{images/spectrum/our-step-points.pdf}%
            };
            \node[inner sep=0, anchor=south west] at (0, 0) {%
                \includegraphics[height=0.5\unit]{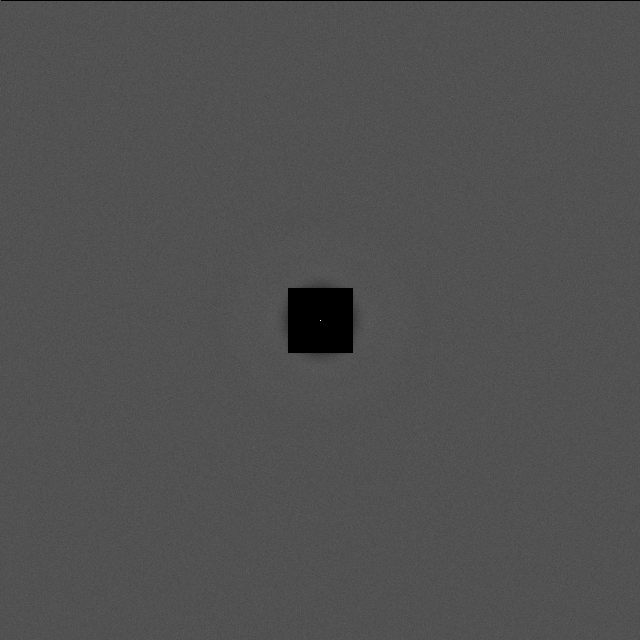}%
            };
        \end{tikzpicture}&
        \begin{tikzpicture}
            \node[inner sep=0, anchor=south west] at (0, 0.51\unit) {%
                \includegraphics[height=0.5\unit]{images/spectrum/heck-points.pdf}%
            };
            \node[inner sep=0, anchor=south west] at (0, 0) {%
                \includegraphics[height=0.5\unit]{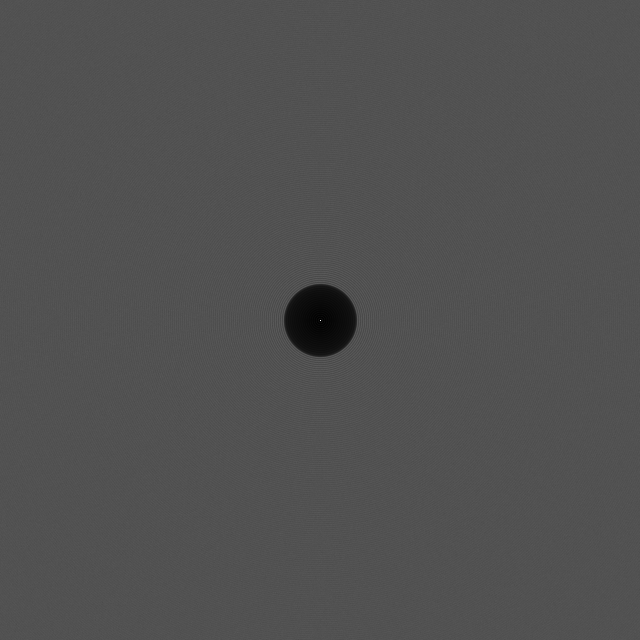}%
            };
        \end{tikzpicture}&
        \includegraphics[height=1.01\unit]{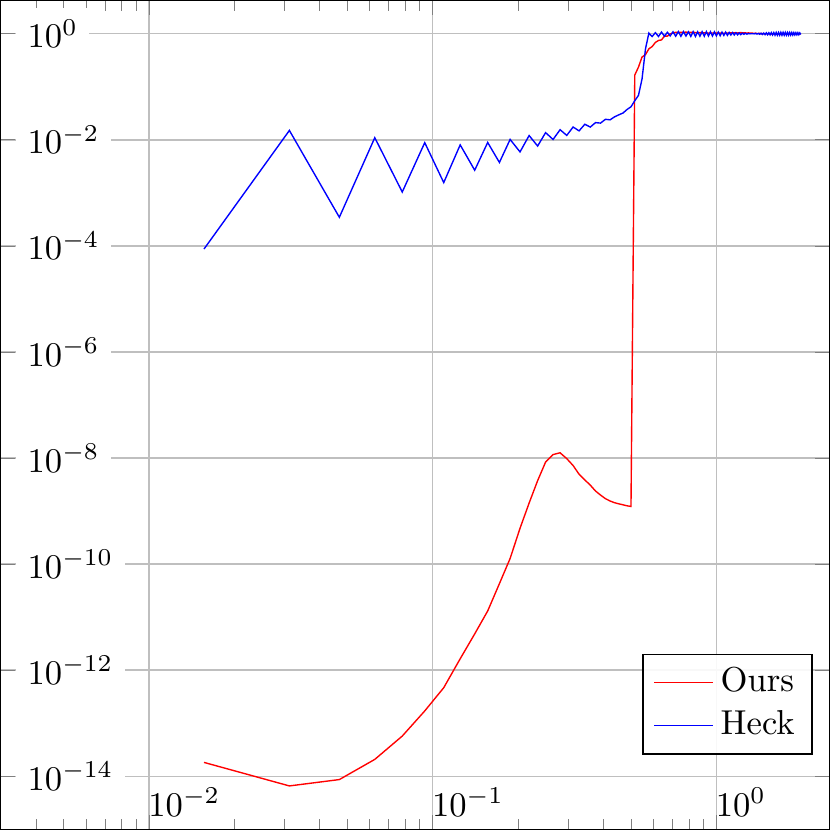}\\[1mm]
        (a) Our Step & (b) Heck & (c) Radial Power Spectra
    \end{tabular*}}\vspace{-9pt}
    \caption{
        A step function obtained by truncating the noise energy function at a frequency of $\sqrt{N}/2$. Please note the logarithmic scale, and that our radial power for low frequencies is smaller than $10^{-8}$. 
        The step of Heck et al. \protect{\shortcite{Heck2013}} is shown in (b) for comparison. 
    }
    \label{fig:step}
\end{figure}
%----------------------------------------

\section{Step Blue Noise}
To test our assumption of optimization economics in Sec.4.3 in the main paper, we use Eq.~(23) to truncate the the energy term at a certain frequency $f_\mathrm{max}$, and minimized the series sum using a gradient descent algorithm.
For values of \mbox{$f_\mathrm{max} < \sqrt{N}/2$} we obtained an almost perfect step, as illustrated in Fig.~\ref{fig:step}.

In this setup, we are optimizing a set of \mbox{}$(2\sqrt{N})^2/2 = 2N$ distinct frequencies, where dividing by two comes from the origin symmetry of the spectrum. Now, for 2D points, this is exactly the same dimensions as the input point sets, hence the optimization goes more smoothly towards a minimum, and reaches a noise floor orders of magnitudes below Heck et al. \shortcite{Heck2013}, where a substantial energy of the optimization algorithm is devoted towards flattening the high-frequency range.
Note, however, that the spectrum still maintains the exponentially-shaped profile, though not fully converged in this demonstration.
Another advantage of our method is the automatic handling of the domain corners that are just ignored in \cite{Heck2013}. 
Our method offers a neat analytical solution that scales linearly with dimensions, thanks to the separability of the Gaussian, compared to an exponential growth in Zhou et al. \shortcite{Zhou2012}, where frequencies are handled individually.
We actually tested the step in higher dimensions, and it scales smoothly.
The width essentially shrinks quickly, as expected, but the actual low-frequency volume is maintained.

\begin{figure}
  \setlength{\unit}{0.14\columnwidth}        % This will make all the gaps 0.05 unit
  {\scriptsize
    \begin{tabular*}{1\columnwidth}{@{}c@{\extracolsep{\fill}}c@{\extracolsep{\fill}}c@{\extracolsep{\fill}}c@{}}
    \includegraphics[height=1\unit]{images/spectrum/1d/16-0.pdf}&%
    \includegraphics[height=1\unit]{images/spectrum/1d/16-1.pdf}&%
    \includegraphics[height=1\unit]{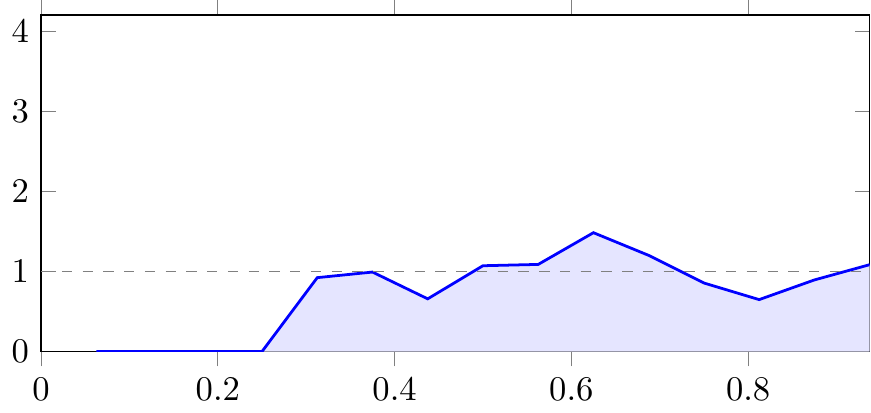}&%
    \includegraphics[height=1\unit]{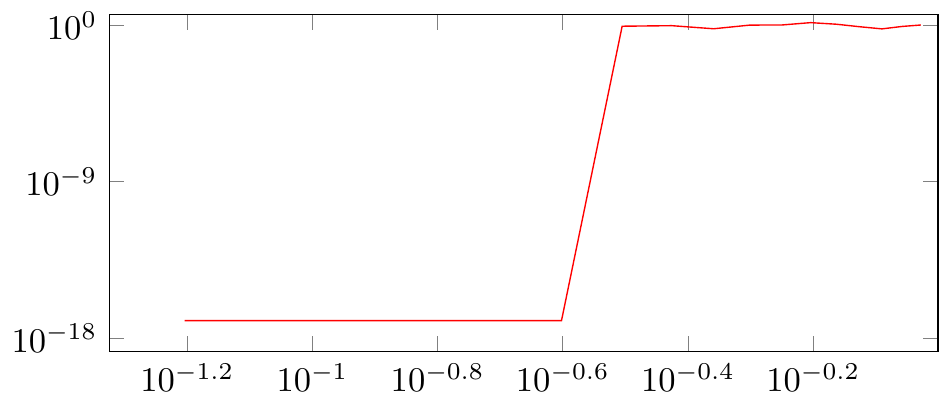}\\[1mm]
    (a) & (b) & (c) & (d)
    \end{tabular*}}\vspace{-9pt}
  \centering
    \caption{
        An example of 1D toroidal blue noise sets showing (a, b) two instances and (c, d) the average periodogram.
        To understand the meaning of this 1D spectrum, one may think of the points as beads on a piece of string wound around a circle of an equal circumference. Now think of a vector from the center of the circle to the center of mass of the points. The magnitude of this vector is precisely the power of frequency 1. For frequency 2, the circle is half-diameter, and the string is wound twice; and so one.
    }
    \label{fig:1D BN}
\end{figure}

One interesting observation is the gradual disappearance of the primary peak in the spectrum with the increase of dimensions.
To understand this, we try to see it in the lower dimensions where it is easier to visualize.
As a proof of concept, we experimented with 1D blue noise, which is quite helpful in understanding.
The problem with 1D blue noise is that pushing any point away from another would inevitably bring it closer to another point on the other side: there are strictly two neighbors that cannot be escaped. This makes any optimization of the BN energy quickly descend towards the inevitable global minimum of a regular grid.
We managed, however, to obtain a blue noise profile in 1D, as illustrated in Fig.~\ref{fig:1D BN}, by using the spectrum truncation idea discussed above. Our experiments so far worked only with a very narrow band, but they provide the first demonstration of 1D blue noise to the best of our knowledge.

%----------------------------------------

%----------------------------------------

Back to the observation about the peaks, when we go 2D, each point may have differently arranged neighbors, and the points may push each other into the larger domain now, until they reach a separation distance where pushing each point would again start to conflict with another point, hence the term \emph{conflict radius} for Poisson disk radius. The average number of neighbors, however, jumps to 6 in 2D for a dense distribution, not only the polar four, and the blue-noise energy is therefore distributed over a number of neighbors that grows faster than the linear growth of the poles. This is closely related to the long-established \emph{kissing-number problem} in mathematics, and we get 13, 24, etc. nearest neighbors of each point that share the conflict energy.
In the frequency domain, the noise energy at the conflict radius is distributed over all the discrete frequencies corresponding to that radius, and we may understand the attenuation of the peak by noting that the surface area of a hyper-sphere grows much faster than its volume, hence we should not expect a concentration of energy at the surface of the excavated low-frequency region as we find in one or two dimensions.

\section{On Radial Power Spectrum\label{app:radial power estimation}}

The efficient and accurate evaluation of periodograms was a corner stone in our work.
Following the guidelines of Schl\"{o}emer et al. \shortcite{Schloemer11PSA}, we built a PSA-like tool on the GPA.
Thanks to the huge computational power of modern GPU's, we were able to compute and process hundreds of thousands spectra to compute smooth yet accurate averages.

Iterating through the Cartesian product of frequencies in an arbitrary number $d$ of dimensions is tricky to code.
For that, we used a trick of iterating linearly to $(2f_\mathrm{max}+1)^d$, the volume of the scanned frequency range, and encode the current loop index in base $2f_\mathrm{max}+1$.
Each digit represents the frequency at the respective dimension.

The second challenge we faced is that the effective frequency range becomes very small in high dimensions, giving a very low resolution for the histogram if we consider the bin size of 1 or 2 used in PSA.
The digital encoding inspired us with a nice solution: each distinct combination of digits represents a distinct radial frequency, and the different permutations of these digits along dimensions give a reasonable support for averaging.
Thus, for high dimensions, we consider the exact radial frequencies in radial average plots.

\begin{figure}
  \setlength{\unit}{0.49\columnwidth}        % This will make all the gaps 0.05 unit
  \centering
  {\scriptsize
    \begin{tabular*}{1\columnwidth}{@{}c@{\extracolsep{\fill}}c@{}}
        \includegraphics[width=1\unit]{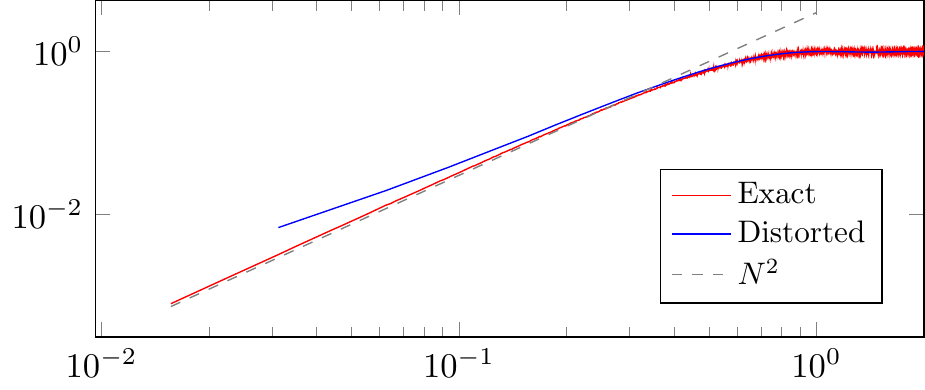}&%
        \includegraphics[width=1\unit]{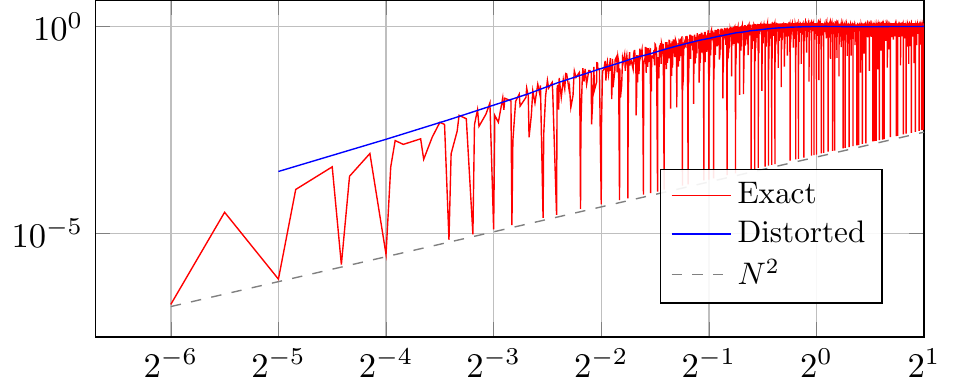}\\[1mm]
        (a) Stratified & (b) Owen-Scrambled Sobol
    \end{tabular*}}\vspace{-9pt}
  \caption{\label{fig:rp correct}%
    Two examples demonstrating the effect of averaging on radial power plots, showing an exact plot on the top versus a distorted one using an annulus width of 2, the default setting of PSA.
  }\vspace{-6pt}
\end{figure}

Trying this also for 2D lead us to learn a very important lesson: for the best understanding, use the exact radial frequency, and do not average them over annulia as proposed by Ulichney \shortcite{Ulichney87Digital,Ulichney88Dithering} and followed up by Schl\"{o}emer et al. \shortcite{Schloemer11PSA}.
The key insight to this is that the periodogram is actually a Fourier series, and is therefore inherently discretised; then averaging at concentric annuli would impose significant distortion, specially in the low frequency range, which is arguably the most important. Intuitively, an annulus width of 1 is a huge distance at radius 1 compared to radius 40, for example, and would shift the center of the frequency bin in a misleading way.
Fig.~\ref{fig:rp correct} shows two examples in log-log scale.
Note the wealth of details concealed behind the average Owen's spectrum.
We also note that, being cellular too, Owen-scrambled Sobol follows the polynomial spectral profile observed in stratified and BNOT distributions.

%----------------------------------------
%----------------------------------------

%----------------------------------------

%----------------------------------------
\begin{figure*}
  \setlength{\unit}{0.33\textwidth}
  \centering
  {\footnotesize
    \begin{tabular*}{1\textwidth}{@{}c@{\extracolsep{\fill}}c@{\extracolsep{\fill}}c@{}}
        \includegraphics[width=1\unit]{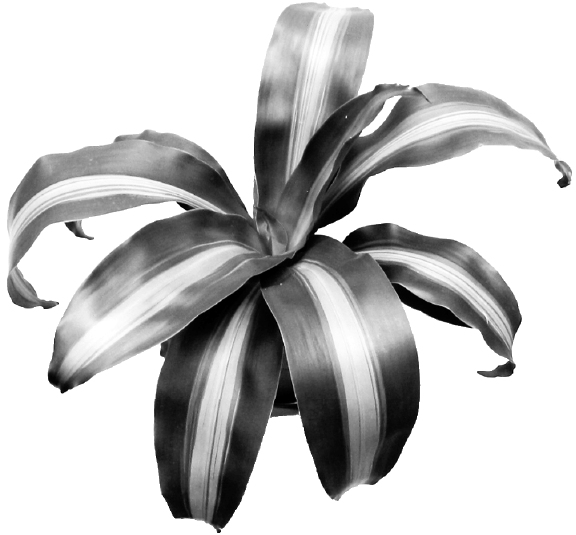}&%
        \includegraphics[width=1\unit]{images/appendix/flower-kdm.pdf}&%
        \includegraphics[width=1\unit]{images/appendix/flower-plus.pdf}\\[1mm]
        (a) Input & (b) KDM & (c) GBN (Ours)
    \end{tabular*}}\vspace{-9pt}
  \caption{\label{fig:flower}%
        Stippling comparison of the flower test image \protect{\cite{Secord2002Weighted}}.
  }\vspace{-6pt}
\end{figure*}
%----------------------------------------

%----------------------------------------
\begin{figure*}
  \setlength{\unit}{0.8\textwidth}
  \centering
  {\footnotesize
  \begin{tikzpicture}%
    \node[inner sep=0, anchor=south west] at (0.8\unit, 0.31\unit) {%
        \frame{\includegraphics[width=0.2\unit]{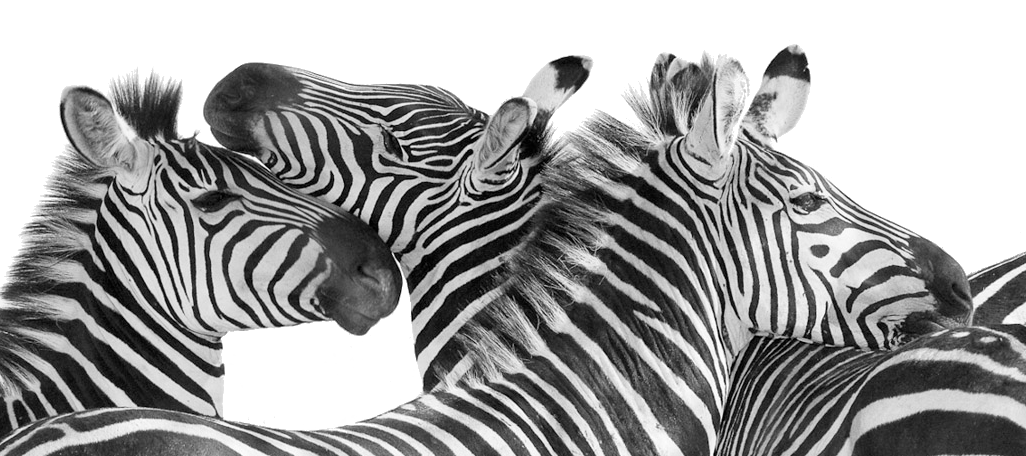}}%
    };%
    \node[inner sep=0, anchor=south west] at (0, 0) {%
        \includegraphics[width=1\unit]{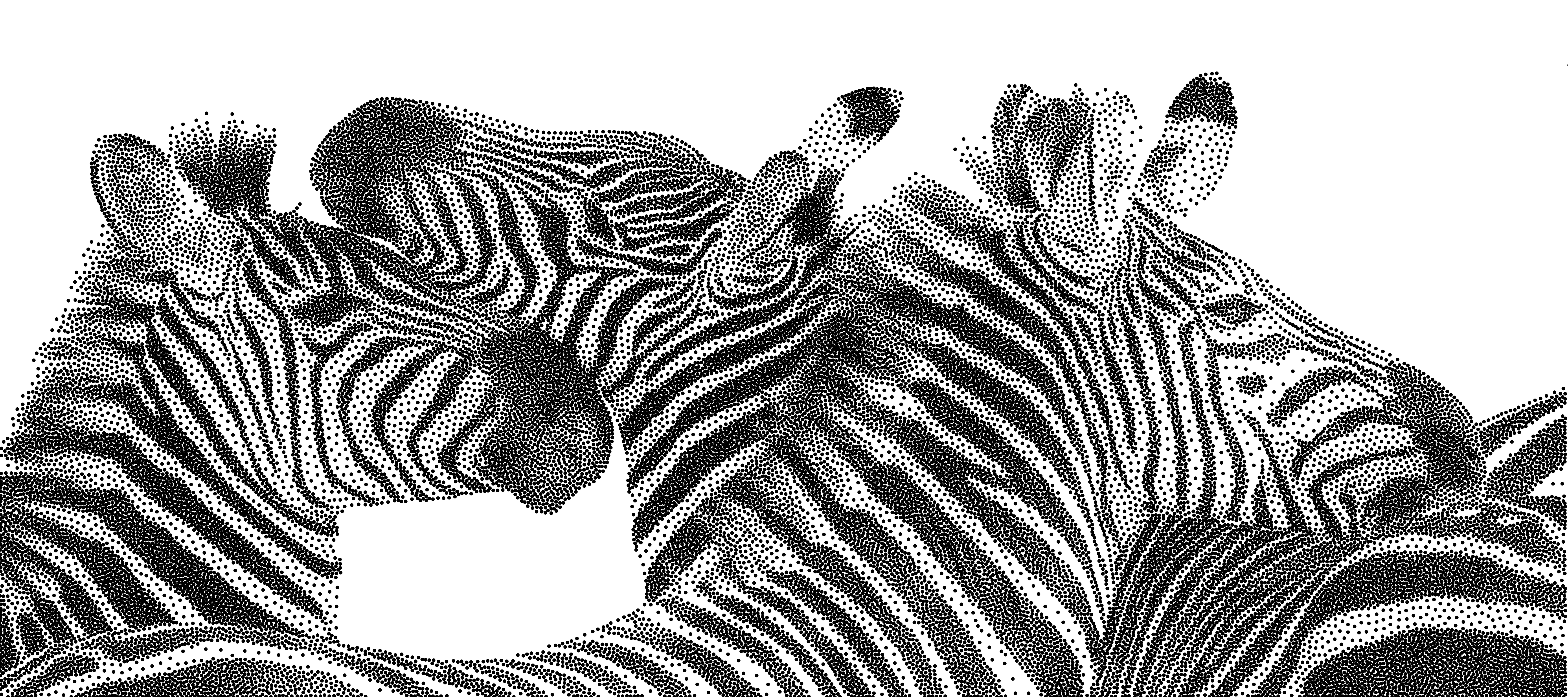}%
    };%
  \end{tikzpicture}\\[2mm]%
  (a) BNOT\\%
  \includegraphics[width=1\unit]{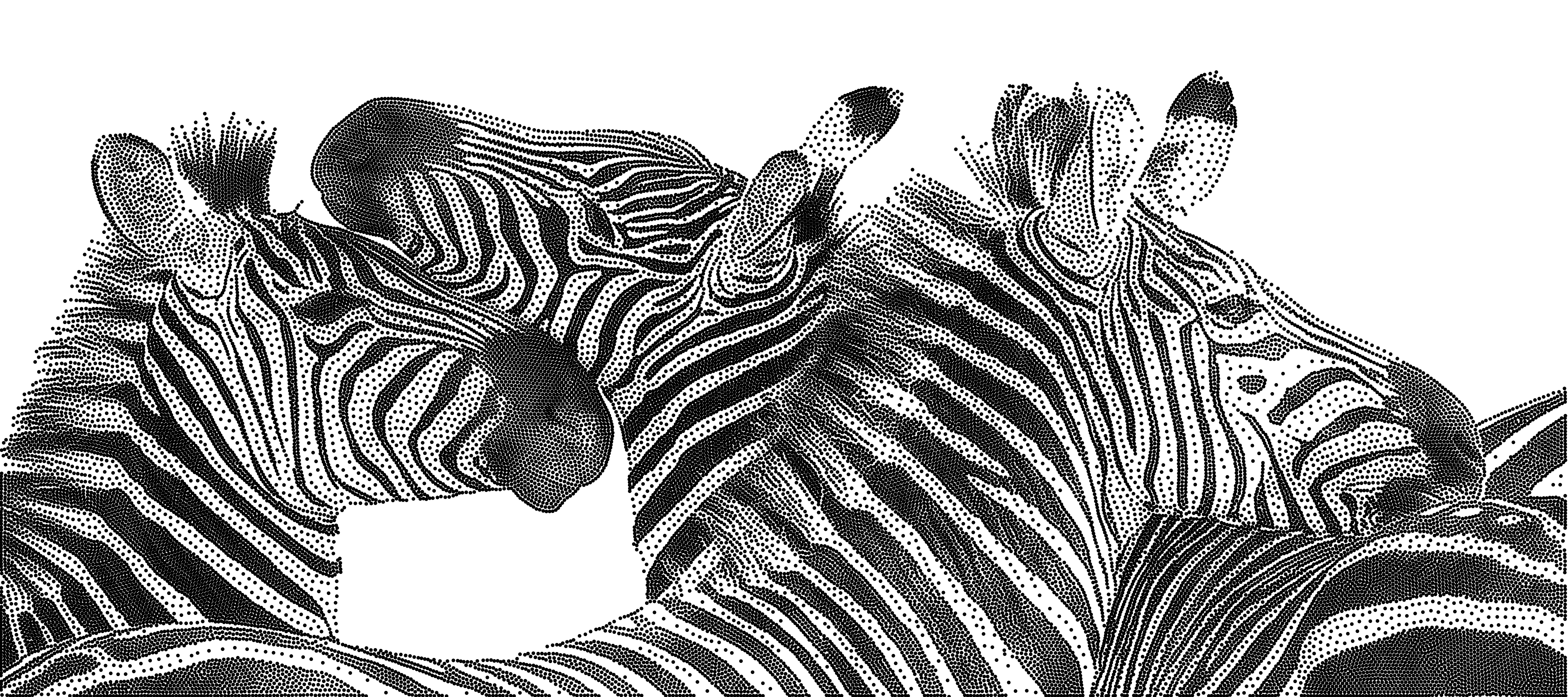}\\[2mm]%
  (b) GBN (Ours)}
  \caption{\label{fig:zebra}%
        Stippling comparison of the zebra test image \protect{\cite{Goes12Blue}}.
  }\vspace{-6pt}
\end{figure*}
%----------------------------------------

\begin{figure*}
  \setlength{\unit}{0.33\textwidth}
  {\scriptsize
  \begin{tabular*}{1\textwidth}{@{}c@{\extracolsep{\fill}}c@{\extracolsep{\fill}}c@{}}
      \centering
      \includegraphics[width=1\unit]{images/appendix/memorial-plus-lowquality.pdf}&%
      \includegraphics[width=1\unit]{images/appendix/memorial-plus-100k.pdf}&%
      \includegraphics[width=1\unit]{images/appendix/memorial-bnot.pdf}\\
      \includegraphics[width=1\unit]{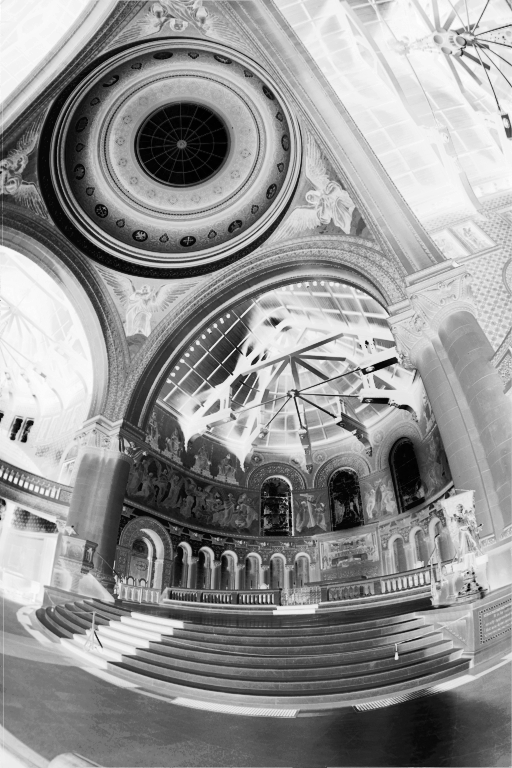}&%
      \includegraphics[width=1\unit]{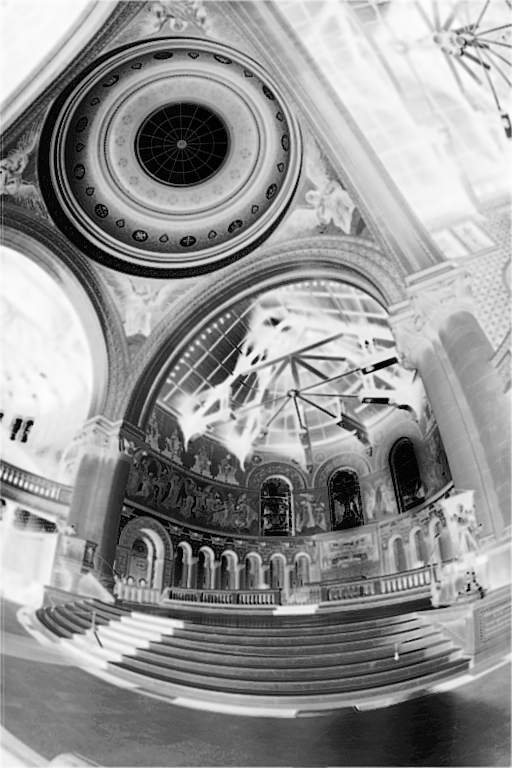}&%
      \includegraphics[width=1\unit]{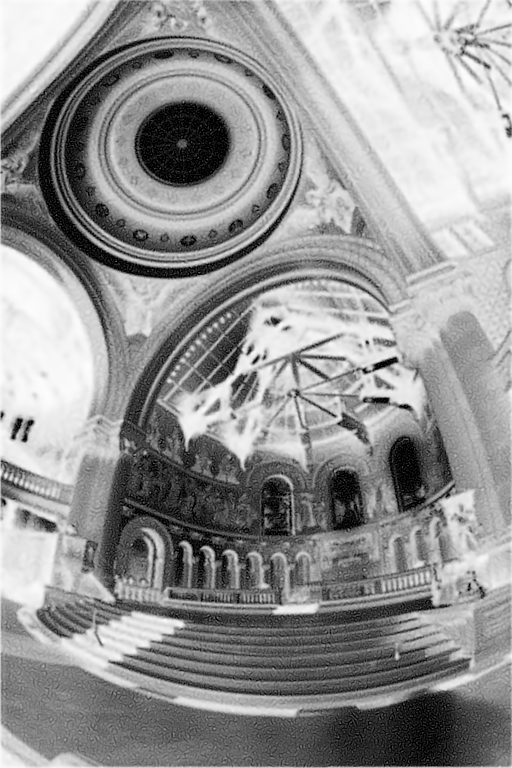}\\[1mm]
      (a) Input & (b) GBN (Ours) & (c) BNOT \protect{\cite{Goes12Blue}}
  \end{tabular*}}\vspace{-9pt}
  \caption{\label{fig:reconstruction}%
        Adaptive sampling and reconstruction comparing (b) kernel-based optimization, using our algorithm, with cellular-based approach, represented by BNOT, using 100K points and a 512$\times$768 input image.
        The reconstructed density maps are shown in the bottom row, along with (a) the input image used as a density map.
        The top-left plot shows our stippling after 300 iterations, which takes about the same time as BNOT, and already reaches a comparable or arguably better quality. In (b), we take 100k iterations to see how far it goes.
  }
\end{figure*}

\section{Additional Results}
We show more stippling results in Fig.~\ref{fig:flower} and Fig.~\ref{fig:zebra} using different methods including BNOT, KDM, and our GBN method.
In Fig.~\ref{fig:reconstruction} we show a comparison of stippling and reconstruction results between our method and BNOT.

\end{document}

%% file: images/integration_tex/gaussians-2d.tex
\definecolor{GBN}{HTML}{ef476f} %red
\definecolor{GBNstep}{HTML}{9c6644} % earth
\definecolor{Owen}{HTML}{118ab2} %blue
\definecolor{BNOT}{HTML}{ffd166} % yellow
\definecolor{Start}{HTML}{c77dff} % purple
\definecolor{random}{HTML}{06d6a0} % green
\definecolor{FPO}{HTML}{f4acb7} % pink
\definecolor{INV}{HTML}{ff8500} % orange
\definecolor{Lloyd}{HTML}{73d2de} % lightblue
\begin{tikzpicture}
	\begin{axis}[
		width = 0.5\linewidth,
		height = 0.35\linewidth,
		ymode=log,
		xmode=log,
		grid=major,
		log basis x={2},
		log basis y={10},
		xmax=2^13,
		xlabel style={at={(0.5,0.05)}},
		ylabel style={at={(0.08,0.5)}},
		xlabel={Number of Points},
		ylabel={Variance},
		cycle list name=color list,
		axis x line*=bottom,
		axis y line*=left,
		legend cell align={left},
		legend style={
		    nodes={scale=1, transform shape},
		    fill opacity=1,
		    draw opacity=1,
		    text opacity=1,
		    at={(0.05,0.15)},
		    anchor=south west
		},
	]
    \addplot[ line width=0.75pt, GBN] table { % GBN
           1 0.37980051505055367
           2 0.18820791442605251
           4 0.08920984816730026
           8 0.03929314048186586
          16 0.01476879594561977
          32 0.00374962406559671
          64 0.00035109886072057
         128 0.00000675817430478
         256 0.00000000920560836
         512 0.00000000000121953
        1024 0.00000000000006512
        2048 0.00000000000002405
        4096 0.00000000000001130
    };\addlegendentry{GBN}
    % \addplot table { % GBN latinized
    %       1 0.38496617761449495
    %       2 0.10865158100088761
    %       4 0.07308706462842759
    %       8 0.03351753460687310
    %       16 0.01468281599982292
    %       32 0.00412260286231478
    %       64 0.00059785335847180
    %      128 0.00006890989984991
    %      256 0.00001251849510281
    %      512 0.00000242025061586
    %     1024 0.00000046031048528
    %     2048 0.00000008436381544
    %     4096 0.00000001517914926
    % };
    \addplot[ line width=0.75pt, GBNstep] table { % GBN Step
           1 0.38450671582787923
           2 0.18923217879825226
           4 0.08951864346297919
           8 0.04002789236399736
          16 0.01488406180648435
          32 0.00675031351287068
          64 0.00102704555792231
         128 0.00020120231256419
         256 0.00000265331266398
         512 0.00000000925072040
        1024 0.00000000000005458
        2048 0.00000000000000005
        4096 0.00000000000000001
    };\addlegendentry{GBN step}
    \addplot[ line width=0.75pt, Owen] table {%Owen
           1 0.38380027389664377
           2 0.18982717989256131
           4 0.08900421199560527
           8 0.04011971384249912
          16 0.01727471103674178
          32 0.00665626927813117
          64 0.00206401422124975
         128 0.00052027035353444
         256 0.00011183511908724
         512 0.00002045974688213
        1024 0.00000338804696081
        2048 0.00000054137265370
        4096 0.00000008096067111
    };\addlegendentry{Owen}
    \addplot[ line width=0.75pt, BNOT] table {%BNOT
           1 0.38537144754591429
           2 0.19103337790783083
           4 0.08963432038954802
           8 0.03926359437366204
          16 0.01477568622374067
          32 0.00376825672717397
          64 0.00034512700066232
         128 0.00001267841303760
         256 0.00000087629069093
         512 0.00000009816997704
        1024 0.00000001170257819
        2048 0.00000000147494426
        4096 0.00000000018195621
    };\addlegendentry{BNOT}
    \addplot[ line width=0.75pt, Start] table { % jtr
           1 0.38387504538832244
          16 0.01855862144446574
          64 0.00258289553016326
         144 0.00064464324862183
         256 0.00022497653310241
         400 0.00009606946867718
         576 0.00004772062593671
         784 0.00002631082850601
        1024 0.00001563884535148
        1296 0.00000971054289930
        1600 0.00000640220759363
        1936 0.00000441610263947
        2304 0.00000312903244062
        2704 0.00000224870901072
        3136 0.00000169945105083
        3600 0.00000127775221763
        4096 0.00000099270387025
    };\addlegendentry{Stratified}
    \addplot[ line width=0.75pt, random] table { % white
           1 0.38203527617566891
           2 0.19445662587974630
           4 0.09538341394400718
           8 0.04813063073962488
          16 0.02433160636789287
          32 0.01215979895610718
          64 0.00603525709674345
         128 0.00299709366809383
         256 0.00151400410979802
         512 0.00075481817141586
        1024 0.00037545108394006
        2048 0.00018680782900455
        4096 0.00009297267398063
    };\addlegendentry{Random}
    \addplot[ line width=0.75pt, FPO] table { % fpo
           1 0.38556452624162609
           2 0.18832559358365973
           4 0.08963640770007223
           8 0.03874272648969787
          16 0.01496845680261888
          32 0.00371312839476909
          64 0.00036945345849574
         128 0.00004292807407153
         256 0.00001404293808889
         512 0.00000607559302313
        1024 0.00000294048103321
        2048 0.00000148663542353
        4096 0.00000081503770169
    };\addlegendentry{FPO}
    \addplot[gray, dashed, line width=0.75pt, domain=1:4096] {x^(-1)};
    \addlegendentry{$N^{-1}$}
% 	\legend{GBN, GBN Latinized, GBN Step, Owen, BNOT, Stratified, Random, FPO, $N^{-1}$}
	\end{axis}
	\node[inner sep=0, anchor=south west] at (0.35\linewidth, 0.2\linewidth) {
        \includegraphics[height=0.07\linewidth]{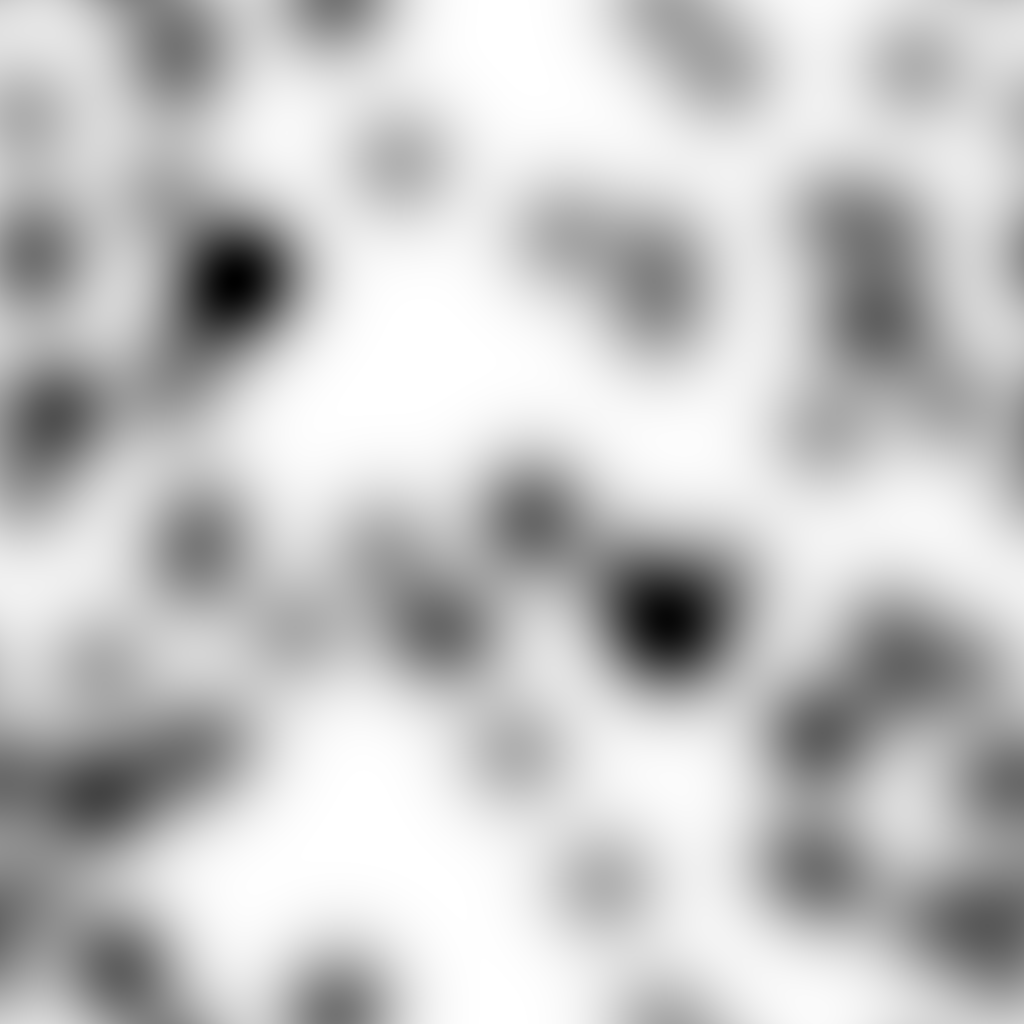}};
    \draw (0.35\linewidth,0.2\linewidth) rectangle (0.42\linewidth,0.27\linewidth);
\end{tikzpicture}

%% file: images/integration_tex/edge-2d.tex
\definecolor{GBN}{HTML}{ef476f} %red
\definecolor{GBNstep}{HTML}{9c6644} % earth
\definecolor{Owen}{HTML}{118ab2} %blue
\definecolor{BNOT}{HTML}{ffd166} % yellow
\definecolor{Start}{HTML}{c77dff} % purple
\definecolor{random}{HTML}{06d6a0} % green
\definecolor{FPO}{HTML}{f4acb7} % pink
\definecolor{INV}{HTML}{ff8500} % orange
\definecolor{Lloyd}{HTML}{73d2de} % lightblue
\begin{tikzpicture}
	\begin{axis}[
		width = 0.5\linewidth,
		height = 0.35\linewidth,
		ymode=log,
		xmode=log,
		grid=major,
		log basis x={2},
		log basis y={10},
		xmax=2^13,
		xlabel style={at={(0.5,0.05)}},
		ylabel style={at={(0.08,0.5)}},
		xlabel={Number of Points},
		ylabel={Variance},
		cycle list name=color list,
		axis x line*=bottom,
		axis y line*=left,
		legend cell align={left},
		legend style={
		    nodes={scale=1, transform shape},
		    fill opacity=1,
		    draw opacity=1,
		    text opacity=1,
		    at={(0.05,0.15)},
		    anchor=south west
		},
	]
    \addplot[ line width=0.75pt, GBN] table { % GBN
           1 0.16545109674238803
           2 0.04893332271210149
           4 0.02505248616985354
           8 0.00556761434858282
          16 0.00249777539676814
          32 0.00084023337651966
          64 0.00030506135084346
         128 0.00010863944319617
         256 0.00003835551820090
         512 0.00001349251574368
        1024 0.00000489728991141
        2048 0.00000167569918518
        4096 0.00000059554679052
    };
    \addlegendentry{GBN}
    % \addplot table { % GBN Latinized
    %       %1 0.00000000000000000
    %       2 0.02550973896819743
    %       4 0.01173946093240388
    %       8 0.00373335391197003
    %       16 0.00142357283146510
    %       32 0.00051295535869954
    %       64 0.00017996030706006
    %      128 0.00006267459926808
    %      256 0.00002203775097396
    %      512 0.00000782054476801
    %     1024 0.00000272140537942
    %     2048 0.00000093882074392
    %     4096 0.00000033835175421
    % };
    \addplot [ line width=0.75pt, Owen] table { % Owen
           1 0.16469470471472473
           2 0.04643470297123951
           4 0.01479803251700149
           8 0.00498538014991969
          16 0.00172140413373834
          32 0.00060184304518482
          64 0.00021162613634969
         128 0.00007500585956483
         256 0.00002606048663642
         512 0.00000927837573542
        1024 0.00000327012281000
        2048 0.00000115169422169
        4096 0.00000041528829070
    };
    \addlegendentry{Owen}
    \addplot [ line width=0.75pt, BNOT] table { % BNOT
           1 0.16694494294093894
           2 0.05088732376019663
           4 0.02103661093525958
           8 0.00562770728686645
          16 0.00247522226618510
          32 0.00085737833288118
          64 0.00030731053446151
         128 0.00010811373043112
         256 0.00003779305380773
         512 0.00001329580820953
        1024 0.00000464548643114
        2048 0.00000165041011361
        4096 0.00000059758803733
    };
    \addlegendentry{BNOT}
    \addplot [ line width=0.75pt, Start] table { % jittered
           1 0.16766261757252748
          16 0.00248177090378667
          64 0.00030620350706061
         144 0.00008885879385237
         256 0.00003795363634787
         400 0.00001935442311681
         576 0.00001097240228075
         784 0.00000690456372199
        1024 0.00000471992459307
        1296 0.00000329672590039
        1600 0.00000239256943380
        1936 0.00000182747919620
        2304 0.00000137870954138
        2704 0.00000108404360336
        3136 0.00000087681036151
        3600 0.00000072023291226
        4096 0.00000058494296164
    };
    \addlegendentry{Stratified}
    \addplot[ line width=0.75pt, random] table { % white
           1 0.16727305884463042
           2 0.08295965134303473
           4 0.04240194273352432
           8 0.02045362116370625
          16 0.01064931101960319
          32 0.00543317425517985
          64 0.00265214488469482
         128 0.00130998063982440
         256 0.00063775954900203
         512 0.00032597014535695
        1024 0.00015656656008021
        2048 0.00007899417920543
        4096 0.00003970529151797
    };
    \addlegendentry{Random}
    \addplot[gray, dashed, line width=0.75pt, domain=1:4096] {0.5 * x^(-1)};
    \addlegendentry{$N^{-1}$}
% 	\legend{GBN, GBN Latinized, Owen, BNOT, Stratified, Random, $N^{-1}$}
	\end{axis}
	\node[inner sep=0, anchor=south west] at (0.35\linewidth, 0.2\linewidth) {
        \includegraphics[height=0.07\linewidth]{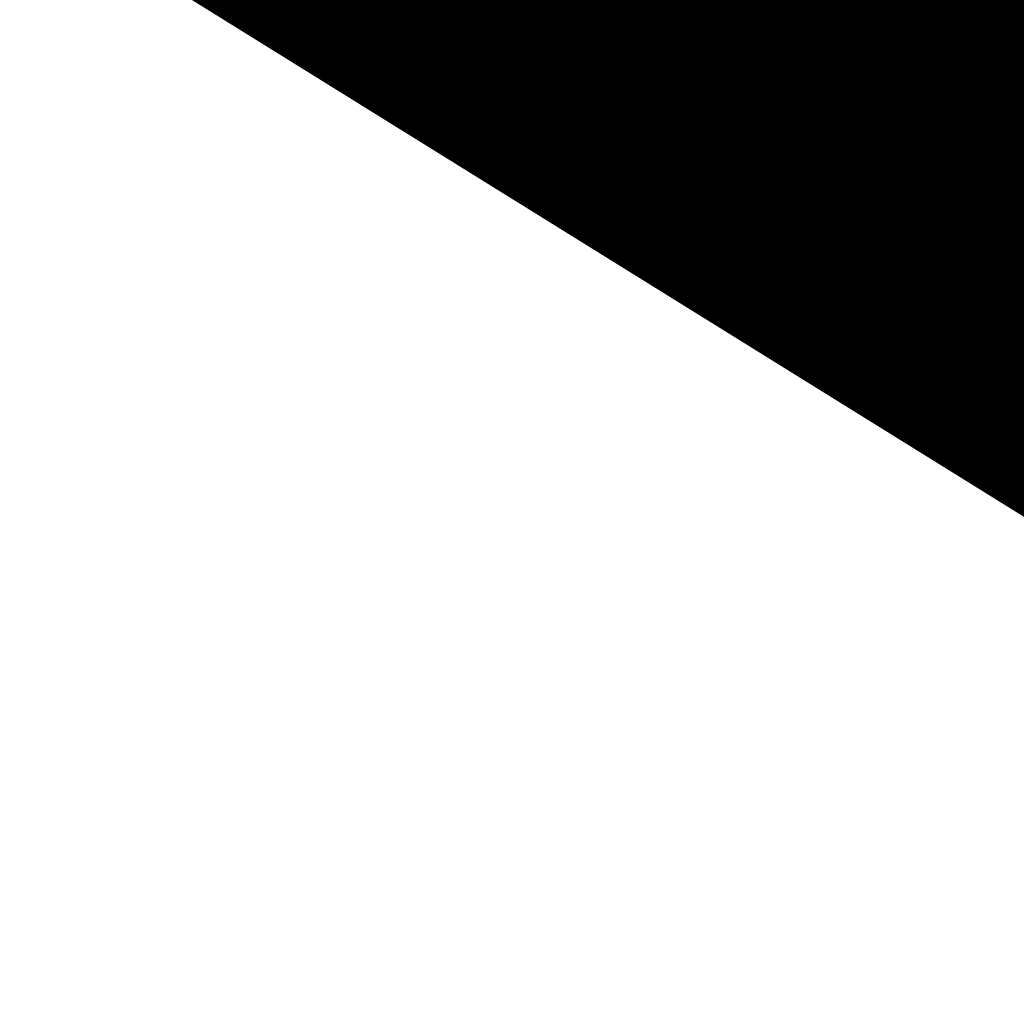}};
    \draw (0.35\linewidth,0.2\linewidth) rectangle (0.42\linewidth,0.27\linewidth);
\end{tikzpicture}

%% file: images/integration_tex/gaussians-3d.tex
\definecolor{GBN}{HTML}{ef476f} %red
\definecolor{GBNstep}{HTML}{9c6644} % earth
\definecolor{Owen}{HTML}{118ab2} %blue
\definecolor{BNOT}{HTML}{ffd166} % yellow
\definecolor{Start}{HTML}{c77dff} % purple
\definecolor{random}{HTML}{06d6a0} % green
\definecolor{FPO}{HTML}{f4acb7} % pink
\definecolor{INV}{HTML}{ff8500} % orange
\definecolor{Lloyd}{HTML}{73d2de} % lightblue
\begin{tikzpicture}
	\begin{axis}[
		width = 0.5\linewidth,
		height = 0.35\linewidth,
		ymode=log,
		xmode=log,
		grid=major,
		log basis x={2},
		log basis y={10},
		xmax=2^13,
		xlabel style={at={(0.5,0.05)}},
		ylabel style={at={(0.08,0.5)}},
		xlabel={Number of Points},
		ylabel={Variance},
		cycle list name=color list,
		axis x line*=bottom,
		axis y line*=left,
		legend cell align={left},
		legend style={
		    nodes={scale=1, transform shape},
		    fill opacity=1,
		    draw opacity=1,
		    text opacity=1,
		    at={(0.05,0.06)},
		    anchor=south west
		},
	]
    \addplot[ line width=0.75pt, GBN] table { % GBN 1.0
           8 0.03538391446695838
          16 0.00897413759276991
          32 0.00091297641417130
          64 0.00006195724995800
         128 0.00000107428979702
         256 0.00000000313976832
         512 0.00000000000173791
        1024 0.00000000000016670
        2048 0.00000000000009153
        4096 0.00000000000005996
    };\addlegendentry{GBN$_{\sigma=1.0}$}
    % \addplot table { % GBN 1.0 Latinized
    %       8 0.03961400696331527
    %       16 0.01157511000629919
    %       32 0.00176143216178690
    %       64 0.00025712823289866
    %      128 0.00004334536811348
    %      256 0.00000902717112490
    %      512 0.00000174937676206
    %     1024 0.00000035949865890
    %     2048 0.00000007174181226
    %     4096 0.00000001398109855
    % };
    \addplot[ line width=0.75pt, dashed, GBN] table { % GBN 0.7
           8 0.03421887760358428
          16 0.00547707251473760
          32 0.00037168419895960
          64 0.00000866030438451
         128 0.00000006263215945
         256 0.00000000288422473
         512 0.00000000060318439
        1024 0.00000000022148311
        2048 0.00000000009560323
        4096 0.00000000004838123
    };\addlegendentry{GBN$_{\sigma=0.7}$}
    \addplot[ line width=0.75pt, dash dot, GBN] table { % GBN 1.5
           8 0.05915429444182890
          16 0.01911290321299669
          32 0.00528660774941790
          64 0.00112669287188773
         128 0.00011613615184337
         256 0.00000401096885564
         512 0.00000003993253627
        1024 0.00000000004041783
        2048 0.00000000000000682
        4096 0.00000000000000193
    };\addlegendentry{GBN$_{\sigma=1.5}$}
    \addplot[ line width=0.75pt, Owen] table { % Owen
           8 0.05509043981068115
          16 0.01887765268694004
          32 0.00679214204739511
          64 0.00202557232131380
         128 0.00051056520467289
         256 0.00013679514158629
         512 0.00002704371463873
        1024 0.00000597773109982
        2048 0.00000101676888542
        4096 0.00000018689294821
    };\addlegendentry{Owen}
    \addplot[ line width=0.75pt, Lloyd] table { % Lloyd
           8 0.03312116719839615
          16 0.00574878209642062
          32 0.00049210832948155
          64 0.00003968695807199
         128 0.00001592020364022
         256 0.00000670043864235
         512 0.00000310434869483
        1024 0.00000158123290835
        2048 0.00000080127377372
        4096 0.00000048185389500
    };\addlegendentry{Lloyd}
    \addplot[ line width=0.75pt, Start] table { % jittered
           8 0.06783378364676642
          27 0.01375120095346194
          64 0.00405241864533993
         125 0.00141391357501886
         216 0.00060880479573870
         343 0.00029429140250736
         512 0.00015726874934486
         729 0.00008932131861623
        1000 0.00005583826395327
        1331 0.00003362946927310
        1728 0.00002152313051848
        2197 0.00001436491430588
        2744 0.00001021190910736
        3375 0.00000733478478240
        4096 0.00000524954427093
    };\addlegendentry{Stratified}
    \addplot[ line width=0.75pt, random] table { % white
           8 0.08213960043809805
          16 0.03896581136980035
          32 0.01950505007467701
          64 0.00983692236738654
         128 0.00549581425486470
         256 0.00249803261253423
         512 0.00132404192305738
        1024 0.00062596447282535
        2048 0.00031769870982499
        4096 0.00014819281598251
    };\addlegendentry{Random}
    \addplot[gray, dashed, line width=0.75pt, domain=8:4096] {1 * x^(-1)};
    \addlegendentry{$N^{-1}$}
% 	\legend{GBN$_{\sigma=1.0}$, GBN$_{\sigma=1.0}$ Latinized, GBN$_{\sigma=0.7}$, GBN$_{\sigma=1.5}$, Owen, Lloyd, Stratified, Random, $N^{-1}$}
	\end{axis}
\end{tikzpicture}

%% file: images/integration_tex/edge-3d.tex
\definecolor{GBN}{HTML}{ef476f} %red
\definecolor{GBNstep}{HTML}{9c6644} % earth
\definecolor{Owen}{HTML}{118ab2} %blue
\definecolor{BNOT}{HTML}{ffd166} % yellow
\definecolor{Start}{HTML}{c77dff} % purple
\definecolor{random}{HTML}{06d6a0} % green
\definecolor{FPO}{HTML}{f4acb7} % pink
\definecolor{INV}{HTML}{ff8500} % orange
\definecolor{Lloyd}{HTML}{73d2de} % lightblue
\begin{tikzpicture}
	\begin{axis}[
		width = 0.5\linewidth,
		height = 0.35\linewidth,
		ymode=log,
		xmode=log,
		grid=major,
		log basis x={2},
		log basis y={10},
		xmax=2^13,
		xlabel style={at={(0.5,0.05)}},
		ylabel style={at={(0.08,0.5)}},
		xlabel={Number of Points},
		ylabel={Variance},
		cycle list name=color list,
		axis x line*=bottom,
		axis y line*=left,
		legend cell align={left},
		legend style={
		    nodes={scale=1, transform shape},
		    fill opacity=1,
		    draw opacity=1,
		    text opacity=1,
		    at={(0.05,0.05)},
		    anchor=south west
		},
	]
    \addplot[ line width=0.75pt, GBN] table { % GBN 1.0
           8 0.01328532812500000
          16 0.00365780859375000
          32 0.00158046777343750
          64 0.00063173095703125
         128 0.00024718627929687
         256 0.00009838800048828
         512 0.00003947428894043
        1024 0.00001643066215515
        2048 0.00000633462619781
        4096 0.00000247709077597
    };
    \addlegendentry{GBN}
    % \addplot table { % GBN Latinized
    %       8 0.00602607812500000
    %       16 0.00249687890625000
    %       32 0.00095808984375000
    %       64 0.00037820532226563
    %      128 0.00014995587158203
    %      256 0.00006071704101563
    %      512 0.00002409267044067
    %     1024 0.00000994317245483
    %     2048 0.00000379065227509
    %     4096 0.00000150451672077
    % };
    \addplot[ line width=0.75pt, Owen] table { % Owen
           8 0.00619657812500000
          16 0.00247274218750000
          32 0.00094230664062500
          64 0.00039549121093750
         128 0.00015609899902344
         256 0.00006050201416016
         512 0.00002396483993530
        1024 0.00000959684085846
        2048 0.00000386580562592
        4096 0.00000156344765425
    };
    \addlegendentry{Owen}
    \addplot[ line width=0.75pt, Lloyd] table { % Lloyd
           8 0.01379754687500000
          16 0.00653287109375000
          32 0.00187539453125000
          64 0.00055951904296875
         128 0.00022969860839844
         256 0.00009263873291016
         512 0.00004065803146362
        1024 0.00001607214736938
        2048 0.00000609430480003
        4096 0.00000270787984133
    };
    \addlegendentry{Lloyd}
    \addplot[ line width=0.75pt, Start] table { % jittered
           8 0.01015904687500000
          27 0.00193593415637860
          64 0.00060098681640625
         125 0.00024952793600000
         216 0.00012237092764060
         343 0.00006619466378805
         512 0.00003764866638184
         729 0.00002482958221138
        1000 0.00001597613400000
        1331 0.00001080780227156
        1728 0.00000745355601370
        2197 0.00000554173948047
        2744 0.00000401996656899
        3375 0.00000311032283128
        4096 0.00000241674113274
    };
    \addlegendentry{Stratified}
    \addplot[ line width=0.75pt, random] table { % white
           8 0.02131109375000000
          16 0.01032893750000000
          32 0.00494408789062500
          64 0.00244689233398437
         128 0.00128671929931641
         256 0.00059095463562012
         512 0.00031944279098511
        1024 0.00015340117931366
        2048 0.00007945428156853
        4096 0.00004014813977480
    };
    \addlegendentry{Random}
    \addplot[gray, dashed, line width=0.75pt, domain=8:4096] {0.25 * x^(-1)};
    \addlegendentry{$N^{-1}$}
    \addplot[gray, dash dot, line width=0.75pt, domain=8:4096] {0.05 * x^(-1.32)};
    \addlegendentry{$N^{-1.32}$}
% 	\legend{GBN, GBN Latinized, Owen, Lloyd, Stratified, Random, $N^{-1}$, $N^{-1.32}$}
	\end{axis}
\end{tikzpicture}

%% file: images/integration_tex/gaussians-8d.tex
\definecolor{GBN}{HTML}{ef476f} %red
\definecolor{GBNstep}{HTML}{9c6644} % earth
\definecolor{Owen}{HTML}{118ab2} %blue
\definecolor{BNOT}{HTML}{ffd166} % yellow
\definecolor{Start}{HTML}{c77dff} % purple
\definecolor{random}{HTML}{06d6a0} % green
\definecolor{FPO}{HTML}{f4acb7} % pink
\definecolor{INV}{HTML}{ff8500} % orange
\definecolor{Lloyd}{HTML}{73d2de} % lightblue
\begin{tikzpicture}
	\begin{axis}[
		width = 0.5\linewidth,
		height = 0.35\linewidth,
		ymode=log,
		xmode=log,
		grid=major,
		log basis x={2},
		log basis y={10},
		xmax=2^13,
		xlabel style={at={(0.5,0.05)}},
		ylabel style={at={(0.08,0.5)}},
		xlabel={Number of Points},
		ylabel={Variance},
		cycle list name=color list,
		axis x line*=bottom,
		axis y line*=left,
		legend cell align={left},
		legend style={
		    nodes={scale=1, transform shape},
		    fill opacity=1,
		    draw opacity=1,
		    text opacity=1,
		    at={(0.05,0.15)},
		    anchor=south west
		},
	]
    \addplot [ line width=0.75pt, GBN]table { % gaussian 1.0
           8 0.00056768116760676
          16 0.00019012897883712
          32 0.00000660242257450
          64 0.00000006395048430
         128 0.00000002011064318
         256 0.00000000032657006
         512 0.00000000000380009
        1024 0.00000000000014889
        2048 0.00000000000001290
        4096 0.00000000000000313
    };\addlegendentry{GBN}
    % \addplot table { % gaussian latinized
    %       8 0.00056427293389813
    %       16 0.00020514021330044
    %       32 0.00001601082866909
    %       64 0.00000216165973953
    %      128 0.00000054568124416
    %      256 0.00000013670196349
    %      512 0.00000002710955325
    %     1024 0.00000000626652422
    %     2048 0.00000000143814244
    %     4096 0.00000000033011414
    % };
    \addplot[ line width=0.75pt, Owen] table { % Owen
           8 0.00078902044768004
          16 0.00036650368032738
          32 0.00010749906854762
          64 0.00001330208048284
         128 0.00000202948970596
         256 0.00000053791079328
         512 0.00000013987240006
        1024 0.00000001732439952
        2048 0.00000000357687497
        4096 0.00000000077617731
    };\addlegendentry{Owen}
    \addplot[ line width=0.75pt, random] table { % white
           8 0.27322992833852492
          16 0.12835653571950555
          32 0.07136599097049032
          64 0.03473340663179634
         128 0.01703478173210126
         256 0.00827196726138030
         512 0.00421670555161789
        1024 0.00208439362173163
        2048 0.00101768213930740
        4096 0.00054919056759758
    };\addlegendentry{Random}
    \addplot[gray, dashed, line width=0.75pt, domain=8:4096] {80*x^(-1)};
    \addlegendentry{$N^{-1}$}
% 	\legend{GBN, GBN Latinized, Owen, Random, $N^{-1}$}
	\end{axis}
\end{tikzpicture}

%% file: images/integration_tex/edge-8d.tex
\definecolor{GBN}{HTML}{ef476f} %red
\definecolor{GBNstep}{HTML}{9c6644} % earth
\definecolor{Owen}{HTML}{118ab2} %blue
\definecolor{BNOT}{HTML}{ffd166} % yellow
\definecolor{Start}{HTML}{c77dff} % purple
\definecolor{random}{HTML}{06d6a0} % green
\definecolor{FPO}{HTML}{f4acb7} % pink
\definecolor{INV}{HTML}{ff8500} % orange
\definecolor{Lloyd}{HTML}{73d2de} % lightblue
\begin{tikzpicture}
	\begin{axis}[
		width = 0.5\linewidth,
		height = 0.35\linewidth,
		ymode=log,
		xmode=log,
		grid=major,
		log basis x={2},
		log basis y={10},
		xmax=2^13,
		xlabel style={at={(0.5,0.05)}},
		ylabel style={at={(0.08,0.5)}},
		xlabel={Number of Points},
		ylabel={Variance},
		cycle list name=color list,
		axis x line*=bottom,
		axis y line*=left,
		legend cell align={left},
		legend style={
		    nodes={scale=1, transform shape},
		    fill opacity=1,
		    draw opacity=1,
		    text opacity=1,
		    at={(0.05,0.15)},
		    anchor=south west
		},
	]
    \addplot[ line width=0.75pt, GBN] table { % gaussian 1.0
           8 0.01294976562500000
          16 0.00623683007812500
          32 0.00298664726562500
          64 0.00143848613281250
         128 0.00069544210815430
         256 0.00035296681976318
         512 0.00014636217765808
        1024 0.00006746410970688
        2048 0.00003197910127640
        4096 0.00001520294991136
    };
    \addlegendentry{GBN}
    % \addplot table { % gaussian 1.0 Latinized
    %       8 0.00899102343750000
    %       16 0.00426245039062500
    %       32 0.00185780732421875
    %       64 0.00090345354003906
    %      128 0.00043540823364258
    %      256 0.00020498700256348
    %      512 0.00009975181732178
    %     1024 0.00004670519266129
    %     2048 0.00002183412561417
    %     4096 0.00001016788625121
    % };
    \addplot[ line width=0.75pt, Owen] table { % Owen
           8 0.01181556875000000
          16 0.00567934960937500
          32 0.00224603710937500
          64 0.00077741062011719
         128 0.00036608466796875
         256 0.00017050649871826
         512 0.00009320860137939
        1024 0.00003986104536057
        2048 0.00001948834369183
        4096 0.00000727913227677
    };
    \addlegendentry{Owen}
    \addplot[ line width=0.75pt, random] table { % white
           8 0.02067073437500000
          16 0.01018173554687500
          32 0.00524775048828125
          64 0.00259381953125000
         128 0.00127651981811523
         256 0.00062311420898438
         512 0.00031788607177734
        1024 0.00015652116756439
        2048 0.00007795085284710
        4096 0.00004068290759325
    };
    \addlegendentry{Random}
    \addplot[gray, dashed, line width=0.75pt, domain=8:4096] {0.2*x^(-1)};
    \addlegendentry{$N^{-1}$}
% 	\legend{GBN, GBN Latinized, Owen, Random, $N^{-1}$}
	\end{axis}
\end{tikzpicture}